\documentclass[aps,prb,twocolumn,amsmath,amssymb,superscriptaddress,floatfix,scrartcl,nofootinbib]{revtex4-1}

\usepackage{amssymb,ulem}
\usepackage{filecontents}
\usepackage{color}
\usepackage{graphicx}
\usepackage{srcltx}
\usepackage{mathrsfs}
\usepackage{relsize}
\usepackage[unicode=true,pdfusetitle,
 bookmarks=false,colorlinks=true,citecolor=blue,urlcolor=blue,linkcolor=red]{hyperref}

\usepackage{braket}
\newcommand{\R}{\mathbb{R}}
\newcommand{\Z}{\mathbb{Z}}
\newcommand{\wt}{\widetilde}

\begin{document}

\title{Bosonic topological phases of matter: bulk-boundary correspondence, SPT invariants and gauging}

\date{\today}

\author{Apoorv Tiwari}
\affiliation{Perimeter Institute for Theoretical Physics,
31 Caroline Street North, Waterloo, Ontario N2L 2Y5, Canada}
\affiliation{Department of Physics, University of Illinois at Urbana-Champaign, 1110 West Green St, Urbana IL 61801}
\author{Xiao Chen}
\affiliation{Kavli Institute of Theoretical Physics, Santa Barbara, CA 93106, USA }
\author{Ken Shiozaki}
\affiliation{Condensed Matter Theory Laboratory, RIKEN, Wako, Saitama, 351-0198, Japan}
\author{Shinsei Ryu}
\affiliation{James Franck Institute and Kadanoff Center for Theoretical Physics, University of Chicago, IL 60637, USA}

\begin{abstract}
\noindent We analyze $2+1d$ and $3+1d$ Bosonic Symmetry Protected
Topological (SPT) phases of matter protected by onsite symmetry group $G$
by using dual bulk and boundary approaches.
In the bulk we study an effective field theory which upon coupling to a
background flat $G$ gauge field furnishes a purely topological response theory.
The response action evaluated on certain manifolds, with appropriate choice of
background gauge field, defines a set of SPT topological invariants.
Further, SPTs can be gauged by summing over all isomorphism classes of flat $G$ gauge fields to obtain Dijkgraaf-Witten topological $G$ gauge theories. 
These topological gauge theories can be ungauged by first introducing and then proliferating defects that
spoils the gauge symmetry.
This mechanism is related to anyon condensation in $2+1d$ and condensing bosonic gauge charges in $3+1d$.
In the dual boundary approach, we study $1+1d$ and  $2+1d$ quantum field
theories that have $G$ 't-Hooft anomalies
that can be precisely cancelled by (the response theory of) the corresponding bulk SPT.
We show how to construct/compute topological invariants for the bulk SPTs directly from
the boundary theories.
Further we sum over boundary partition functions with different background
gauge fields to construct $G$-characters
that generate topological data for the bulk topological gauge theory.
Finally,
we study a $2+1d$ quantum field theory with
a mixed $\Z_2^{T/R} \times U(1)$ anomaly
where
$\Z_2^{T/R}$ is time-reversal/reflection symmetry, and  
the $U(1)$ could be a 0-form or 1-form symmetry depending on the choice of time reversal/reflection action.
We briefly discuss the bulk effective action and topological response for a
theory in $3+1d$ that cancels this anomaly.
This signals the existence of SPTs in $3+1d$ protected by 0,1-form $U(1)\times \Z_{2}^{T,R}$.  
\end{abstract}
\maketitle

\tableofcontents
\section{Introduction}

\noindent Over the last several years the classification and characterization of
gapped quantum phases of matter
have become an important pursuit in the field of condensed matter physics.
The rather vast landscape of gapped phases of matter can be organized according to
(i) the type of microscopic matter, i.e., fermionic or bosonic;
(ii) global symmetries which could act onsite or on spacetime indices or on
both;
(iii) gauge symmetries, i.e., manifestation of constraints or conserved charges;
(iv) entanglement patterns, i.e., broadly speaking short-ranged or long-range
entangled matter.
\cite{fradkin2013field, chiu2016classification, zeng2015quantum, senthil2015symmetry}

\bigskip \noindent A sub-class of the above gapped quantum phases that have 
gained importance due to both theoretical and experimental reasons in the recent past years
are short-range entangled phases of matter with global symmetries, also known as
symmetry protected topological phases of matter or SPTs.\cite{HasanKane10,
  QiZhangreview11} Such phases of matter cannot be connected to the trivial
product state (trivial insulator) (or to one another) by a symmetric adiabatic deformation that preserves the gap. 
Equivalence classes of Bosonic SPTs have been classified using
group cohomology \cite{chen2013symmetry}
and the equivariant cobordism group
\cite{kapustin2014symmetry,kapustin2014bosonic}.
Non-interacting fermionic phases of matter have been
classified using tools in homotopy theory \cite{schnyder2008classification,
  kitaev2009periodic}.
Interacting fermionic phases have been studied using
super group cohomology \cite{gu2014symmetry}
and spin cobordism \cite{kapustin2015fermionic, freed2016reflection}
(see also
\cite{gaiotto2016spin,wang2017towards,bhardwaj2017state,cheng2017loop})
respectively.

\paragraph{Bulk response theories, etc.}
\bigskip \noindent For this work we limit ourselves to bosonic SPT phases.
Except for Sec.\ \ref{CP_x_U(1)_SPT}, we only consider the simplest case of
phases protected by discrete abelian global symmetry $G$.
In $d+1$ dimensions, such phases of matter are classified by group cohomology.
Each distinct phase can be labelled by a group cocycle \cite{chen2013symmetry} 
\begin{align}
\omega\in H^{d+1}_{\text{group}}(G,U(1)).
\end{align}
It is expected that the low-energy and long-wavelength physics of each phase may
be captured by an invertible topological quantum field theory
(TQFT)\cite{freed2016reflection, kapustin2014symmetry}
whose Euclidean partition function we will denote by $\mathcal Z^q[N]$ where $q$ is
representative of $\omega$
and $N$ is a compact and oriented $d+1$-dimensional manifold.
A device one uses in these classification approaches is to probe the phase of
matter by coupling it to a background flat $G$ gauge field.
In the presence of background $G$ field the partition function takes the form 
\begin{align}
\mathcal Z^{q}[N,A]=e^{iI^{q}[N,A]}.
\end{align}
When the correlation length of the system is much shorter than the system size,
$I^q[N,A]$ is expected to be almost insensitive to smooth deformations of the
background
configuration $A$ and manifold $N$.
In fact in the zero correlation length limit we expect $I^{q}[N,A]$ to be a
topological term.
It is expected \cite{kapustin2014symmetry, kapustin2014symmetry} that the response theory $I^q[N,A]$
only depends
on the cobordism class of $[N,A]\in \Omega^{SO}_{d+1}(BG)$,
where $\Omega^{SO}_{d+1}(BG)$ is the oriented cobordism group.
More precisely, $(N_1,A_1)$ and $(N_2,A_2)$ are said to be cobordant if there
exists an oriented $d+2$ manifold $X$ with a $G$-bundle
that can interpolate between $(N_1,A_1)$ and $(N_2,A_2)$.
Since SPT phases are short-range entangled and symmetry preserving, they have a
unique ground state.
Consequently, the modulus of the partition function $|\mathcal Z^q[N,A]|=1$.
The topological invariants for SPTs are provided by the set $\left\{\mathcal Z^{q}[N,A]\right\}$ where $[N,A]$ are the generators of $\Omega^{SO}_{d+1}(BG)$. \cite{kapustin2014symmetry, kapustin2014bosonic, wen2015construction, shiozaki2016many, shiozaki2017matrix}

\bigskip \noindent  In addition to probing an SPT phase with a background $G$
gauge field, one could further sum over all flat $G$-fields
which is known as `orbifolding' or gauging -$G$\cite{dijkgraaf1989operator,
  ginsparg1988applied}.
Upon gauging, different SPTs map to distinct topological gauge theories
known as Dijkgraaf-Witten theories \cite{dijkgraaf1990topological}
or their spin analogues \cite{gaiotto2016spin, cheng2017loop}.
The partition function can be computed as
\begin{align}
\mathcal Z_{\text{DW}}^{q}[N]\propto \sum_{A\in H^{1}(N,G)}e^{iI^{q}[N,A]}.
\end{align}
Clearly, different $d+1$-cocycles furnish distinct Dijkgraaf-Witten theories.
These can be distinguished by the partition functions they furnish on
topologically non-trivial manifolds.
For example the manifolds generating the cobordism group described above could
be used as theoretical devices to distinguish different theories.
Alternately, it is useful to consider Dijkgraaf-Witten theory in the presence of background defects/sources such as
\begin{align}
\mathcal Z_{\text{DW}}^{q}[N,J_{qp}]\propto \sum_{A\in H^{1}(N,G)}e^{iI^{q}[N,A]+\int_{N}J_{qp}\wedge A}
\end{align}
where the quasiparticle current $J_{qp}$ is a $d-1$-form $\delta$ function supported on a closed $1$
manifold $L\subset N$ such that $\int_{N}A\wedge J_{qp}=\oint_{L}A$.
One could also introduce quasivortices `$J_{qv}$' that source $A$ in the sense
that $\oint dA=\oint J_{qv}\in G$.
Distinct Dijkgraaf-Witten theories assign different topological invariants to
linked configurations of multiple quasi-vortices.
Hence after gauging, these topological invariants may also be used to
distinguish the parent SPT phases.
\cite{levin2012braiding, wang2014braiding, lin2015loop, tiwari2016wilson, chen2016bulk}

\bigskip \noindent The $G$-symmetry can be `ungauged' within Dijkgraaf-Witten
theory by gauging a dual symmetry $\widehat{G}=\text{Rep}(G)$
which is generated by the quasiparticle configurations.
Physically this implies proliferating worldlines of quasiparticles and
destroying the gauge symmetry.
Practically ungauging involves summing over different configurations of $J_{qp}$
with an appropriate weight.
As expected, ungauging $G$ gets us back to what we had before gauging $G$ which
was a $G$-SPT labelled by `$q$':
\cite{vafa1989quantum, bhardwaj2017finite, bhardwaj2017state, kapustin2017fermionic} 
\begin{align}
\sum_{J_{qp}}\mathcal Z_{\text{DW}}^q[N,J_{qp}]e^{-i\int_{N}J_{qp}\wedge A}\propto e^{iI^{q}[N,A]}.
\end{align}

\paragraph{Anomalous boundary theories}
\bigskip \noindent 
Besides being distinguished by bulk response to flat $G$-bundles, SPTs have
interesting boundary (surface) theories.
It is known that $d$-dimensional surfaces of $d+1$-dimensional SPTs protected by
$G$ symmetry support a quantum field theory with a $G$-'t-Hooft anomaly,
\cite{sule2013symmetry, hsieh2014symmetry, hsieh2016global, kapustin2014symmetry, kapustin2014anomalous, kapustin2014anomalies, han2017boundary} 
i.e., a quantum field theory with a global $G$ symmetry that cannot be promoted
to a gauge symmetry at the quantum level \cite{hooft1980naturalness}
on an intrinsically $d$-dimensional manifold.
More precisely, let $M$ be a $d$-manifold and $A$ a flat $G$-bundle,
then the partition function of a theory with a possible 't-Hooft anomaly is non gauge-invariant
\begin{align}
Z^{q}[M,A]\neq  Z^{q}[M, A+\delta A].
\label{`t_Hooft}
\end{align} 
Here, $\delta A$ is a 0-form gauge transformation of $A$.
Usually the strategy when confronted with such ambiguities in quantum field
theory
is to look for local counter terms that make the partition function unambiguous,
i.e., to look for a functional $\mathcal L^q_{c.t.}(A)$ built from local $G$-bundle data such that 
\begin{align}
Z^q_{\text{reg}}[M,A]:= Z^q[M,A]e^{i\int_{M}\mathcal L^q_{c.t.}(A)}
\end{align} 
is gauge invariant.
For theories with 't-Hooft anomalies, no such local counter-term can be
constructed.
In fact one needs a $d+1$-manifold $N$ ($\partial N=M$) which houses the SPT to
construct a well-defined partition function
which takes the form
\begin{align}
\mathcal Z^q[N,A]=Z^q[M,A]e^{iI^{q}[N,A]}.
\end{align}
Somewhat imprecisely, we use `$A$' both for the lifted
$G$-bundle on $N$ as well
as its restriction to $\partial N=M$.

\bigskip \noindent
An alternate diagnostic of the 't-Hooft anomaly and the one we will consider in
this paper is an obstruction
to gauging or orbifolding $G$.
We will show that it is impossible to find any local gauge-invariant counterterm $\mathcal L_{c.t.}[A]$ such that 
\begin{align}
Z_{\text{orb}}[M]:=\sum_{[A]} Z_{\text{reg}}[M,A]=\sum_{[A]}Z[M,A]e^{i\int_{M}\mathcal L_{c.t.}(A)}
\nonumber 
\end{align} 
is invariant under the group of diffeomorphisms $Diff(M)$. We will in particular be interested in the large diffeomorphisms of $M$.\cite{ryu2012interacting, sule2013symmetry, hsieh2014symmetry, hsieh2016global}

\paragraph{Bulk-boundary correspondence}
\bigskip \noindent 
We note that an 't-Hooft anomaly is a strong non-perturbative constraint in the
sense that 't-Hooft anomalies are conserved along
the renormalization group flows.
Although this is a strong constraint, it by no means uniquely specifies the
surface theory on $M$.
Broadly speaking there are three distinct possibilities that can saturate the
't-Hooft anomaly.
The anomaly may be saturated by a quantum field theory that (i) spontaneously
breaks $G$ symmetry;
(ii) is gapless with a non-local action of $G$;
(iii) is gapped and supports non-trivial (fractionalized) excitations that
cannot be realized
on an intrinsically $d$ dimnensional manifold with $G$ symmetry.
\cite{vishwanath2013physics, chen2015anomalous, burnell2014exactly, wang2013boson, fidkowski2015realizing}

\bigskip \noindent Using the anomaly matching criteria, once we establish that a
certain quantum field theory
with partition function $Z^q[\partial N=M]$ is a suitable candidate for the
surface/edge theory for an SPT $\mathcal Z^{q}[M]$,
we proceed to explore the bulk-boundary correspondence.
We do so in two related but distinct ways.
(i) We construct SPT topological invariants directly from a surface/edge
computation using the recently studied\cite{shiozaki2016many}
cut and glue approach,
and (ii) We construct topological data corresponding to the Dijkgraaf-Witten
topological gauge theory directly from the surface/edge theories.
The latter is done by first constructing twisted partition functions $Z^q[M,A]$ and
then summing them up into $G$-invariant characters
that are representative of bulk excitations.
These methods have been well known for $2+1d$ topological phases and their
$1+1d$ boundaries
\cite{witten1989quantum, dijkgraaf1991quasi, wen1992theory, hatsugai1993chern,
  cappelli2002thermal, cappelli1997modular, cappelli2010chiral,
  cappelli2011partition}
and were recently generalized to $3+1d$ topological phases and their $2+1d$
surfaces
\cite{chen2016bulk, chen2017gauging}.
Here we provide a procedure to construct such $G$-characters by directly
implementing cohomology twists instead of explicitly computing twisted partition
functions.
$G$-characters are defined in such a way that they transform projectively under large
diffeomorphisms (modular transformations) of $M$,
and the projective phases encode the relevant topological data. 

\bigskip \noindent  Finally we switch directions and consider a bosonic quantum
field theory
in $2+1$-dimensions with $\Z_2^{T,R} \times U(1)_p$ symmetry. 
Here $\Z_2^{T,R}$ refers to time reversal or $\Z_2$-reflection symmetry and by $U(1)_p$ we mean a $p$-form $U(1)$ symmetry that may be gauged by
coupling to $p+1$-form flat $U(1)$ gauge field.
We specifically consider the cases $p=0,1$ and show that for certain action of
$\Z_2^{T,R} \times U(1)_p$,
there is an 't-Hooft anomaly that can be cancelled by a $3+1d$ invertible topological field theory.
This signals the existence of bosonic SPTs in
$3+1$-dimensions protected by $\Z_2^{T,R} \times U(1)_p$.
We propose bulk candidate effective field theories for these phases of matter.

\subsection{Plan for the paper}
\noindent Before getting into the details, let us briefly describe the plan for the rest of the paper.

\bigskip \noindent
In Sec.\ \ref{2+1d} and \ref{3+1d},
we study bosonic topological phases of matter with global discrete abelian
symmetry $G$ in $2+1$ and $3+1$-dimensions respectively.
We study these phases and their gauged versions by analyzing the bulk directly
and from a complimentary viewpoint,
by analyzing their gapless boundary theories.
In Sec.\ \ref{generalities},
we briefly comment on how this generalizes to $d+1$-dimensions.
\begin{center}
		\small{\bf Bulk analysis}\\
	\end{center}
We begin with an invertible TQFT that can describe bosonic $G$-SPT phases with
topologically distinct realizations of $G$ symmetry
labelled by `$q$'. We carry out the following steps: 
\begin{itemize}
\item Couple to a background $G$ gauge field $A$ on a closed, oriented $d+1$-dimensional manifold to compute distinct topological response theories 
\begin{align}
\mathcal Z^q[N,A]=e^{iI^{q}[N,A]}.
\end{align}
\item In general $I^{q}[N,A]\in \R/2\pi\Z$ and the set
  $\left\{e^{iI^{q}[N,A]}\right\}$ of $U(1)$ phases for all $[N,A]$ that generate $\Omega_{d+1}(BG)$
  form the set of SPT topological invariants, i.e., they differentiate different
  SPT phases.
  For a discrete abelian group $G$ which is always isomorphic to
  $\prod_{i=1}^{k}\Z_{n_i}$,
  the topological invariants turn out to be a combination of
  partition functions on
  lens spaces and three-torus with appropriate flat $G$ bundles in
  $2+1$-dimensions
  and (lens space $\times$ a one-sphere) and the four-torus with appropriate $G$-bundles in $3+1$-dimensions. We compute these topological invariants. 
\item Gauge $G$ by summing over flat $G$ bundles to obtain the partition
  function for a $G$-topological gauge theory, i.e., Dijkgraaf-Witten theory.
\item Introduce quasi-particle sources within Dijkgraaf-Witten theory that generate a dual symmetry $\hat{G}$ and finally ungauge $G$ by gauging $\hat{G}$ to return to the SPT phase.
\end{itemize}
\begin{center}
		\small{\bf Boundary analysis}\\
	\end{center}
To compliment the bulk analysis we study a class of simple models that describe
possible edges/surfaces for $G$-bosonic SPTs.
The following computations verify this fact:
\begin{itemize}
\item We couple the boundary theory to a background $G$ gauge field and compute `twisted partition functions' $Z^q[M,A]$.
\item Take the aforementioned approach and try to gauge $G$. We treat
  gauge-ability of $G$ as a diagnostic for a trivial/non-trivial bulk and  show that the 't-Hooft anomaly matches with the gauge anomaly of the SPT
  response theory on an open $d+1$ manifold
  confirming that this model indeed describes the surface of an SPT.
\item Once it is established that the theory describes the boundary of an
  SPT,
  the SPT invariants can be constructed directly from the surface theory
  following a cut and glue construction
  whose calculation essentially restricts to the boundary theory computation.
\item Furthermore $G$-orbifold characters can be constructed from
  the `twisted partition functions'.
 Modular transformations of these characters reproduce the topological data corresponding to the bulk topological gauge theory obtained by gauging the bulk SPT.
\end{itemize}
\begin{center}
		\small{\bf SPT protected by $\Z_2^{T,R}\times U(1)$ symmetry in $3+1d$ }\\
	\end{center}
In Sec.\ \ref{CP_x_U(1)_SPT} we study surface theory for $3+1d$ SPTs protected
by $\Z_2^{T,R} \times U(1)_p$ for the case $p=0,1$.
We show that under certain action of $\Z_2^{T,R}\times U(1)_p$, there is a 't-Hooft
anomaly on the surface.
We construct bulk effective field theories corresponding to such phases.

\begin{center}
		{\bf Notations }\\
	\end{center}
  Before getting to the main text we briefly summarize the notations we use.
  We will be working with topological phases on a $d+1$-dimensional bulk
  manifold $N$ which is always compact and oriented.
  When we discuss purely bulk physics then we often consider $N$ to be closed.
  However when we consider edge/surface physics we consider $N$ to be an open
  manifold such that $\partial N=M$.
  We will denote background $G$-gauge fields by $A$.
  These may be both in the bulk or on the boundary.
  When mentioned (for e.g., during the gauging procedure) we will promote $A$ to
  be dynamical.
  By $\widehat{G}$, we imply the group Pontrjagin dual to $G$, i.e.,
  $\widehat{G}=\left\{\mu: G\to U(1)\right\}$.
  For discrete abelian groups, $\widehat{G}\simeq G$. 
 
\bigskip 
\begin{tabular}{ |p{1.7cm}||p{6cm}|  }
 \hline
 \multicolumn{2}{|c|}{Bulk notations} \\
 \hline
 Notation & Description and comments\\
 \hline \hline
 $\mathcal Z^{q}[N,A]$   & SPT partition function on $N$ with background $G$ bundle $A$. `$q$' labels a $d+1$-cocycle $\omega\in H^{d+1}_{\text{group}}(G,U(1))$.\\
 \hline
 $I^{q}[N,A]$ & SPT response theory.  \\
 \hline
 $\mathcal Z_{\text{DW}}^{q}[N]$ & Dijkgraaf-Witten partition function for $q\in H^{d+1}_{\text{group}}(G,U(1))$ obtained by gauging $q$-SPT. \\ 
 \hline
\end{tabular}

 \bigskip 
 \begin{tabular}{ |p{1.7cm}||p{6cm}|  }
 \hline 
 \multicolumn{2}{|c|}{Boundary notations} \\
 \hline
 Notation & Description and comments\\
 \hline \hline
 $Z^q[M,A]$   & Partition function for QFT describing surface of $q$-SPT on $d$-manifold $M$ in the presence of background $G$ bundle $A$.\\ \hline
 $Z^{q,\epsilon}[M,A]$&  Partition function with discrete torsion phase $\epsilon\in H^{d}(G,U(1))$. Physically $\epsilon$ labels a $d$-dimensional -SPT. \\ \hline
 $Z_{\text{orb}}^{q}[M]$ & Partition function obtained by starting from $Z^q[M,A]$ and orbifolding-$G$.\\ \hline
 $\chi^{q}_{\mu,\lambda_1,\ldots,\lambda_{d-1}}$    & Orbifold characters constructed by summing twisted partition functions $Z^q[M,A]$. These can be used to compute topological data for bulk Dijkgraaf-Witten theory.\\
 \hline
\end{tabular}

\section{$2+1d$ topological phases and their $1+1d$ edges}
\label{2+1d}
\subsection{Bulk physics}
\noindent {\bf SPT effective field theories:}
It is known that SPTs with unitary onsite symmetry can be modeled by $BF$
theories with distinct symmetry actions.
\cite{kapustin2014anomalies, lu2012theory, ye2016topological}
For example $G=\Z_n^k$-SPTs in $2+1d$ may be modeled by $k$-copies of $BF$ theory at `level' one:  
\begin{align}
\mathcal S[a,b]=\int_{N}\sum_{I,J=1}^{k}\frac{\delta_{IJ}}{2\pi}b^{I}\wedge da^{J}+ \cdots, 
\end{align}
where $a^{I}$ and $b^{I}$ are $U(1)$-connections subject to the flux
quantization conditions
$\oint_S da, \oint_S d b \in 2 \pi \Z$ for $S \in Z_2(N;\Z)$.
By `$\cdots$' we imply other non-topological symmetry preserving terms that we
ignore in the limit of zero correlation length.
This theory is trivial, in the sense its partition function $\mathcal
Z[N]=1$\cite{witten2003sl}
on any closed 3-manifold $N$.
However it can be coupled to a flat background $G$ gauge field $A^{I}$ in
topologically distinct ways
which correspond to various SPT actions
\begin{align}
\mathcal S^q[a,b,A]=\int_{N} 
\frac{\delta_{IJ}}{2\pi}b^{I}\wedge da^{J}+\mathcal S^{q}_{\text{cpl}}[a,b,A].
\label{2+1_spt_action}
\end{align} 
Here,
$\mathcal{S}^{q}_{\text{cpl}}[a,b,A]$ is the part of the action involving
coupling to sources $A^{I}$.
 Flat $G$ gauge fields are characterized by their holonomies, or
equivalently, $A\in H^{1}(N,G)$.
$G$-SPTs are classified by group cohomology and can be labelled by a 3-cocycle
$\omega\in H_{\text{group}}^{3}(G,U(1))$.
Here `$q$' is meant to be a representative of $\omega$. When $G=\left(\mathbb Z_{n}\right)^k$ there are three classes of 3-cocycles 
\begin{align}
    \omega_{\text{type-I}}({\bf{a}},{\bf{b}},{\bf{c}}) &=\exp{\left\{\frac{2\pi iq_{I}}{n^2}a^{I}\left(b^{I}+c^{I}-[b^{I}+c^{I}]\right)\right\}},
    \nonumber \\
    \omega_{\text{type-II}}({\bf{a}},{\bf{b}},{\bf{c}}) &=\exp{\left\{\frac{2\pi iq_{IJ}}{n^2}a^{I}\left(b^{J}+c^{J}-[b^{J}+c^{J}]\right)\right\}},
    \nonumber \\
    \omega_{\text{type-III}}({\bf{a}},{\bf{b}},{\bf{c}}) &=\exp{\left\{\frac{2\pi iq_{IJK}}{n}a^Ib^Jc^K \right\}},
    \label{2+1_cocycles}
\end{align}
where ${\bf{a}}=(a^1,a^2,\ldots,a^k)$, etc., ${\bf{a}},{\bf{b}},{\bf{c}}\in \Z_n^k$
and $[a^{I}+b^{I}]:= a^{I} +b^{I} \ \text{mod} \ n$.
These different families of cocycles are called type-I,II,III respectively \cite{propitius1995topological}. The parameters $q_{I},q_{IJ},q_{IJK}$ take values in $\Z \ \text{mod} \ n\Z$, hence  
\begin{align}
H^{3}_{\text{group}}[(\mathbb Z_n)^k,U(1)]
=&\; (\mathbb Z_{n})^{\left[
\left( \begin{array}{c}
k \\
1  \end{array} \right)
+
\left( \begin{array}{c}
k \\
2  \end{array} \right)
+
\left( \begin{array}{c}
k \\
3  \end{array} \right)
\right]}.
\end{align}
Any $G$ SPT is prescribed by the set of $\Z_n$ parameter $q=\left\{q_{I},q_{IJ},q_{IJK}\right\}\in H_{\text{group}}^{3}(G,U(1))$. Different coupling terms corresponding to different families of 3-cocycles take the form
\begin{align}
  \mathcal S_{\text{cpl}}^{q_{I}}[a,b,A]&= -\frac{1}{2\pi}\int_{N}A^{I}\wedge (db^{I}+q_{I}da^{I}),
                                          \nonumber \\
  \mathcal S_{\text{cpl}}^{q_{IJ}}[a,b,A]&= -\frac{1}{2\pi}\int_{N}A^{I}\wedge (db^{I}+q_{IJ}da^{J}),
                                           \nonumber \\
  \mathcal S_{\text{cpl}}^{q_{IJK}}[a,b,A]&=-\frac{1}{2\pi}\int_{N}A^{I}\wedge (db^{I}+\frac{n^2q_{IJK}}{2\pi}a^{J}\wedge a^{K}), 
\end{align}
where $I,J,K$ are not summed over. Integrating over $a^{I},b^{I}$ one obtains a response theory in terms of background $G$-bundle:
\begin{align}
  \mathcal Z^q[N,A]=&\; \int \mathcal D[a,b]e^{i \mathcal{S}^q[a,b,N,A]}=:e^{iI^{q}[N,A]}.
\label{type_response}
\end{align} 
The response theories $I^{q}[N,A]$ take the form  
\begin{align}
  I^{q_I}[N,A]&= -\frac{q_{I}}{2\pi}\int_{N}A^{I}\wedge dA^{I},
                 \nonumber \\
  I^{q_{IJ}}[N,A]&=-\frac{q_{IJ}}{2\pi}\int_{N}A^{I}\wedge dA^{J},
                    \nonumber \\
I^{q_{IJK}}[N,A]&= -\frac{q_{IJK}n^2}{4\pi^2}\int_{N}A^{I}\wedge A^{J} \wedge A^{K}. 
\label{type_I_II_III_response}
\end{align} 
The relation between SPT response theories \eqref{type_I_II_III_response}
and the respective cocycles \eqref{2+1_cocycles} can be seen most clearly within
a simplicial construction.
(See App.\ \ref{relation}.) 

\bigskip\noindent {\bf Topological invariants for SPTs:}
SPT topological invariants are a set of $U(1)$-valued quantities that
can distinguish different phases.
These are supplied by the partition functions $\mathcal
Z^{q}[N,A]$ which are pure $U(1)$ phases $e^{iI^{q}[N,A]}$.
Here, $[N,A]$ are the set of generators of
$\Omega_{3}^{SO}(BG)$,
the oriented equivariant cobordism group over the classifying space of $G$.
For $G=\Z_n^{k}$, we will confirm that the lens space $L(n,1)$ and three-torus
$T^3$
with appropriate flat $G$-bundles are sufficient to detect and classify $G$-SPTs.
Let us compute the partition functions on these manifolds. 

\begin{itemize}
\item{\bf{Type-I and type-II cocycles:}}
  SPTs with type-I and type-II symmetry action
  can be distinguished by their partition functions
  on lens space ($L(n,1)$) with an appropriate background $G$-bundle.
  The topology of Lens space is captured by the torsion part
  of its homology groups
\begin{align}
H_{1}(L(n,1),\Z)=H^{2}(L(n,1),\Z)=\Z_n.
\end{align}
Then $[A]\in \text{Tor}(H^{2}(L(n,1),\Z))$. The Chern-Simons term which appears in the type-I response theory $I^{q_{I}}[N,A]$ evaluates to 
\begin{align}
e^{I^{q_{I}}[L(n,1),[A]]}=&\;\exp\left\{\frac{-iq_{I}}{2\pi}\int_{L(n,1)}A^{I}\wedge dA^{I}\right\}\nonumber \\
 =&\;\exp\left\{-iq_I\oint_{C_{A}}A\right\}\nonumber \\
=&\; \exp\left\{-iq_Ia_I\oint_{C_{1}}A\right\}\nonumber \\
 =&\; \exp\left\{-\frac{2\pi i q_Ia_I^{2}}{n}\right\}, 
\label{lens_space_type_1}
\end{align}
where $C_{A}\in H_{1}(L(n,1),\Z)$ is Poincare dual to $[A]\in\text{Tor}\left(H^{2}(L(n,1),\Z)\right)$.  Further we have chosen the configuration $[A]$ such that $C_{A}=a_IC_1$ where $C_{1}$ is the generator of $H_{1}(L(n,1),\Z)$. Hence the SPT invariant is
\begin{align}
e^{-iI^{q_I}[L(n,1),[A]]}=e^{\frac{2\pi iq_I a_I^2}{n}}. 
\end{align}
The SPT invariant with type-II response theory \eqref{type_I_II_III_response} can be computed similarly. 
\begin{align}
e^{iI^{q_{IJ}}[L(n,1),[A]]}=&\;\exp\left\{-\frac{iq_{IJ}}{2\pi}\int_{L(n,1)}A^{I}\wedge dA^{J}\right\}\nonumber \\
 =&\;\exp\left\{-iq_{IJ}\oint_{C_{A}}A^I\right\}\nonumber \\
=&\; \exp\left\{-iq_{IJ}a_J\oint_{C_{1}}A^I\right\}\nonumber \\
 =&\; \exp\left\{-\frac{2\pi i q_{IJ}a_Ia_{J}}{n}\right\}. 
\label{type_2}
\end{align}
\item {\bf{Type-III cocycles:}} SPTs with type-III response theories can be detected on $T^3$ with a background $G$ bundle
\begin{align}
e^{iI^{q_{IJK}}[T^3,A]}=&\; \exp{\left\{ -\frac{in^2q_{IJK}}{4\pi^2}\int_{ T^3}A^{I}\wedge A^J \wedge A^K\right\}}  \nonumber \\
=&\; \exp{\left\{-\frac{2\pi i q_{IJK}}{n}\epsilon^{ijk}a_{I,i}b_{J,j}c_{K,k}\right\}}
\label{torus_type_3}
\end{align}
where ${\bf{a}}_I=(a_{I,1},a_{I,2},a_{I,3})$
are the holonomies around the three cycles of $T^3$.
\end{itemize}
Summarizing, the complete set of invariants for bosonic SPTs protected by $G=\Z_n^k$ are
\begin{align}
&\;\left\{e^{-iI^{q}[L(n,1),A]},e^{-iI^{q}[T^3,A]}\right\}\nonumber \\=&\;\left\{e^{\frac{2\pi i}{n}(q_{I}a_{I^2}+q_{IJ}a_Ia_{J})} 
, e^{2\pi i \frac{q_{IJK}}{n}\epsilon^{ijk}a_{I,i}b_{J,j}c_{K,k}}\right\}
\end{align}
\noindent
More generally, if $G=\prod_{I=1}^{k}\Z_{n_I}$, then the 
SPTs classified by parameters $\left\{q_{I},q_{IJ},q_{IJK}\right\}$ parametrizing type-I,II,III kind of responses respectively can be detected on $\left\{L(n_{I},1),L(\text{gcd}(n_{I},n_{J}),1),T^{3}\right\}$ respectively.\cite{wen2014symmetry, wang2015field, tantivasadakarn2017dimensional}

\bigskip \noindent {\bf Topological gauge theories from gauging SPTs :}
Gauging of SPTs can be carried out by first computing the response to flat
$G$-bundles \eqref{type_response}
and then summing over all flat bundles with the appropriate normalization. By this procedure, one obtains the well known Dijkgraaf-Witten topological gauge theory labelled by $q\in H^{3}_{\text{group}}(G,\R/2\pi \Z)$:
\begin{align}
  \mathcal Z_{\text{DW}}^{q}[N]&=
                                 \frac{1}{|H^{0}(N,G)|}\sum_{A\in H^{1}(N,G)}e^{iI^q[N,A]}
                                 \nonumber \\
&= \frac{1}{n^k}\int \prod_{I=1}^{k}\mathcal D[A^I,B^I]e^{i\int_{N}\frac{n\delta_{IJ}}{2\pi}B^{I}\wedge dA^{J}+iI^{q}[N,A]
}
\label{gauging_2+1_spt}
\end{align}
where in the second line we have specialized to $G=\Z_n^{k}$ and written the gauged SPT action in the familiar
continuum form as a `twisted' multicomponent $BF$ theory. $A^{I},B^{I}$ are
1-form $U(1)$ connections.
Integrating over $B^{I}$ imposes that $A$ is
a flat $G$-bundle and takes us back to the original expression.
Since $(1/2\pi) dB^{I}$ is a 2-form with integral periods we can write 
\begin{align}
\frac{1}{2\pi}dB^{I}=d\beta^{I} + \sum_{j\in \text{Free}(H^{2}(N,\Z))}m^{I}_j \lambda_{j}
\end{align} 
where $m_j\in \Z$ and $\lambda_j$ is a basis on the space of integral harmonic 2-forms. Then, integrating over $B^I$, we get
\begin{align}
  \mathcal Z^{q}_{\text{DW}}&=
                              \frac{1}{n^k}\int\prod_{I=1}^{k}\mathcal D[A^{I},\beta^{I}] e^{\frac{in\delta_{IJ}}{2\pi}\int_{N}\beta^{I}\wedge F^{J}_{A}}
                              \nonumber \\ 
                            & \times \prod_{j}\left[\sum_{m^I_j\in\Z}e^{in\delta_{IJ}m^I_{j}\int \lambda_j\wedge A^J}\right]e^{iI^q[N,A] }
                              \nonumber \\
 & = \frac{1}{n^k}\int \prod_{I=1}^{k}\mathcal D[A^{I}] \delta(nF^I_A) \prod_{j}\left[\sum_{m_j^I\in\Z}e^{inm^I_{j}\int \lambda_j\wedge A^I}\right]
   e^{iI^q[A,N] }
   \nonumber \\
                            &= \frac{1}{n^k}\int
  \prod_{I=1}^{k}                            \mathcal D[A^{I}]\delta(nF^I_{A})\delta(\oint_{L_j} A^I\in \frac{2\pi}{n}\Z) e^{iI^q[N,A] }
                              \nonumber \\
&= \frac{1}{n^k}\sum_{A\in H^{1}(N,\Z_n^k)}e^{iI^{q}[N,A]}. 
\label{integrating_B}
\end{align} 
The sum over $\beta^{I}$ fixes $nF^{I}_{A}=0$ which implies that $F^{I}_{A}=0$ unless $\text{Tor}(H^{2}(N,\Z))\neq 0$. The sum over $m_{j}$ sets holonomy of $A^I$ to be a multiple of $2\pi /n$ along $L_j$ the 1-cycle poincare dual to $\lambda_j$. In other words $[A]\in H^{1}(M,\Z_n^k)$, a flat $\Z_n^k$-gauge field. Let us take a look at few examples:
\begin{itemize}
\item {\bf{Type-I and type-II cocycles:}}
  Consider a 3-manifold $N$ with vanishing torsion. Then since $dA^{I}=0$, we
  get $I^{q_I}[N,A]=I^{q_{IJ}}[N,A]=0$.
 Therefore 
\begin{align}
  \mathcal Z^{q}_{\text{DW}}[N]&=
                                 \frac{1}{|G|}
                                 \sum_{[A]\in H^{1}(N,G)}1 \nonumber \\
&= |G|^{b_1(N)-1}, 
\label{torus_type_1}
\end{align}
where $b_1(N)$ refers to the 1st Betti number of $N$. If $N=S^1\times M$, the partition function evaluates to 
\begin{align}
\mathcal Z^{q}_{\text{DW}}[M\times S^1]\equiv\text{GSD}[M]=|G|^{b_1(M)}
\end{align}
where $\text{GSD}[M]$ denotes the groundstate degeneracy on $M$.
Similarly, the gauged partition function for type-I and type-II cocycle on for $G=\Z_{n}$ and $G=\Z_{n}^2$ respectively  can be evaluated on $L(n,1)$ using \eqref{lens_space_type_1} and \eqref{type_2}
\begin{align}
  \mathcal Z^{q_I}_{\text{DW}}[L(n,1)]=&\;\frac{1}{n}\sum_{a_{I}=0}^{n-1}e^{\frac{2\pi i q_Ia_I^2}{n}},
                                         \nonumber \\
\mathcal Z^{q_{IJ}}_{\text{DW}}[L(n,1)]=&\;\frac{1}{n^2}\sum_{a_{I},a_{J}=0}^{n-1}e^{\frac{2\pi i q_{IJ}a_Ia_{J}}{n}},
\end{align}
which vanish if $(n,q_{I})$ or $(n,q_{IJ})$ are coprime respectively.
\item {\bf{Type-III cocycles:}}
  The partition function on $T^3$ for type-III cocycle can be computed using \eqref{torus_type_3}
 \begin{align}
\mathcal Z_{\text{DW}}^{q}[T^3] 
=&\;\frac{1}{|G|}\sum_{{\bf{a}},{\bf{b}},{\bf{c}}\in \Z_n^3}e^{\frac{2\pi i q_{IJK}}{n}\epsilon^{ijk}a_{I,i}b_{J,j}c_{K,k}} \nonumber \\
=:&\; \text{GSD}[T^2] < |G|^2
\label{type3_partn_fn}
\end{align}
For $G=\Z_2^3$, $q_{123}=1$, \eqref{type3_partn_fn} evaluates to $\mathcal Z^q_{\mathrm{DW}}[T^3]=22=\text{GSD}[T^2]$\cite{propitius1995topological}. Groundstates on a torus can be labelled by the spectrum of Wilson operators in a topological gauge theory, therefore this implies that there are 22 independent Wilson operators. The total quantum dimension is the same for different Dijkgraaf-Witten theories corresponding to the same $G$, hence we obtain
\begin{align}
|G|^2=\sum_{i=1}^{\text{GSD}[T^2]}d_i^2.
\end{align}
If $\text{GSD}[T^2]< |G|^2$ there must be at least a single Wilson operator with quantum dimension greater than 1. This is a way to see that type-III theory has non-abelian excitations even though $G$ is an abelian group \cite{propitius1995topological}. A dual approach based on analyzing Wilson operators directly in the continuum theory may also be used to compute this groundstate degeneracy.\cite{he2017field}  
\end{itemize}

\begin{figure}[bt]
\centering
\includegraphics[scale=0.30]{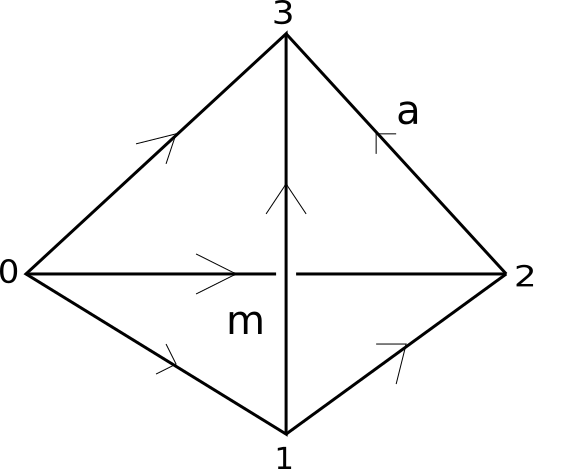}
\\
\caption{
Triangulation of a three-torus containing one 0-simplex, three 1-simplices, three 2-simplices and six 3-simplices.  
}
\label{pairing_fig}
\end{figure}

\noindent {\bf{Ungauging and anyon condensation:}} Let us consider the continuum formulation of Dijkgraaf-Witten theory \eqref{gauging_2+1_spt} in the presence of quasiparticle sources $J_{qp}$
\begin{align}
\mathcal Z_{\text{DW}}^{q}[N,J_{qp}]=&\;\frac{1}{|H^{0}(N,\Z_n^k)|}\sum_{A\in H^{1}(N,G)}e^{iI^q(N,A)+i\int_N J^I_{qp}\cup A^I} \nonumber \\
=&\; \frac{1}{n^k}\int \prod_{I=1}^{k}\mathcal D[A^I,B^I]\exp\left\{\int_{N}\frac{in}{2\pi}B^{I}\wedge dA^{I} \right. \nonumber \\
&\; \left. +iI^{q}[N,A]+i\int_{N}J^I_{qp}\wedge A^I
\right\}
\end{align}
where the background fields $J^{I}_{qp}$ are 2-form fields with integral periods \cite{bauer2005class}. Since $\oint A^{I}\in \left(2\pi \Z\right)/n$, the periods of $J_{qp}$ only make sense modulo $n$, more precisely $J_{qp}\in H^{2}(N,\widehat{G})$ where $\widehat{G}=\text{Rep}(G)\simeq G$. There is a perfect pairing 
\begin{align}
\int_N: H^{1}(N,G)\times H^{2}(N,\widehat{G})\to \R/2\pi \Z
\label{pairing}
\end{align}
that is realized by wedge product followed by integration. For a simplicial
definition, consider a 3-simplex as in
Fig.\ {\ref{pairing_fig}} $\int_{\Delta}J_{qp}\cup A= J_{qp}[012](A[23])=m(a)=\frac{2\pi ma}{n}$. 

\noindent
$J_{qp}$ generates a 1-form $\widehat{G}$ symmetry.  To see this, we follow the procedure standard in Hamiltonian quantization of gauge systems. Let $N=M\times S^{1}$. We define a charge operator $\mathcal Q^{I}(\lambda^{I})$ corresponding to $\widehat{G}$ symmetry 
\begin{align}
\delta_{J^{I}_{qp}}S=\int_{N}\delta J^I_{qp}\wedge A^I
\Rightarrow \mathcal Q^I(\lambda^I):=\frac{1}{2\pi}\int_M\lambda^{I}\wedge A^{I}
\end{align} 
where $\mathcal Q^{I}(\lambda^{I})$
is the charge operator that generates the 1-form gauge transformation and $\lambda \in \Omega_{\Z}^{1}(M)$ parametrizes the transformation. Then the 1-form symmetry acts as
\begin{align}
\mathcal Q^{I}(\lambda^{I}):&\; J^{I}_{qv}\mapsto J^{I}_{qv}+d\lambda^{I}; \nonumber \\ 
:&\; B^{I}\mapsto B^{I}-\lambda^{I}.
\end{align}
Gauging this dual 1-form symmetry means summing over $J^{I}_{qp}\in H^{2}(M,\widehat{G})$. Let us call the partition function after gauging the 1 -form $\widehat{G}$ symmetry $\mathcal Z_{\text{DW}/\hat{G}}^q$. Then
\begin{align}
  \mathcal Z_{\text{DW}/\widehat{G}}^{q}[N,\hat{A}]=&
                                                      \sum_{J_{qp}}e^{-i\int_{N} J^{I}_{qp}\wedge \hat{A^{I}} } \,
                                                      \mathcal Z_{\text{DW}}^q[N,J_{qp}] \nonumber \\
                                                    &=
                                                      \sum_{J_{qp}}\sum_{A} e^{i\int_{N}J^I_{qp}\wedge (A^I-\hat{A}^I)+iI^q[N,A]}
                                                      \nonumber \\
                                                    &= e^{iI^q[N,\hat{A}]}. 
\end{align}
Hence gauging the dual $\hat{G}$ 1-form global symmetry is equivalent to un-gauging $G$. The symmetry is generated by the world-line of $A$ and may be understood as anyon condensation. \cite{bais2009condensate, kong2014anyon, neupert2016boson, burnell2017anyon}
 

\subsection{Edge physics}
\noindent
Consider the $1+1d$ bosonic conformal field theory
on two-dimensional spacetime $M$
described by the action
\begin{align}
S[M]=\int_{ M}\sum_{I=1}^{k}\left[\frac{1}{4\pi}\partial_x\phi^{1,I}\partial_{t}\phi^{2,I}-\mathcal H(\phi^{1,I},\phi^{2,I})\right]
\label{1+1d_theory}
\end{align}  
where $\phi^{1,I},\phi^{2,I}: M\mapsto \mathbb R/2\pi \mathbb Z$. $\mathcal H$ denotes the Hamiltonian which we shall set to $\mathcal H=1/4\pi \sum_{I,i}(\partial_x\phi^{i,I})^2$.
The action \eqref{1+1d_theory} is invariant under different realizations of
global 0-form $\mathbb Z^k_n$-symmetry.
It is well-known that edge theories for $G$-SPTs suffer
from a $G$ 't-Hooft anomaly, i.e.,
there is an obstruction to promoting the global $G$-symmetry to a gauge symmetry.
A diagnostic of this anomaly that we will use is modular invariance. Modular invariance is a consistency criteria for a healthy quantum field theory. The idea is as follows: consider putting a quantum field theory on a manifold $M$. Then we require that the partition function be invariant under large diffeomorphisms of $M$. {\footnote{Large diffeomorphisms are those diffeomorphisms that are not path connected to the identity diffeomorphism in the space of diffeomorphisms. These form a group known as the mapping class group of $M$ which we shall abbreviate $MCG(M)$. Then 
\begin{align}
MCG(M)=\pi_0\left[\text{Diff}(M)\right]
\end{align} 
} }  
We will be particularly interested in $M=T^2$ for which $MCG(T^2)=SL(2,\mathbb Z)$ which has two generators $S,T$ with the action
\begin{align}
S:\left( \begin{array}{c}
 t \\
 x \end{array} \right) \mapsto &\; 
 \left( \begin{array}{c}
 -x \\
          t \end{array} \right),
  \nonumber \\
 T:\left( \begin{array}{c}
 t \\
 x \end{array} \right) \mapsto &\;
 \left( \begin{array}{c}
 t+x \\
          x \end{array} \right).
\end{align}
A modular invariant partition function is one for which
\begin{align}
Z[U M]=Z[M]; \quad U\in MCG(M)
\end{align}
A diagnostic for a theory with a global or 't-Hooft anomaly is the inexistence of a modular invariant partition function for the gauged (or orbifolded) theory\cite{ryu2012interacting, sule2013symmetry, hsieh2014symmetry, hsieh2016global}. To be more precise the partition function of the gauged theory takes the form
\begin{align}
Z_{\text{orb}}[M]=\frac{1}{|H^{0}(M,G)|}\sum_{A\in H^{1}(M,G)}\theta(A)Z[M,A]
\label{gauged_pf}
\end{align}
where $Z[M,A]$ is the `twisted' partition function computed in the presence of background flat $G$ gauge field $A\in H^{1}(M,G)$. 
In case a theory admits distinct $G$ actions we will denote by `$q$' a specific
realization of $G$-symmetry. We label a `twisted' partition function with this choice of symmetry action by $Z^q[M,A]$. In \eqref{gauged_pf}, the different
twisted sectors are weighted by $\theta(A)$ where $\theta$ is a function
$\theta: H^{1}(M,G)\to U(1)$ as a set.
More precisely we must think of $\theta(A)$ as a counterterm built from local
gauge data $A$ paired with the manifold, 
$
\theta(A)=\exp\left\{i\int_{M}\mathcal L_{c.t.}(A)\right\}.
$
Generally there might be inequivalent choices of $\theta$ that furnish modular
invariant partition functions. More precisely $\theta(A)$ as well as
$\theta(A)\epsilon (A)$ may be used to construct modular invariants. Here
$\epsilon$ is the discrete torsion phase classified by
$H^{2}_{\text{group}}(G,\R/2\pi \Z)$
(see App.\ \ref{discrete_torsion} for details). 

\bigskip \noindent
The theory has a 't-Hooft anomaly if there does not exist any gauge invariant $\theta(A)$ such that
\begin{align}
Z_{\text{orb}}[UT^2] = Z_{\text{orb}}[T^2]; \quad U=S,T
\end{align}
We will see that the theory \eqref{2+1_spt_action} introduced earlier exactly cancels the 't-Hooft anomaly of  \eqref{1+1d_theory} when $M=\partial N$ and the SPT effective action \eqref{2+1_spt_action} lives on $N$. Hence the 't-Hooft anomalies discussed here are prescribed by the same data $`q'\in H^{3}_{\text{group}}(G,\R/2\pi \Z)$ as $2+1d$ SPTs. Since the anomaly of the $1+1d$ theory is cancelled by the bulk $2+1d$ SPT, together they may be coupled consistently to a background $G$ gauge field and gauged. In other words
\begin{align}
\mathcal Z_{\text{DW}}^{q}[N]=\frac{1}{|H^{0}(N,G)|}\sum_{[A]\in H^{1}(N,G)}Z^{q}\left[M,A\right]\mathcal Z^{q}\Big[N|_{\partial N=M},A\Big]
\end{align} 
is the partition function for a healthy $G$ gauge theory
which is the Dijkgraaf-Witten theory labelled by 3-cocycle  $`q\text{'} \in H_{\text{group}}^3(G,\mathbb R/2\pi \mathbb Z)$. 

\bigskip \noindent Let us consider the case of $G=\Z_n^2$. We choose the simple case of $\Z_n^2$ to avoid dealing with orbifolding type-III cocycles which appear for $G=\Z_{n}^k$ when $k\geq 3$. Type-III cocycles are quite subtle for several reasons and we will mostly leave them out of our discussion. Since $H^{3}_{\text{group}}(\Z_n^2,\R/2\pi \Z)=\Z_n^{3}\simeq (q_1,q_2,q_{12})$ there could be three distinct kinds of $G$ actions and combinations thereof. Let us denote these by $\hat{g}_1,\hat{g}_{2},\hat{g}_{12}$ respectively. Explicitly their action on \eqref{1+1d_theory} is   
\begin{align}
\hat{g}_I:\left[ \begin{array}{c}
 \phi^{1,I} \\
 \phi^{2,I}  \end{array} \right] \mapsto &\; 
 \left[ \begin{array}{c}
 \phi^{1,I} \\
 \phi^{2,I}  \end{array} \right] +\frac{2\pi}{n}
 \left[ \begin{array}{c}
1 \\
q_{I}  \end{array} \right] \nonumber \\
 \hat{g}_{IJ}:\left[ \begin{array}{c}
 \phi^{1,I} \\
 \phi^{2,J}  \end{array} \right] \mapsto &\; 
 \left[ \begin{array}{c}
 \phi^{1,I} \\
  \phi^{2,J}  \end{array} \right] + \frac{2\pi}{n}
 \left[ \begin{array}{c}
1 \\
q_{IJ}  \end{array} \right]; I<J
\end{align}
We follow the canonical formalism in order to gauge the global $G$ symmetry. The
first step is to compute twisted partition functions $Z^{q}[M,A]$. Since $A$ is
flat it is characterized by holonomies along homology cycles in $M$ i.e $[A]\in
\text{Hom}\left[H_1(M,\mathbb Z),G\right]$. Let us fix $M=T^{2}$, then
$[A]\simeq ({\bf{a}},{\bf{b}})$ where ${\bf{a}},{\bf{b}}\in G$ are the
holonomies along the time and space cycle respectively.
The partition functions in the twisted sectors are
\begin{align}
  Z^q[T^2,A]=Z^q_{\bf{a},{\bf{b}}}:=\text{Tr}_{\mathcal H^q_{\bf{b}}}
  \left[\hat{\bf{a}}\,
  e^{2\pi i \tau_1 P-2\pi \tau_2 H}\right],
\label{twist_pf_2d}
\end{align}
where
$\tau=\tau_1+i\tau_2$ is the modular parameter of the flat spacetime torus,
$H,P$ are the Hamiltonian and the momentum, respectively, and 
we have defined the twisted Hilbert space
$\mathcal H^q_{\bf{b}}$
which satisfies the twisted boundary conditions 
\begin{align}
\left( \begin{array}{c}
 \phi^{1,J} \\
 \phi^{2,J}  \end{array} \right)(x+L)=
 \left( \begin{array}{c}
 \phi^{1,J} \\
 \phi^{2,J}  \end{array} \right)(x)+\frac{2\pi}{n}
 \left( \begin{array}{c}
 b_{J} \\
          q_{J}b_{J}+q_{IJ}b_{J}  \end{array} \right).
\end{align}
Let us define charge operators 
\begin{align}
Q^{i,I}:=\frac{1}{2\pi}\int dx \partial_{x}\phi^{\bar{i},I}; \quad i,\bar{i}\in 1,2; \quad  i\neq \bar{i}
\end{align}
which implement $U(1)$ transformations
\begin{align}
e^{i\lambda Q^{i,I}}: \phi^{i,I}\to \phi^{i,I}+\lambda.
\end{align}
Then $\hat{\bf{a}}$ appearing in \eqref{twist_pf_2d} takes the form
\begin{align}
\hat{\bf{a}}:=\exp\left\{\frac{2\pi i}{n}\left[a_IQ^{1,I} +a_1q_1Q^{2,1}+(a_2q_2+a_1q_{12})Q^{2,2} \right]\right\}.
\end{align}
These twisted partition functions can be computed using standard methods in conformal field theory (see for example\cite{francesco2012conformal, blumenhagen2012introduction, sule2013symmetry}). We will mainly be interested in modular properties of the twisted partition functions.
\begin{align}
  T:Z_{{\bf{a}},{\bf{b}}}^q(\tau)&\mapsto
                                   Z^q_{{\bf{a}},{\bf{b}}}(\tau+1)\nonumber \\
&= T^q_{{\bf{a}},{\bf{b}}} Z^{q}_{{\bf{a}}+{\bf{b}},{\bf{b}}}(\tau) \nonumber \\
                                 &=
                                   e^{-\frac{2\pi i}{n^2} \left[\sum_{I}q_Ib_I^2+q_{12}b_1b_2\right]}
     Z^q_{{\bf{a}}+{\bf{b}},{\bf{b}}}(\tau),
     \nonumber \\ 
  S: Z_{{\bf{a}},{\bf{b}}}^q(\tau) &\mapsto
                                     Z^q_{{\bf{a}},{\bf{b}}}(-1/\tau)\nonumber \\
&= S^{q}_{{\bf{a}},{\bf{b}}} Z^q_{-{\bf{b}},{\bf{a}}}(\tau)  \nonumber \\
&= e^{\frac{2\pi i}{n^2}\left[2\sum_{I}q_Ia_Ib_I+q_{12}(a_1b_2+b_1a_2)\right]} Z^q_{-{\bf{b}},{\bf{a}}}(\tau). 
\label{modular_2+1}
\end{align}
Under large gauge transformations, $Z^{q}_{{\bf{a}},{\bf{b}}}$ transforms as
\begin{align}
  Z^{q}_{{\bf{a}}+n{\bf{e}}_1,{\bf{b}}}(\tau)&= e^{\frac{2\pi i(q_1b_1+q_{12}b_2)}{n}} Z^{q}_{{\bf{a}},{\bf{b}}}(\tau),
                                                \nonumber \\
  Z^{q}_{{\bf{a}}+n{\bf{e}}_2,{\bf{b}}}(\tau)&= e^{\frac{2\pi i(q_2b_2+q_{12}b_1)}{n}} Z^{q}_{{\bf{a}},{\bf{b}}}(\tau),
                                                \nonumber \\
  Z^{q}_{{\bf{a}},{\bf{b}}+n{\bf{e}}_1}(\tau)&= e^{\frac{2\pi i(q_1a_1+q_{12}a_2)}{n}} Z^{q}_{{\bf{a}},{\bf{b}}}(\tau),
                                                \nonumber \\
   Z^{q}_{{\bf{a}},{\bf{b}}+n{\bf{e}}_2}(\tau)&= e^{\frac{2\pi i(q_2a_2+q_{12}a_1)}{n}} Z^{q}_{{\bf{a}},{\bf{b}}}(\tau),
\label{large_gauge_1+1}
\end{align}
where ${\bf{e}}_1=(1,0)$ and ${\bf{e}}_2=(0,1)$.

\bigskip \noindent {\bf{Gauging trivial symmetry action:}}
Let us first consider the partition functions twisted
by trivial symmetry action, i.e., $q=0$.
For this trivial case an equal weight sum over all twisted sectors is modular invariant
\begin{align}
Z^0_{\text{orb}}(\tau)=\frac{1}{|G|}\sum_{{\bf{a}},{\bf{b}}\in G}Z^0_{{\bf{a}},{\bf{b}}}(\tau).
\end{align} 
More generally, we may introduce a $U(1)$ valued function $\epsilon:G^2\to U(1)$ to obtain a partition function
\begin{align}
Z^{0,\epsilon}(\tau)=\frac{1}{|G|}\sum_{{\bf{a}},{\bf{b}}\in G}\epsilon({\bf{a}},{\bf{b}})Z^0_{{\bf{a}},{\bf{b}}}(\tau).
\label{discrete}
\end{align}
Modular invariance and factorizability of the partition function at higher genus
impose several constraints on $\epsilon$ such that distinct choices of $\epsilon$ are classified by $H^{2}_{\text{group}}(G,U(1))$ as 
\begin{align}
\epsilon({\bf{a}},{\bf{b}})=\frac{c({\bf{a}},{\bf{b}})}{c({\bf{b}},{\bf{a}})}
\end{align}
where $[c]\in
H_{\text{group}}^{2}(G,U(1))$.\cite{vafa1986modular,gaberdiel2000discrete} (see
App.\ \ref{discrete_torsion} for details).
 Bosonic SPTs in $1+1d$ protected by $G$ symmetry are also classified by $H^{2}_{\text{group}}(G,U(1))$. The partition function for SPT described by $[c]\in H^{2}_{\text{group}}(G,U(1))$ on a 2-torus with flat $G$ gauge field $A$ evaluates to  
\begin{align}
Z_{{\bf{a}},{\bf{b}}}^{c}=c({\bf{a}},{\bf{b}})/c({\bf{b}},{\bf{a}})=\epsilon({\bf{a}},{\bf{b}}).
\end{align}
Therefore the freedom of adding a discrete torsion phase while constructing a modular invariant partition function is equivalent to adding a $1+1d$ $G$-SPT. This is ofcourse expected since a $1+1d$ SPT is perfectly consistent on a closed $2$-manifold and therefore should not contribute to the anomaly.
Hence the anomaly on the boundary of a $2+1d$ SPT is insensitive to pasting of a $1+1d$ SPT protected by $G$ (or more generally $H$ such that $G\subset H$). 

\bigskip \noindent {\bf{Gauging non-trivial symmetry action:}} Now let us try to gauge $G$ for the action where $q\neq 0$. We mentioned earlier that this is related to non-trivial $q\in H_{\text{group}}^{3}(G,\mathbb R/2\pi \mathbb Z)$. Using \eqref{modular_2+1} we obtain the following conditions from requiring modular invariance 
\begin{align}
  \theta({\bf{a}},{\bf{b}})=&\; e^{\frac{2\pi i}{n^2} \left[\sum_{I}q_Ib_I^2+q_{12}b_1b_2\right]} \theta({\bf{a}}+{\bf{b}},{\bf{b}}),
                              \nonumber \\
\theta({\bf{a}},{\bf{b}})=&\; e^{-\frac{2\pi i}{n^2}\left[\sum_{I}2q_Ia_Ib_I+q_{12}(a_1b_2+b_1a_2)\right]}\theta(-{\bf{b}},{\bf{a}}). 
\label{c.t._properties}
\end{align} 
Using the first equation above, it can be seen that 
\begin{align}
  \theta({\bf{a}} + n{\bf{e}}_1,{\bf{e}}_1)=&\;e^{2\pi iq_1/n} \theta({\bf{a}},{\bf{e}}_1),
                                              \nonumber \\
  \theta({\bf{a}} + n{\bf{e}}_2,{\bf{e}}_2)=&\;e^{2\pi iq_2/n} \theta({\bf{a}},{\bf{e}}_2),
                                              \nonumber \\
\theta({\bf{a}} + n({\bf{e}}_1+{\bf{e}}_2),{\bf{e}}_1+{\bf{e}}_2)=&\;e^{\frac{2\pi i\left(q_1+q_2+q_{12}\right)}{n}} \theta({\bf{a}},{\bf{e}}_1+{\bf{e}}_2). 
\end{align}
We interpret $\theta({\bf{a}},{\bf{b}})$ as a local counter-term needed to make the partition function modular invariant. That is $\theta({\bf{a}},{\bf{b}})=e^{iS_{c.t.}[{\bf{a}},{\bf{b}}]}$. 
We learn that requiring modular invariance forces us to choose a counter-term which is not invariant under-large gauge transformations ${\bf{a}}\mapsto {\bf{a}} + n{\bf{e}}_I$ and ${\bf{a}}\mapsto {\bf{a}} + n{\bf{e}}_1+n{\bf{e}}_2$. Hence there is a conflict between gauge invariance and modular invariance which is a diagnostic of a 't-Hooft anomaly.  
There is no way to preserve both modular invariance and gauge invariance
for \eqref{1+1d_theory} when $q\neq 0$. This implies that \eqref{1+1d_theory}
suffers from a 't-Hooft anomaly and cannot be promoted to a $G$-gauge theory.
We can however couple it to a TQFT in $2+1d$ that cancels the 't-Hooft anomaly
of the $1+1d$ theory \eqref{1+1d_theory}.
Above we constructed an invertible TFT \eqref{2+1_spt_action} that exactly cancels the boundary anomaly. To see this, we compute the following response action
\begin{align}
I^{q}[D^{2}_{\bf{a}}\times S^{1}_{\bf{b}},A]=-\int_{D^{2}_{\bf{a}}\times S^{1}_{\bf{b}}}\left[\frac{q_I}{2\pi}A^{I}\wedge dA^{I}+\frac{q_{IJ}}{2\pi}A^{I} \wedge dA^{J}\right].
\label{anomaly_check}
\end{align} 
By $D^{2}_{\bf{a}}\times S^{1}_{\bf{b}}$, we denote the configuration where $N=D^{2}_{xy}\times S_t^{1}$, and the $G$ gauge field has a symmetry defect puncturing $D^2$ such that  
\begin{align}
\oint_{S_t^1}A^{I}=&\; \frac{2\pi}{n}b_{I}; \qquad \oint_{\partial D^{2}_{xy}}A^{I}= \frac{2\pi }{n}a_{I}
\label{field_conf}
\end{align}
Note that this is not a flat field configuration as it is sourced by a extrinsic symmetry defect. Then the partition function for an SPT described by \eqref{1+1d_theory} evaluates to
\begin{align}
  \mathcal Z^q[D^{2}_{\bf{a}}\times S^{1}_{\bf{b}},A]&=
 e^{iI^{q}[D^{2}_{\bf{a}}\times S^{1}_{\bf{b}},A] } \nonumber \\
&= e^{-\sum_{I}\frac{2\pi iq_{I}a_Ib_{I}}{n^2}- \frac{2\pi iq_{IJ}(a_{I}b_{J}+a_{J}b_{I})}{n^2}}
\end{align} 
which exactly satisfies the properties \eqref{c.t._properties} and hence cancels the modular anomaly of the $1+1d$ theory. Furthermore it transforms under large gauge transformations in an opposite way to \eqref{large_gauge_1+1}. Hence coupled to an invertible TFT in the bulk, \eqref{1+1d_theory} is healthy. 

\bigskip \noindent Further, \eqref{anomaly_check} is anomaly-free on a closed manifold and the global $G$ symmetry can be gauged to obtain DW theory with topological order. This topological order is characterized by some data such as braiding phases and topological spin. It has long been known that the topological data of the bulk TQFT can be extracted directly from the $1+1d$ edge theory. \cite{witten1989quantum, dijkgraaf1991quasi, wen1992theory, hatsugai1993chern, cappelli2002thermal, cappelli1997modular, cappelli2010chiral, cappelli2011partition}

\bigskip \noindent {\bf{$G$-characters and topological data:}}
In order to obtain topological data corresponding to the bulk topological gauge theory directly from the edge theory, we will construct a complete set of characters.  
\begin{align}
\chi_{\mu,{\bf{a}}}=\frac{1}{\sqrt{|G|}}\sum_{{\bf{b}}\in G}\mu({\bf{b}})Z^{q}_{{\bf{b}},{\bf{a}}}(\tau)
\end{align}
where $\mu\in \text{Rep}(G)$.
For example if $G=\Z_n$,
then explicitly $\mu(b)=e^{\frac{2\pi i\mu b}{n}}$. Each character constructed from the edge theory corresponds to an excitation within the bulk topological gauge theory. These characters form a projective representation of the mapping class group $SL(2,\Z)$ and the $S$ and $T$ matrices of projective phases encode bulk topological data
\begin{align}
  \mathcal S\chi_{{\bf{\mu}},{\bf{a}}}=&\;\sum_{{\bf{\mu'}},{\bf{a'}}}S_{({\bf{\mu}},{\bf{a}}),({\bf{\mu'}},{\bf{a'}})}\chi_{{\bf{\mu'}},{\bf{a'}}},
                           \nonumber \\
\mathcal T\chi_{{\bf{\mu}},{\bf{a}}}=&\;\sum_{{\bf{\mu'}},{\bf{a'}}}T_{({\bf{\mu}},{\bf{a}}),({\bf{\mu'}},{\bf{a'}})}\chi_{{\bf{\mu'}},{\bf{a'}}} \nonumber \\
=&\; \exp\left\{2\pi ih_{{\bf{\mu}},{\bf{a}}}\right\}\chi_{{\bf{\mu}},{\bf{a}}}.
\label{modular_chi}
\end{align}
Notice the action of $\mathcal T$ is diagonal and the eigenvalue of
$\chi_{{\bf{\mu}},{\bf{a}}}$,
$\exp 2\pi i h_{{\bf{\mu}},{\bf{a}}}$,
is the topological
spin of the bulk excitation corresponding to $\chi_{{\bf{\mu}},{\bf{a}}}$
via the bulk-boundary correspondence. 
Instead of directly evaluating the partition function in the twisted sector $Z_{{\bf{b}},{\bf{a}}}^q(\tau)$ (labelled by ${\bf{b}}$, ${\bf{a}}$) and extracting the $\mathcal{S}$ and $\mathcal{T}$ matrices from it \cite{dijkgraaf1991quasi, cappelli1997modular, cappelli2010chiral, cappelli2011partition, chen2017orbifolding}, we can construct $\bar{Z}_{{\bf{b}},{\bf{a}}}^q(\tau)$ from $Z_{{\bf{b}},{\bf{a}}}^0(\tau)$ in the following way,
\begin{align}
  \bar{Z}^{q}_{{\bf{b}},{\bf{a}}}(\tau):=&\; \gamma_{{\bf{a}}}^{q}({\bf{b}})Z^{0}_{{\bf{b}},{\bf{a}}}(\tau),
                             \nonumber \\
\bar{\chi}_{{\bf{\mu}},{\bf{a}}}=&\; \frac{1}{\sqrt{|G|}}\sum_{{\bf{b}}\in G}{\bf{\mu}}({\bf{b}})\bar{Z}^{q}_{{\bf{b}},{\bf{a}}}(\tau),
\label{Zbar_defn}
\end{align}
where $Z^0_{{\bf{b}},{\bf{a}}}(\tau)$ is the twisted partition function for the
trivial SPT phase.
In $\bar{Z}_{{\bf{b}},{\bf{a}}}^q(\tau)$, the interesting topological data is encoded in $\gamma_{{\bf{a}}}^q({\bf{b}})$, which  has the important algebraic property
\begin{align}
\gamma_{{\bf{a}}}^{q}({\bf{b}})\gamma_{{\bf{a}}}^{q}({\bf{c}})=\beta_{{\bf{a}}}^{q}({\bf{b}},{\bf{c}})\gamma_{{\bf{a}}}^{q}({\bf{b}}+{\bf{c}}).
\end{align}
The group 2-cocycle $\beta_{{\bf{a}}}^q \in C_{\text{group}}^2(\Z_n,U(1))$ is obtained from
$\omega_q({\bf{a}},{\bf{b}},{\bf{c}})$
[Eq.\ \eqref{2+1_cocycles}] by taking an slant product, i.e.,
$\beta_{{\bf{a}}}({\bf{b}},{\bf{c}})=i_{\bf{a}}\omega({\bf{a}},{\bf{b}},{\bf{c}})$
(for details, see App.\ \ref{group_cohomology}).
Explicitly,
$\beta^q_{\bf{a}}$ and $\gamma^q_{\bf{a}}$
take the form
\begin{align}
\beta_{{\bf{a}}}^q({\bf{b}},{\bf{c}}) &= \exp{\left\{\frac{2 \pi i }{n^2}\sum_{I}a_I (b_I+c_I - [b_I+c_I])\right\}}\nonumber \\
                                      &\;
                                        \quad \times \exp{\left\{ \frac{2\pi iq_{IJ}}{n^2}a_I (b_J+c_J - [b_J+c_J]) \right\}},
                                         \nonumber \\
  \gamma_{{\bf{a}}}^q({\bf{b}})&=
                                 \exp
                                 \left\{
                                 \frac{2\pi i}{n^2}(\sum_{I}q_{I}a_{I}b_{I}+
  q_{IJ}a_{I}b_{J})\right\}.
  \end{align}
  $\bar{Z}^{q}_{{\bf{b}},{\bf{a}}}$ in \eqref{Zbar_defn} is easier to
  work with than $Z_{{\bf{b}},{\bf{a}}}^q$
  since we do not need to evaluate the twisted partition function
$Z^{0}_{{\bf b},{\bf a}}$ directly, which may sometimes be tedious.
  Further, $\bar{Z}^{q}_{{\bf b},{\bf a}}(\tau)$
    and $Z^{q}_{{\bf b},{\bf a}}(\tau)$ have the same
properties under modular and large gauge transformation,
which is all we require.
It is straightforward to check that modular matrices computed from $\bar{\chi}_{\mu,a}$ match up with \eqref{modular_chi},\cite{sule2013symmetry, chen2017orbifolding} 
\begin{align}
  \bar{T}_{({\bf{\mu}},{\bf{a}}),({\bf{\mu}}',{\bf{a}}')}=&\;\delta_{{\bf{\mu}},{\bf{\mu}}'}\delta_{{\bf{a}},{\bf{a}}'}{\bf{\mu}}({\bf{a}})\gamma^{q}_{{\bf{a}}}({\bf{a}}),
                                \nonumber \\
\bar{S}_{({\bf{\mu}},{\bf{a}}),({\bf{\mu}}',{\bf{a}}')}=&\; \frac{1}{n}\mu({\bf{a}}'){\mu'}^{-1}(-{\bf{a}})\gamma^{q}_{{\bf{a}}}({\bf{a}}')\gamma^{q}_{{\bf{a}}'}({\bf{a}}).
\end{align}

\bigskip \noindent {\bf{SPT invariants from edge theory:}} Next we show that the
SPT invariants for type-I and type-II SPTs can be computed directly from the
edge theory \eqref{1+1d_theory}. Let us consider an SPT protected by $G=\Z_n^k$
with symmetry action described by some combination of type-I and type-II
3-cocycles `$q$'. Then such SPTs can be distinguished by their partition
functions on lens space. In \cite{shiozaki2016many}, it was shown that the Lens
space partition function may be simulated by an expectation value of a non-local
partial rotation operation on
the groundstate on $S^2$. Let the theory \eqref{2+1_spt_action} be defined on
$N=S^2 \times S^{1}$, where $S^2$ is the spatial manifold. The theory has a
unique groundstate $|GS_{S^2}^q\rangle$. The partition function on lens space
may be simulated as
\begin{align}
\mathcal Z^{q}[L(n,1),A]=\langle GS^q_{S^2} |\hat{C}_{n,D}({\bf{a}})|GS^q_{S^2} \rangle
\end{align}
where $\hat{C}_{n,D}({\bf{a}})$ is an operator that implements a partial
$n$-fold rotation on a disc like subregion $D\subset S^2$ followed by the symmetry operation $\hat{{\bf{a}}}$. To motivate this definition, we recall the fact that lens space may be constructed from the surgery \cite{jeffrey1992chern}
\begin{align}
L(n,1)=[D^2\times S^1] \sqcup_{\varphi} [D^2 \times S^1]
\end{align}
where $\sqcup_{\varphi}$ denotes gluing the boundaries $\partial [D^2\times S^1]=T^2$ via the large diffeomorphism $\varphi=ST^nS$. In \cite{shiozaki2016many}, it was shown that $\hat{C}_{n,D}$ corresponds to the same diffeomorphism $\varphi$. Then the lens space partition function with background field holonomy ${\bf{a}} \in G$ around the torsion cycle may be computed as  
\begin{align}
\mathcal Z^q[L(n,1),A]=&\; \langle GS^q |\hat{C}_{n,D}({\bf{a}})|GS^{q}\rangle \nonumber \\ 
=&\; \frac{\text{Tr}_{\mathcal H^q(D)}\left[\hat{C}_{n,D}({\bf{a}})\rho_{D}\right] }{\text{Tr}_{\mathcal H^q(D)} \left[\rho_D\right]}
\end{align}
where we have traced out the disc-like region $\bar{D}$ compliment to $D$. We
denote the Hilbert space on $D$ (respectively $\partial D$) for the SPT
described by 3-cocycle `$q$'$\in H^{3}_{\text{group}}(G,U(1))$ as $\mathcal
H^q(D)$ (respectively $\mathcal H^{q}(\partial D)$).
The reduced density matrix on $\rho_{D}$ is given by
the thermal density matrix on $\partial D$ at 
inverse temperature $\xi$,
which is related to the bulk correlation length 
\cite{qi2012general, wong2017note}
\begin{align}
\rho_{D}=\frac{e^{-\xi\hat{H}_{\partial D}}}{\text{Tr}_{\mathcal H^{q}(\partial D)}\left[e^{-\xi\hat{H}_{\partial D}}\right]}.
\end{align}
Then the lens space partition function may be evaluated as
\begin{align}
\mathcal Z^{q}[L(n,1),A]=&\; \frac{\text{Tr}_{\mathcal H^q(\partial D)}\left[\hat{C}_{n,\partial D}({\bf{a}})e^{-\xi \hat{H}_{\partial D}}\right]}
{\text{Tr}_{\mathcal H^q(\partial D)}\left[e^{-\xi \hat{H}_{\partial D}}\right]} \nonumber \\
=&\; \frac{\text{Tr}_{\mathcal H^q(\partial D)}\left[\hat{\bf{a}}e^{-\frac{i\hat{P}L}{n}-\xi \hat{H}_{\partial D}}\right]}
{\text{Tr}_{\mathcal H^q(\partial D)}\left[e^{-\xi \hat{H}_{\partial D}}\right]}
 \nonumber \\
=&\; \frac{ Z^{q}_{({\bf{a}},0)}\left(\frac{i\xi}{L}-\frac{1}{n}\right) }{Z^q_{(0,0)}\left(\frac{i\xi}{L}\right)} \nonumber \\
=&\; \frac{\sum_{\bf{b}}\left(ST^nS\right)_{({\bf{a}},0)}^{({\bf{b}}_{\tau},{\bf{b}}_x)}Z^q_{({\bf{b}}_{\tau},{\bf{b}}_x)} \left(-\frac{1}{n}+\frac{iL}{\xi n^2}\right)}
{\sum_{{\bf{b}}}S_{(0,0)}^{({\bf{b}}_{\tau},{\bf{b}}_x)}Z^q_{({\bf{b}}_{\tau},{\bf{b}}_x)} \left(\frac{iL}{\xi}\right)} \nonumber \\
=&\; e^{\frac{2\pi i (q_{I}a_I^{2}+q_{IJ}a_Ia_J)}{n}}\frac{ Z^{q}_{(-{\bf{a}},0)}\left(-\frac{1}{n}+\frac{iL}{\xi n^2}\right)}{Z^{q}_{(0,0)} \left(\frac{iL}{\xi}\right)} \nonumber \\
=&\; e^{\frac{2\pi i (q_Ia_I^{2}+q_{IJ}a_Ia_J)}{n}} \Big(1+\mathcal O(e^{-L/\xi})\Big).
\label{lens_sim}
\end{align}
In the last line we have taken the limit
where the inverse temperature $\xi$ is
much smaller than $L$, 
the circumference of $\partial D$ ($\xi/L\to 0$).
Hence we can read off the SPT invariant
\begin{align}
\mathcal Z^q[L(n,1),A]=e^{\frac{2\pi i}{n} \left(q_Ia_I^{2}+q_{IJ}a_Ia_J\right)}.
\end{align}


\section{$3+1d$ topological phases and their $2+1d$ gapless surfaces}
\label{3+1d}
\subsection{Bulk physics}

\noindent {\bf {SPT effective actions:}} Similar to the $2+1$-dimensional case, $3+1d$ SPTs can be modeled by multiple copies of level 1 $BF$ theories with topologically distinct coupling to a flat background $G$ bundle. SPT phases with this symmetry are classified by $H^{4}_{\text{group}}(G,U(1))$. For example consider $G=\mathbb Z_{n}^k$ bosonic SPTs which can be modeled by the following effective field theories\cite{chen2013symmetry, kapustin2014anomalous, wang2015field, ye2016topological}
\begin{align}
\mathcal S^q(a,b,A)=\int_{N}\frac{\delta_{IJ}}{2\pi}b^I\wedge da^J + \mathcal S_{\text{cpl}}^q(a,b,A) 
\label{SPT_eff_ac}
\end{align}
where $a$ and $b$ are
1-form and 2-form $U(1)$ gauge field,  
$I,J=1,\ldots,k$,
and $q$ denotes the representative $\omega\in H^{4}_{\text{group}}(G,U(1))$, For $G=\left(\mathbb Z_n\right)^k$, 
\begin{align}
H^{4}_{\text{group}}[(\mathbb Z_n)^k,U(1)]
=&\; (\mathbb Z_{n})^{\left[
2\times \left( \begin{array}{c}
k \\
2  \end{array} \right)
+
\left( \begin{array}{c}
k \\
3  \end{array} \right)
+
\left( \begin{array}{c}
k \\
4  \end{array} \right)
\right]}
\end{align}
Different 4-cocycles $[\omega]\in H^{4}(G,U(1))$ are
of three kinds named `type-II,III,IV' which explicitly take the form
\begin{align}
  \omega_{\text{type-II}}({\bf{a}},{\bf{b}},{\bf{c}},{\bf{d}})
  &= e^{\frac{2\pi iq_{IJ}}{n^{2}}a^{I}b^{J}\left(c^{J}+d^{J}-[c^J+d^J]\right)},
    \nonumber \\
  \omega_{\text{type-III}}({\bf{a}},{\bf{b}},{\bf{c}},{\bf{d}})
  &= e^{\frac{2\pi iq_{IJK}}{n^{2}}a^{I}b^{J}\left(c^{K}+d^{K}-[c^{K}+d^{K}]\right) },
    \nonumber \\
  \omega_{\text{type-IV}}({\bf{a}},{\bf{b}},{\bf{c}},{\bf{d}})
  &= e^{\frac{2\pi i q_{IJKL}}{n}a^{I}b^{J}c^{K}d^{L}},
\label{four_cocycles}
\end{align}
where $[a^I+b^I]$ denotes addition modulo $n$.
Here $q=\left\{q_{IJ},q_{IJK},q_{IJKL}\right\}$ are a set of parameters valued in $\Z \ \text{mod} \ n\Z$ that label different SPTs. Then distinct SPT effective field theories differ in how they couple to the background flat $G$ gauge field. The coupling terms corresponding to different cocycle types take the form
\begin{align}
\mathcal S^{q_{IJ}}_{\text{cpl}}(a,b,A)&=-\frac{1}{2\pi}\int_{N} A^{I} \wedge \left(b^I+\frac{nq_{IJ}}{2\pi}
                                          a^{J}\wedge d a^J\right),
                                          \nonumber \\
\mathcal S^{q_{IJK}}_{\text{cpl}}(a,b,A)&=-\frac{1}{2\pi} \int_{N} A^{I} \wedge \left(b^I+\frac{nq_{IJK}}{2\pi}
                                          a^{J}\wedge d a^K\right),
                                          \nonumber \\
\mathcal S^{q_{IJK}}_{\text{cpl}}(a,b,A)&=-\frac{1}{2\pi}\int_{N} A^{I} \wedge \left(b^I+\frac{n^3q_{IJKL}}{4\pi^2}
a^{J}\wedge a^K \wedge a^{L}\right).
\label{cpl_terms_3+1}
\end{align}
Generally, the coupling to background field $A^{I}$ may involve a combination of type-II,III,IV terms for some choice of `$q$'. For simplicity we will treat these terms separately. The response theory can be obtained by integrating over the matter fields $a,b$.
\begin{align}
e^{iI^q[N,A]}&= \int \mathcal D[\left\{a,b\right\}]e^{i \mathcal{S}^q(a,b,N,A)}. 
\end{align}
The different response theories are
\begin{align}
  e^{iI^{q_{IJ}}[N,A]}
                        &= \exp\left\{-\frac{inq_{IJ}}{4\pi^{2}}\int_{N}A^{I}\wedge A^{J}\wedge dA^{J}\right\},
                          \nonumber \\
  e^{iI^{q_{IJK}}[N,A]}
                        &= \exp\left\{-\frac{inq_{IJK}}{4\pi^{2}}\int_{N}A^{I}\wedge A^{J}\wedge dA^{K}\right\},
                          \nonumber \\
  e^{iI^{q_{IJKL}}[N,A]}&= \exp\left\{-\frac{in^3q_{IJKL}}{8\pi^{3}}\int_{N}A^{I}\wedge A^{J}\wedge A^{K} \wedge A^{L}\right\}.
                          \label{respacito}
\end{align} 
In the above, $I,J,K,L$ are not summed over.

\bigskip \noindent {\bf {Topological invariants for SPTs:}} It was recently shown \cite{kapustin2014symmetry, kapustin2014bosonic}, that bosonic SPTs are classified by the equivariant cobordism group and SPT topological invariants are the set
$\left\{e^{-iI^{q}[N,A]}\right\}$
of $U(1)$ phases where $[N,A]$ are the set of generators of the equivariant cobordism group $\Omega^{SO}_{4}(BG)$.
For $G=\prod_{I}\Z_{n_I}$, the generating manifolds for type-II,III,IV terms parametrized by $\left\{q_{IJ},q_{IJK},q_{IJKL}\right\}$ are $\left\{L(\text{gcd}(n_{I},n_{J}),1)\times S^1, L(\text{gcd}(n_{I},n_{J},n_{K}),1)\times S^1,T^{4}\right\}$ respectively, equipped with some appropriate $G$-bundle\cite{tantivasadakarn2017dimensional}. Here we compute invariants for $G=\Z_n^k$ for which $L(n,1)\times S^1$ and $T^4$ suffice. Generalization to other discrete abelian groups is straightforward. 
\begin{itemize}
\item {\bf{Type-II and type-III cocycles:}} Type-II and type-III cocycles can be detected on $N=L(n,1)\times S^1$. Let $S\in \text{Tor}\left(H_{2}(N,\Z)\right)$ be Poincare dual to the generator of $A^{J}\in \text{Tor}\left(H^{2}(N,\Z)\right)$. Then we obtain
\begin{align}
e^{iI^{q_{IJ}}[N,A]}=&\; \exp\left\{-\frac{inq_{IJ}}{4\pi^{2}}\int_{N}A^{I}\wedge A^{J}\wedge dA^{J}\right\} \nonumber \\
=&\; \exp\left\{-\frac{inq_{IJ}}{2\pi}\int_{S=S^{1}\times C_{A^J}}A^{I}\wedge A^{J}\right\} \nonumber \\
=&\; \exp\left\{-\frac{inq_{IJ}a_J}{2\pi}\int_{S=S^{1}\times C_{1}}A^{I}\wedge A^{J}\right\} \nonumber \\
=&\; \exp\left\{-\frac{2\pi i q_{IJ} }{n}a_J(b_Ia_J-a_Ib_J)\right\} 
\label{lens_type_2_resp}
\end{align}
where we have decomposed the $S=S^1\times C$ where $C$ is the torsion 1-cycle in $N$. $(a_I,b_I)$ are the $\Z_n$ holonomies along $C_1$ and $S^1$ for the $I$th flavor of $\Z_n$. The calculation for type-III follows very similarly.
\begin{align}
e^{iI^{q_{IJK}}[N,A]}=&\; \exp\left\{-\frac{inq_{IJK}}{4\pi^{2}}\int_{N}A^{I}\wedge A^{J}\wedge dA^{K}\right\} \nonumber \\
=&\; \exp\left\{-\frac{inq_{IJK}}{2\pi}\int_{S=S^{1}\times C_{A^K}}A^{I}\wedge A^{J}\right\} \nonumber \\
=&\; \exp\left\{-\frac{inq_{IJK}a_K}{2\pi}\int_{S=S^{1}\times C_{1}}A^{I}\wedge A^{J}\right\} \nonumber \\
=&\; \exp\left\{-\frac{i2\pi q_{IJK} }{n}a_K(b_Ia_J-a_Ib_J)\right\}.
\label{lens_type_3_resp}
\end{align}
\item {\bf{Type-IV cocycles:}} Type-IV topological term can be detected on $T^4$ with appropriate background flat $G$-bundle. The response theory evaluates to
\begin{align}
e^{iI^{q_{IJKL}}}=&\; \exp\left\{-\frac{in^3q_{IJKL}}{8\pi^3}\int_{ T^4}A^{I}\wedge A^J \wedge A^K \wedge A^L \right\}\nonumber \\
=&\; \exp\left\{-\frac{2\pi i q_{IJKL}}{n}\epsilon^{ijkl}a_{I,i}b_{J,j}c_{K,k}d_{L,l}\right\}
\label{torus_type_4_resp}
\end{align}
where $I,J,K,L=1,2,3,4$, ${\bf{a}}=(a_1,a_2,a_3,a_4)$, and ${\bf{a}},{\bf{b}},{\bf{c}},{\bf{d}}\in \Z_n^4$ are the holonomies around the three cycles of $T^4$.
\end{itemize} 
The complete set of topological invariants for bosonic SPTs protected by $G=\Z_{n}^k$ then is
\begin{align}
\left\{e^{-iI^{q}[L(n,1)\times S^1,A]}, e^{-iI^{q}[T^4,A]}\right\}
\end{align}
{\bf {Topological gauge theories from Gauging SPTs:}} $3+1d$ SPTs can be gauged by first coupling to a flat bundle as we have done above and then summing over all possible flat bundles. The gauged partition function function on  a manifold $N$ takes the form 
\begin{align}
\mathcal Z_{\text{DW}}^{q}[N]=\frac{1}{|H^{0}(N,G)|} \sum_{[A]}\mathcal Z^{q}[N,A].
\end{align}
The gauged theory is the well-known Dijkgraaf-Witten theory which has
topological order.
The ground-state degeneracy on any 3-manifold $M$ can be computed as
$\mathcal Z_{\text{DW}}^{q}[M\times S^{1}]=\text{GSD}^{q}[M]$.
These theories can be differentiated by the phases they assign to multi-linked configurations of vortices. \cite{wang2014braiding, jiang2014generalized, chen2016bulk, tiwari2016wilson, wang2015non, putrov2017braiding}. These loop braiding statistics may be computed in the bulk by performing modular transformations on the basis of groundstates on a three-torus\cite{jiang2014generalized} and reading off the projective phases in the modular matrices. Alternately they may be computed from the Wilson operator algebra of the Dijkgraaf-Witten theories \cite{tiwari2016wilson} or by directly computing partition functions on manifolds with multi-link vortex defects embedded. Type-II and type-III Dijkgraaf-Witten theories in $3+1d$ assign non-trivial braiding phases to linked three-loop configurations in spacetime or three-loop braiding processes whereas type-IV theory assigns non-trivial phases to linked four-loop configurations. Let us consider a few specific examples
\begin{itemize}
\item {\bf{Type-II and type-III Dijkgraaf-Witten theories:}} 
Consider putting type-II or type-III theory on a manifold $N=M\times S^1$ and gauging. Suppose $\text{Tor}(H_{1}(M),\Z)=0$. Then $I^{q}[N,A]=1$, therefore we get 
\begin{align}
\mathcal Z_{\text{DW}}^{q}[M\times S^1]=&\;\frac{1}{|G|}\sum_{[A]\in H^{1}(N,G)}1 \nonumber \\
=&\; |G|^{b_1(M)}=:\text{GSD}^q\left[M\right]. 
\end{align}
Next if $M$ has torsion, for example if $N=L(n,1)\times S^1$, for type-II cocycle with $G=\Z_n^2$ we get
\begin{align}
\mathcal Z^{q_{IJ}}_{\text{DW}}[N]=&\;\frac{1}{|n|^2}\sum_{a_{I},b_{I}\in \Z_n}e^{-\frac{2\pi i q_{IJ} }{n}a_J(b_Ia_J-a_Ib_J)} \nonumber \\
=:&\; \text{GSD}^{q_{IJ}}\left[L(n,1)\right]
\end{align}
Similarly for type-III cocycle with $G=\Z_n^3$ we get
\begin{align}
\mathcal Z^{q_{IJK}}_{\text{DW}}[N]=&\;\frac{1}{|n|^3}\sum_{a_{I},b_{I}\in \Z_n}e^{-\frac{2\pi i q_{IJK} }{n}a_K(b_Ia_J-a_Ib_J)} \nonumber \\
=:&\; \text{GSD}^{q_{IJK}}\left[L(n,1)\right]
\end{align}
\item Unlike type-II and type-III Dijkgraaf-Witten theories in $3+1d$, for
  Type-IV cocycle, $\text{GSD}^q[T^3]< |G|^{3}$. Similar to type-III cocycle in
  $2+1d$ [Eq.\ \eqref{type3_partn_fn}], this is related to the fact that type-IV DW theory actually has non-abelian excitations. In other words the quantum dimension of some of the quasivortices is greater than one. The partition function for $G=\Z_n^4$ on the four-torus is
\begin{align}
\mathcal Z_{\text{DW}}^{q_{IJKL}}\left[T^4\right]=&\;\frac{1}{n^4}\sum_{{\bf{a}},{\bf{b}},{\bf{c}},{\bf{d}}\in \Z_n^4}e^{-\frac{2\pi i q_{IJKL}}{n}\epsilon^{ijkl}a_{I,i}b_{J,j}c_{K,k}d_{L,l}} \nonumber \\
=&\; \text{GSD}^{q_{IJKL}}\left[T^3\right]
\end{align}
where ${\bf{a}}=(a_1,a_2,a_3,a_4)$, and ${\bf{a}},{\bf{b}},{\bf{c}},{\bf{d}}\in \Z_n^4$ are the holonomies around the three cycles of $T^4$.
\end{itemize}
For some purposes it is convenient to formulate the $G$-gauged theory in the continuum as a coupled $BF$ theory (see for example \cite{tiwari2016wilson})
\begin{align}
\mathcal Z_{\text{DW}}^{q}[N]=&\;\frac{1}{|G|}\sum_{[A]\in H^{1}(N,G)}e^{iI^{q}[A]} \nonumber \\
\xrightarrow{G=\Z_n^k}&\;\frac{1}{n^{k}}\int \prod_{I=1}^{k}\mathcal D[A^{I},B^{I}]e^{\frac{in\delta_{IJ}}{2\pi}\int B^{I}\wedge dA^{J}+iI^{q}[A] }
\end{align}
where $A$ and $B$ are 1-form and 2-form $U(1)$ connections with standard
quantization conditions.
Since $(1/2\pi)dB^{I}\in \Omega^{3}_{\Z}(N)$, we can integrate them out to impose that $A^{I}$ are flat $\Z_n$ gauge fields. The calculation is very similar to \eqref{integrating_B}.

\bigskip \noindent {\bf{Ungauging in the $3+1d$ bulk:}}
More generally one can gauge $G$ in the presence of background quasiparticle sources $J_{qp}\in H^{3}(N,\hat{G})$. The gauged partition function takes the form 
\begin{align}
\mathcal Z_{\text{DW}}^{q}[N,J_{qp}]=&\; \frac{1}{|G|}\int \mathcal D[A,B]e^{\frac{in\delta_{IJ}}{2\pi}\int B^{I}\wedge dA^{J}+iI^{q}[A] +i\int_{N}J^I_{qp}\wedge A^I} \nonumber \\
=&\;\frac{1}{|G|}\sum_{[A]}e^{iI^{q}[N,A]+i\int_{N}J^I_{qp}\wedge A^I  }
\end{align}  
where the background fields $J^{I}_{qp}$ are 3-form fields with integral periods \cite{bauer2005class}. Since $\oint A^{I}\in \left(2\pi \Z\right)/n$, the periods of $J_{qp}$ are only physically distinguishable modulo $n$, more precisely $J_{qp}\in H^{3}(N,\widehat{G})$ where $\widehat{G}=\text{Rep}(G)\simeq G$. There is a perfect pairing 
\begin{align}
\int_N: H^{1}(N,G)\times H^{3}(N,\widehat{G})\to \R/2\pi \Z
\label{pairing}
\end{align}
that is realized by wedge product followed by integration. $J_{qp}$ generates a 2-form $\widehat{G}$ symmetry implemented by the charge operator $\mathcal Q^{I}(\lambda^I)$ corresponding to $\widehat{G}$ symmetry.  
\begin{align}
\mathcal Q^I(\lambda^I):=\frac{1}{2\pi}\int_M\lambda^{I}\wedge A^{I}
\end{align} 
where $\lambda^{I} \in \Omega_{\Z}^{2}(M)$. Then the 2-form symmetry acts
\begin{align}
\mathcal Q^{I}(\lambda^{I}):&\; J^{I}_{qv}\mapsto J^{I}_{qv}+d\lambda^{I}; \nonumber \\ 
:&\; B^{I}\mapsto B^{I}-\lambda^{I}
\end{align}
Gauging this dual 2-form symmetry means summing over $J^{I}_{qp}\in H^{3}(M,\widehat{G})$. Let us call the partition function after gauging the 2 -form $\widehat{G}$ symmetry $\mathcal Z_{\text{DW}/\hat{G}}^q$, then
\begin{align}
 \mathcal Z_{\text{DW}/\widehat{G}}^{q}[N,\hat{A}]
  =&\; \sum_{J_{qp}}e^{-i\int_{N} J_{qp}\wedge \hat{A} }  \mathcal Z_{\text{DW}}^q[N,J_{qp}] \nonumber \\
=&\;\sum_{J_{qp}}\sum_{A} e^{i\int_{N}J_{qp}\wedge (A-\hat{A})+iI^q[N,A]}
 \nonumber \\
=&\; e^{iI^q[N,\hat{A}]}. 
\end{align}
Hence gauging the dual $\widehat{G}$ 1-form global symmetry is equivalent to un-gauging. The symmetry is generated by the world-line of $A$ and may be understood as physically as proliferating or condensing the gauge charge $\sim dB$ which is always bosonic since $[B,B]=0$. Hence this procedure works for all bosonic SPTs protected by onsite symmetry. 
\subsection{Surface physics}
\noindent We model the gapless surface of $3+1d$ bosonic SPTs described by \eqref{SPT_eff_ac} by the following quantum field theory\cite{wu1991topological, balachandran1993edge, amoretti2012three, chen2016bulk}
\begin{align}
S=\int_{ M}\sum_{I=1}^{k}\left[\frac{1}{2\pi }d\zeta^{I}\wedge d\phi^{I}-\mathcal H(\zeta^{I},\phi^{I}) \right].
\label{2+1_action}
\end{align}
Here, $\phi^I:M\to \mathbb  R/2\pi \mathbb Z$ and $\zeta^I$ are 1-form $U(1)$ connections which satisfy Dirac quantization conditions
\begin{align}
\oint_{Z_{1}(M,\Z)} \frac{d\phi^I}{2\pi}\in \Z; \qquad \oint_{Z_2(M,\Z)}\frac{d\zeta^I}{2\pi}\in \Z.
\end{align}
This model has a global 0-form (and 1-form) $U(1)^{k}$ symmetry.
We will however be interested in the discrete subgroup $G=\Z_n^k \subset
U(1)^{k}$. Similar to $1+1d$, we probe the theory by coupling to a flat $G$
gauge field $A\in H^{1}(M,G)$ and use modular invariance of the orbifolded
partition function as a diagnostic for whether the model with a specific action
of $G$ has a 't-Hooft anomaly. In other words we put the theory on $M=T^3$ and
check whether it is possible to construct a partition function upon summing all
twisted sectors (flat $G$ bundles) such that the summed partition function is
invariant under large diffeomorphisms of $M$ as well as large gauge
transformations.
The group of large diffeomorphisms on $M=T^3$, i.e., $MCG(T^3)=SL(3,\mathbb Z)$ which is generated by $U_1,U_2$ with the action
\begin{align}
U_1:\left( \begin{array}{c}
 t \\
 x \\
 y \end{array} \right) \mapsto &\; 
 \left( \begin{array}{c}
 y \\
 t \\
          x \end{array} \right),
  \nonumber \\
 U_2:\left( \begin{array}{c}
 t \\
 x \\
 y \end{array} \right) \mapsto &\;
 \left( \begin{array}{c}
 t+x \\
 x \\
 y \end{array} \right). 
\end{align}
A modular invariant partition function is one for which
\begin{align}
Z[UM]=Z[M]; \quad U\in MCG(T^3)
\end{align}
The diagnostic for a theory with a global or 't-Hooft anomaly will be the
inexistence of a modular invariant partition for the gauged (or orbifolded
theory).
For a review of quantization of \eqref{2+1_action}, 
see App.\ \ref{2+1_action_appendix}.

\bigskip \noindent As with the $1+1d$ case \eqref{1+1d_theory}, we expect \eqref{2+1_action} to accommodate distinct realizations of $G=\Z_n^{k}$ which we label by `$q$'. We will denote partition functions of these models in the presence of a background $G$ bundle $A$ as $Z^{q}[M,A]$. By anomaly matching one can learn that these quantum field theories require a bulk which cancels the anomaly. Such bulk theories would be provided by SPT effective actions \eqref{SPT_eff_ac}. As a warm-up let us consider the simplest $G$ action which is non-anomalous and hence does not require a bulk to support it.

\bigskip \noindent {\bf{ Non-anomalous 0-form $\Z_n$ symmetry:}} A single copy of \eqref{2+1_action} is invariant under 
a global 0-form $U(1)$ symmetry 
\begin{align}
\phi(x)\mapsto \phi(x) + \alpha
\label{trivial_action}
\end{align}
where $\alpha$ is a constant. Gauging this $U(1)$ symmetry implies introducing a flat 1-form $U(1)$ gauge field $A$ and replacing the differential
\begin{align}
d\phi \mapsto D_{A}\phi:= d\phi + A 
\end{align}
with the gauge transformation
\begin{align}
  \phi(x)&\mapsto \phi(x) +\alpha(x),
                  \nonumber \\
A(x) &\mapsto A(x)-d\alpha(x).   
\end{align}
Here we gauge a subgroup $\Z_n\subset U(1)$ by restricting the holonomies of $A$ to $\Z_n$. Then defining $d\tilde{\phi}:=D_{A}\phi$ which obeys the twisted quantization condition
\begin{align}
\oint_{L}\frac{d\tilde{\phi}}{2\pi}\in \Z+ \oint_{L}\frac{A}{2\pi}
\label{twisted_quantization}
\end{align}
i.e., quantizing in the presence of background $A$ implies imposing twisted
boundary condition.
Then the gauging procedure is the same as before;
First we
compute
the partition functions in
the twisted sectors $Z^0[M,A]$ and then sum over them    
\begin{align}
  Z_{\text{orb}}^{0}[M]=\frac{1}{|H^{0}(M,G)|}\sum_{[A]\in H^{1}( M,G)}\theta(A)
  Z^{0}[ M,A].
\label{gauge}
\end{align}
We compute $Z_{\text{orb}}^0[M,A]$ within the canonical formalism. Following \eqref{twisted_quantization} we impose twisted boundary conditions. Let us set $M=T^3$ and the holonomies of $A$ along the $x,y$ cycles be $\lambda_{1,2}$ respectively, then the twisted Hilbert space is defined as
\begin{align}
  \mathcal H_{\lambda_1,\lambda_2}=\left\{
\phi(x,y),\zeta(x,y)
  \Big|\oint_{L_{1,2}}d\phi= \frac{2\pi}{n} \lambda_{1,2} \right\}.
\end{align}
Similarly, we can also twist in the time direction, in the path integral picture, this means coupling to a background $U(1)$ field with non-trivial holonomy in the time-cycle. In the canonical formalism, this is implemented via a global $\mathbb Z_n$ symmetry operator 
\begin{align}
\mathcal G(\lambda_0):=&\; \exp\left\{\frac{2\pi i\lambda_0}{n}  \mathcal Q\right\} \nonumber \\
=&\; \exp\left\{\frac{i\lambda_0}{n} \int_{T^2} d\zeta \right\} =\exp\left\{ \frac{2\pi i \lambda_0\beta_0}{n}\right\} 
\end{align}
where $\beta_0$ is defined in \eqref{zeromode_expansion}. $\mathcal G(\lambda_0)$ implements the transformation $\phi\mapsto \phi+2\pi \lambda_0/n$.
\begin{align}
\mathcal G(\lambda_0):&\;\phi\mapsto \phi+\frac{2\pi \lambda_0}{n}. 
\end{align}
Then
the partition function in the twisted
sectors are computed as
\cite{chen2016bulk} 
\begin{align}
  Z^0_{\lambda_0,\lambda_1,\lambda_2}&\;=
                                       \text{Tr}_{\mathcal H_{\lambda_1,\lambda_2}}\left[\mathcal G(\lambda_0)e^{2\pi iR_0H'}\right] \nonumber \\
&\;=Z_{osc} \sum_{N_{0,1,2}\in\mathbb Z}\exp\left\{
-\frac{\pi \tau_2}{2R_2}N_{0}^2 
  \right. \nonumber \\
&\; \left. 
 -2\pi R_2\tau_2\left(N_1+\frac{\lambda_1}{n}\right)^2-\frac{2\pi R_0R_1}{R_2}\left(N_2+\frac{\lambda_2}{n}\right)^2 
\right. \nonumber \\
&\; \left.
+2\pi i\tau_1 N_0\left(N_1+\frac{\lambda_1}{n}\right) +\frac{2\pi i  N_0\lambda_0}{n} 
\right\}
\label{twist_partn_fn}
\end{align}
As we will mostly be working on $T^3$, we simply label the partition functions with $\lambda_{0,1,2}$, the $G$ holonomies on $T^3$. Under $SL(3,\mathbb Z)$ modular transformations, the twisted sectors transform as
\begin{align}
  U_2Z^0_{\lambda_0,\lambda_1,\lambda_2}=&\;  Z^0_{\lambda_0-\lambda_1,\lambda_1,\lambda_2},
                                           \nonumber \\
  M Z^0_{\lambda_0,\lambda_1,\lambda_2}=&\;  Z^0_{\lambda_0,-\lambda_2,\lambda_1},
                                          \nonumber \\
U_1' Z^0_{\lambda_0,\lambda_1,\lambda_2}=&\;  Z^0_{\lambda_1,\lambda_0,\lambda_2}.
\label{modular_anomaly_free}
\end{align}
A modular invariant partition function may be constructed by taking an equal
weight sum,
i.e., $\theta(A)=1$ in \eqref{gauge} 
\begin{align}
Z_{\text{orb}}=\frac{1}{n}\sum_{\lambda_0,\lambda_1,\lambda_2\in\mathbb Z_n}  Z^0_{\lambda_0,\lambda_1,\lambda_2}
\end{align}
In fact, we need not choose $\theta(A)=1$. We saw in \eqref{discrete}, there was a freedom worth $H^{2}_{\text{group}}(G,U(1))$ in constructing a modular invariant partition function which corresponded to pasting a $1+1d$ $G$ SPT onto \eqref{1+1d_theory} and then gauging. Similarly in $2+1d$, given a modular invariant partition function, we can always find a new one by picking a $[\beta]\in H^{3}(BG,\mathbb R/2\pi \mathbb Z)$ and orbifolding with phase-factors
\begin{align}
Z^{\beta}_{\text{orb}}[M]=\frac{1}{|G|}\sum_{[\mathcal A]\in \text{Map}[M,BG]} e^{i\int_{M}\mathcal A^{*}\beta} Z[M,\mathcal A^{*}EG] 
\end{align}
where $BG$ is the classifying space of $G$, $EG$ is the universal $G$-bundle over $BG$ and the sum is over homotopy classes of maps from $M$ to $BG$. $\exp\left\{i\int_{M}\mathcal A^{*}\beta\right\}$ is the partition function for an $2+1$-dimensional $G$ SPT with background flux $\mathcal A^{*}EG$, hence the freedom of adding a phase corresponds to pasting a $2+1d$ SPT onto \eqref{2+1_action}. 

\bigskip \noindent {\bf{Anomalous symmetry action:}} Let us consider orbifolding $G$ action corresponding to type-II or type-III cocycle. The minimum case where such a symmetry can be implemented is for $G=\Z_n^3$ on three copies of \eqref{2+1_action}. 
\begin{align}
S=\int_M\left[\frac{\delta_{IJ}}{2\pi}d\phi^{I}\wedge d\zeta^{J}-\mathcal H(\phi^I,\zeta^I)\right]
\label{two_copies}
\end{align}
where $I,J=1,2,3$. The simplest $G$ action acts independently on the three copies as \eqref{trivial_action} as described above. Other $G$-actions couple the multiple copies in a non-trivial way and may be labelled by 
$q=\left\{q_{IJ},q_{IJK}\right\}$.
Let us consider
the coupling to background $G$ field $A$
and consider the action
\begin{align}
S=&\; \int_M \sum_{I,J=1,2}
\left[\frac{\delta_{IJ}}{2\pi} d\phi^{I}\wedge d\zeta^J-\mathcal H(\phi^{I},\zeta^{I})\right. \nonumber \\
&\; \left. 
+\frac{1}{2\pi}A^{I}\wedge\left(d\zeta^{I}+\frac{n}{2\pi}q^{IJ} d\phi^I \wedge d\phi^{J}\right. \right. \nonumber \\
&\; \left.\left.+\frac{n}{2\pi}q_{IJK}d\phi^{J}\wedge d\phi^{K}\right) 
\right] 
\label{coupled_two_copies}
\end{align}  
where $q_{IJ},q_{IJK}\in [0,\ldots,n-1]$ are $\Z_n$ valued parameters that
parametrize distinct couplings to the background field.
By inspecting the equations of motion we learn that the fields $\phi^{I}$ and $\zeta^{I}$ satisfy twisted boundary conditions
\begin{align}
  \frac{1}{2\pi}\oint_{L} d\phi^{I} =&\; \frac{1}{2\pi} \oint_{L} A^{I},
                                       \nonumber \\
\frac{1}{2\pi}\oint_{S} d\zeta^{I}=&\; \frac{q_{IJ}n}{4\pi^2}\oint_{S} d\phi^{I}\wedge A^{J} +\frac{q_{IJK}n}{4\pi^2}\oint_{S} d\phi^{J}\wedge A^{K}. 
\end{align}
Upon fixing background $A$ such that 
\begin{align}
\oint_{L_i\in H_1(T^3,Z)}A^I= \frac{2\pi \lambda_{i}^I}{n} 
\end{align}
we define twisted Hilbert spaces as 
\begin{align}
  \mathcal H^{q}_{\lambda^I_1,\lambda^J_2}=&\;\left\{
   \phi^I(x,y),\zeta^I(x,y)
                                             \Big| \oint_{L_{i}}d\phi^I=\frac{2\pi \lambda_i^I}{n}, \right. \nonumber \\
&\; \left.
\oint_{T^2}d\zeta^I=\frac{2\pi}{n} \left(q_{IJ}\epsilon^{ij}\lambda^{I}_{i}\lambda^{J}_{j}
+
q_{IJK}\epsilon^{ij}\lambda^{J}_{i}\lambda^{K}_{j}\right) \right\}
\label{twisths}
\end{align}
\begin{widetext}
The symmetry operators take the form 
\begin{align}
\mathcal G^{q}_{I}(\lambda_0^I)=&\; \exp\left\{\frac{2\pi i\lambda^{I}_0}{n}\mathcal Q^{I}\right\}; \quad \text{where} \ \mathcal Q^I:=\int_{\Sigma}\frac{\delta \mathcal L}{\delta A^I_0} \nonumber \\
=&\; \exp\left\{\frac{2\pi i \lambda_0^{I}}{n} \int_{T^2} \left(d\zeta^{I}+\frac{nq_{IJ}}{2\pi} d\phi^{I}\wedge d\phi^{J}
+\frac{nq_{IJK}}{2\pi} d\phi^{J}\wedge d\phi^{K}
\right)\right\} \nonumber \\
=&\; \exp\left\{\frac{2\pi i \lambda_0^{I}}{n}\Big[\beta_0^{I}+\frac{nq_{IJ}}{2\pi }\epsilon^{ij} \beta_i^I\beta_j^{J}
+\frac{nq_{IJK}}{2\pi }\epsilon^{ij} \beta_i^J\beta_j^{K}
\Big]\right\}
\label{twistso}
\end{align}
Using the twisted Hilbert space \eqref{twisths} and the symmetry operator \eqref{twistso}, the twisted partition functions can be computed
\begin{align}
Z^{q}_{\lambda^I_0,\lambda^I_1,\lambda^I_2}=&\; \text{Tr}_{\mathcal H^q_{\lambda_i^I}}\left[\mathcal G^q_I(\lambda^I_0)e^{2\pi iR_0H'}\right] \nonumber \\
=&\; Z_{osc} 
 \sum_{N^I_{0,1,2}\in\mathbb Z}\exp\sum_{I=1,2,3}\left\{
-\frac{\pi \tau_2}{2R_2}\left[N^I_{0}+\epsilon^{ij}q_{IJ}\left(N_{i}^{I}+\frac{\lambda_i^{I}}{n}\right)\lambda_j^{J}
+\epsilon^{ij}q_{IJK}\left(N_{i}^{J}+\frac{\lambda_i^{J}}{n}\right)\lambda_j^{K}\right]^2 \right. \nonumber \\
&\; \left. 
- 2\pi R_2\tau_2\left(N^I_1+\frac{\lambda^I_1}{n}\right)^2
-\frac{2\pi R_0R_1}{R_2}\left(N^I_2+\frac{\lambda^I_2}{n}\right)^2 
\right. \nonumber \\
&\; + \left.
2\pi i\tau_1 \left(N^I_1+\frac{\lambda^I_1}{n}\right) \left[N^I_{0} 
+\epsilon^{ij}q_{IJ}\left(N_{i}^{I}+\frac{\lambda_i^{I}}{n}\right)\lambda_j^{J}
+\epsilon^{ij}q_{IJK}\left(N_{i}^{J}+\frac{\lambda_i^{J}}{n}\right)\lambda_j^{K}
\right] \right. \nonumber \\
&\; \left.  +
\frac{2\pi i\lambda_0^I}{n}\left[N^I_{0} 
+2\epsilon^{ij}q_{IJ}\left(N_{i}^{I}+\frac{\lambda_i^{I}}{n}\right)\lambda_j^{J}
+2\epsilon^{ij}q_{IJK}\left(N_{i}^{J}+\frac{\lambda_i^{J}}{n}\right)\lambda_j^{K}
\right]
\right\}. 
\label{twistpf}
\end{align}
\end{widetext}
Under large gauge transformations,
the partition functions in the different sectors
transform as
\begin{align}
Z^{q}_{\lambda^{I}_0+n{\bf{e}}_{I},\lambda_1^{I},\lambda^{I}_2}=e^{\frac{2\pi i\epsilon_{ij}}{n}\left(q_{IJ}\lambda^{I}_i\lambda^{J}_j+q_{IJK}\lambda^{J}_i\lambda^{K}_j\right)} 
Z^{q}_{\lambda^{I}_0,\lambda_1^{I},\lambda^{I}_2}.
\label{surf_large_gauge_anomaly}
\end{align}
On the other hand, 
under $SL(3,\mathbb Z)$ modular transformations,
the partition functions in the different sectors transforms as 
\begin{align}
  U_{2}Z^{q}_{\lambda^I_0,\lambda^{I}_1,\lambda^I_2}=&\;
                                                       e^{-\sum_{I}\frac{2\pi i\lambda_{1}^{I}\epsilon^{ij}}{n^2}
\left( 
q_{IJ}\lambda_i^I\lambda^{J}_j
+q_{IJK}\lambda_i^J\lambda^{K}_j\right)}
                                                       Z^{q}_{\lambda_0^I-\lambda^{I}_1,\lambda^I_1,\lambda^{I}_2},
                                                       \nonumber \\
  M Z^{q}_{\lambda^I_0,\lambda^{I}_1,\lambda^I_2}=&\;  Z^{q}_{\lambda^I_0,-\lambda^{I}_2,\lambda^I_1},
                                                    \nonumber \\
U'_1 Z^{q}_{\lambda^I_0,\lambda^{I}_1,\lambda^I_2}=&\; e^{\sum_I\frac{4\pi i\lambda^I_0\epsilon^{ij}}{n^2}\left(q_{IJ}\lambda_i^I\lambda_j^{J}+
q_{IJK}\lambda_i^J\lambda_j^{K}\right)}
 Z^{q}_{-\lambda^I_1,\lambda^{I}_0,\lambda^I_2}. 
\end{align}
Let us try to construct a modular invariant partition function 
\begin{align}
  Z^{q}_{\text{orb}}=\frac{1}{|G|}\sum_{\lambda^I_0,\lambda^I_1,\lambda^I_2\in G} \theta^q(\lambda^I_0,\lambda^I_1,\lambda^I_2)
  Z^{q}_{\lambda^I_0,\lambda^I_1,\lambda^I_2}.
\end{align}
Imposing invariance under $U_2$ transformation, we obtain
\begin{align}
\frac{\theta^{q}(\lambda^I_0,\lambda^I_1,\lambda^I_2)}{\theta^{q}(\lambda^{I}_0-\lambda^{I}_1,\lambda^{I}_1,\lambda^{I}_2)}=e^{\sum_{I}\frac{2\pi i}{n^2}
\left( 
q_{IJ}(\lambda_1^I)^2\lambda^{J}_2
+q_{IJK}(\lambda_1^J)^2\lambda^{K}_2\right)}.
\label{eqn1}
\end{align}
Inspecting the $U_2$ transformation property of $\theta^{q}$, we find the following constraints under large gauge transformations
\begin{align}
  \theta^{q}\left(n({\bf{e}}_I+{\bf{e}}_J),{\bf{e}}_I+{\bf{e}}_J,{\bf{e}}_J\right)=&\;e^{\frac{2\pi iq_{IJ}}{n}}\theta^{q}\left(0,{\bf{e}}_I+{\bf{e}}_J,{\bf{e}}_J\right),
                                                                                     \nonumber \\
  \theta^{q}\left(n({\bf{e}}_I+{\bf{e}}_J),{\bf{e}}_I+{\bf{e}}_J,{\bf{e}}_K\right)=&\;e^{\frac{2\pi iq_{IJK}}{n}}\theta^{q}(0,{\bf{e}}_I+{\bf{e}}_J,{\bf{e}}_K).
                                                                                     \nonumber \\
\end{align}
This shows that there is a conflict between gauge
invariance and modular invariance
when $q\neq 0$
indicating a 't-Hooft anomaly. 

\bigskip \noindent
To show that this 't-Hooft anomaly
for $Z^{q}[M,A]$ is cancelled by a bulk SPT,
consider the following combination of type-II and type-III
response theories \eqref{respacito}:
\begin{align}
I^{q}[N,A]=&\;-\frac{n}{4\pi^2}\int_{N}\Big\{q_{IJ}A^{J}\wedge A^{I}\wedge dA^{I}   \nonumber \\ &\;+q_{IJK}A^{J} \wedge A^{K} \wedge dA^{I}\Big\}.
\end{align}
Let $N=D^{2}\times S^{1}\times S^{1}$ with a $G$ configuration such that the holonomies around the first and second $S^1$ are $\lambda^{I}_1$ and $\lambda^{I}_2$ respectively. Further consider a puncture on $D^2$ such that the holonomy of the gauge field around $\partial D^{2}$ is $\lambda_0^{I}$. We denote this configuration $[N,A]\equiv D^{2}_{\lambda_0^I}\times S^{1}_{\lambda_1^{I}} \times S^{1}_{\lambda_2^{I}}$. The response theory for this background configuration evaluates to 
\begin{align}
e^{iI^{q}[D^{2}_{\lambda_0^I}\times S^{1}_{\lambda_1^{I}} \times S^{1}_{\lambda_2^{I}},A]}=e^{-\frac{2\pi i\epsilon^{ij}}{n^{2}}\left[q_{IJ}\lambda_0^{I}\lambda_{i}^{I}\lambda_{j}^{J}+
q_{IJ}\lambda_0^{I}\lambda_{i}^{J}\lambda_{j}^{K}
 \right]}
\end{align} 
which has the same properties as those required from $\theta^{q}(\lambda^{I}_0,\lambda^{I}_1,\lambda^{I}_2)$ in order to make the gauged theory consistent.

\bigskip \noindent {\bf{$G$-characters and topological data:}} Similar to the $1+1$-dimensional case one can construct $G$ characters from the $2+1d$ surface theory which encode topological data of the bulk topological gauge theory labelled by $[\omega]\in H^{4}_{\text{group}}(G,U(1))$. The characters are constructed as
\begin{align}
\chi^{q}_{\mu^I,\lambda^I_1,\lambda^I_2}=\frac{1}{\sqrt{|G|}}\sum_{\lambda^I_0\in G}\mu^I(\lambda^I_0) Z^{q}_{\lambda^I_0,\lambda^I_1,\lambda^I_2}
\end{align}
where ${\bf{\mu}}\in \text{Rep}(G)$, for $G=\Z_n^k$,
$\mu^I(\lambda^I_0)=\exp\left\{\frac{2\pi i \delta_{IJ}\mu^{I}\lambda^{J}_0 }{n}\right\}$. Instead of working with $Z^{q}_{\lambda^I_0,\lambda^I_1,\lambda^I_2}$, we find it convenient and illustrative to work with  $\bar{Z}^{q}_{\lambda^I_0,\lambda^I_1,\lambda^I_2}$, where
\begin{align}
\bar{Z}^{q}_{\lambda^I_0,\lambda^I_1,\lambda^I_2}:=&\; \gamma_{\lambda^I_1,\lambda^I_2}^{q}(\lambda^I_0)Z^{0}_{\lambda^I_0,\lambda^I_1,\lambda^I_2} 
\end{align}
where $\gamma_{\lambda^I_1,\lambda^I_2}^{q}$
is a projective $G$ representation which satisfies
\begin{align}
\gamma_{\lambda^I_1,\lambda^I_2}^{q}(\lambda^I_0)\gamma_{\lambda^I_1,\lambda^I_2}^{q}(\lambda_{0}^{I'})=&\; \beta_{\lambda^I_1,\lambda^I_2}^{q}(\lambda^I_0,\lambda_0^{I'})\gamma_{\lambda^I_1,\lambda^I_2}^{q}(\lambda^I_0+\lambda_0^{I'}) \nonumber \\
\text{where} \qquad \beta^{q}_{\lambda^I_1,\lambda^I_2}(\lambda^I_0,\lambda_0^{I'})=&\; i_{\lambda^I_1}i_{\lambda^I_2}\omega^{q}(\lambda^I_1,\lambda^I_2,\lambda^I_0,\lambda_0^{I'}) \nonumber
\end{align}
\\
\noindent We note that $\bar{Z}^{q}_{\lambda^I_0,\lambda^I_1,\lambda^I_2}$ and $Z^{q}_{\lambda^I_0,\lambda^I_1,\lambda^I_2}$ have the same properties under modular and large gauge transformations, hence it will suffice for our purposes to use $\bar{Z}^q$ instead of $ Z^q$. Then we may write
\begin{align}
\bar{\chi}^{q}_{\mu^I,\lambda^I_1,\lambda^I_2}=&\;\frac{1}{\sqrt{|G|}}\sum_{\lambda^I_0\in G}\mu^I(\lambda^I_0)\bar{Z}^{q}_{\lambda^I_0,\lambda^I_1,\lambda^I_2} \nonumber \\
=:&\;  \frac{1}{\sqrt{|G|}}\sum_{\lambda^I_0\in G}\Gamma^{q}_{\mu^I,\lambda^I_1,\lambda^I_2}(\lambda^I_0)Z^{0}_{\lambda^I_0,\lambda^I_1,\lambda^I_2}
\end{align}
where
$\Gamma^q_{\mu^I,\lambda^I_1,\lambda^I_2}(\lambda^I_0):=\mu^I(\lambda^I_0)\gamma^q_{\lambda^I_1,\lambda^I_2}(\lambda^I_0)$. 
For the specific case of type-II and type-III cocycle, 
$\gamma^q_{\lambda^I_1,\lambda^I_2}(\lambda^I_0)$ takes the form
\begin{align}
\gamma^q_{\lambda^I_1,\lambda^I_2}(\lambda^I_0)=\exp\left\{\frac{2\pi i}{n^2}\left(q_{IJ}\lambda_0^{I}\lambda_1^{I}\lambda_2^{J}
+
q_{IJK}\lambda_0^{I}\lambda_1^{J}\lambda_2^{K}
\right)\right\}
\end{align} 
By the bulk boundary correspondence,
the character $\bar{\chi}^q_{\mu^I,\lambda^I_1,\lambda^I_2}$ corresponds to a bulk
excitation with linked fluxes $\lambda^I_1$ and $\lambda^I_2$ and charge $\mu^I$. \cite{moradi2015universal, chen2016bulk}\footnote{We thank Xueda Wen for clarifying this picture}
The dimension of the representation $\text{dim}(\Gamma^q_{\mu^I,\lambda^I_1,\lambda^I_2})$
is the quantum dimension of the excitation corresponding to $\bar{Z}^q_{\lambda_0^{I},\lambda^I_1,\lambda^I_2}$. \\

\noindent The modular $SL(3,\mathbb Z)$ matrices can be computed as 
\begin{widetext}
\begin{align}
U_2\bar{\chi}^q_{\mu^I ,\lambda^I_1,\lambda^I_2}=&\;\frac{1}{\sqrt{|G|}}\sum_{\lambda_0^{I}\in G}\Gamma^q_{\mu^I ,\lambda^I_{1},\lambda^I_{2}}(\lambda^I_{0}) Z^0_{\lambda^I_0+\lambda_1^I,\lambda^I_{1},\lambda^I_2} \nonumber \\
=&\; \frac{1}{\sqrt{|G|}}\sum_{\lambda^I_0}\frac{\Gamma^q_{\mu^I ,\lambda^I_{1},\lambda^I_{2}}(\lambda^I_0)}{\Gamma^q_{\mu^I ,\lambda^I_{1},\lambda^I_{2}}(\lambda^I_{0}+\lambda_1^I)}\Gamma^q_{\mu^I ,\lambda^I_{1},\lambda^I_{2}}(\lambda^I_0+\lambda^I_1)
Z^0_{\lambda^I_0+\lambda^I_1 ,\lambda^I_{1},\lambda^I_{2}} \nonumber \\
=&\; \exp\left\{-\frac{2\pi i\lambda_0^{I}}{n^2}\left(q_{IJ}\lambda_1^{I}\lambda_2^{J}+q_{IJK}\lambda_1^{J}\lambda_2^{K}\right) 
   -\frac{2\pi i\delta_{IJ}\mu^{I} \lambda_{1}^{J} }{n}\right\}\bar{\chi}^q_{\mu^I,\lambda^I_1,\lambda^I_2},
   \nonumber \\
{U'}_1\bar{\chi}^q_{\mu^I, \lambda^I_1,\lambda^I_2}=&\;\frac{1}{\sqrt{|G|}}\sum_{\lambda^I_0}\Gamma^q_{\mu^I,\lambda^I_1,\lambda^I_2}(\lambda^I_0)Z_{\lambda^I_1,\lambda^I_0,\lambda^I_2} \nonumber \\ 
=&\; \frac{1}{|G|}\sum_{\lambda_1^{I'},\mu^{I'}}\Gamma^q_{\mu^I,\lambda^I_1,\lambda^I_2}(\lambda^{I'}_1)\left[\Gamma^q_{\mu^{I'},\lambda^{I'}_1,\lambda^{I}_2}(\lambda^I_1)\right]^{-1}\bar{\chi}^q_{\mu^{I'},\lambda_1^{I'},\lambda^{I}_2}
\end{align}
\end{widetext}
These match with modular matrices computed directly from the orbifold partition functions with twisted symmetry action. The $U_2$ eigenvalues are analogous to topological spin for string operators whereas the projective phases for the $U_1'$ transformation encodes the braiding statistics between string-like and particle like excitations as well braiding of three-strings known as three-loop braiding. \cite{wang2014braiding, jiang2014generalized, chen2016bulk, tiwari2016wilson}

\bigskip \noindent {\bf{SPT invariants from surface computations:}} Above we saw that bosonic SPTs protected by $G=\Z_n^k$ and described by type-II and/or type-III 4-cocycles `$q$'$\in H^{4}_{\text{group}}(G,U(1))$ can be detected by their partition functions on $L(n,1)\times S^1$ with appropriate background $G$-bundle. Now we show that these invariants can be directly computed from the surface theory \eqref{2+1_action}. This computation is based on the fact that the partition function on $L(n,1)\times S^1$ can be simulated by the groundstate expectation value of a partial $C_n$ rotation operation on the spatial manifold $S^2\times S^1$.\cite{shiozaki2016many}

\bigskip \noindent Consider putting the theory \eqref{SPT_eff_ac} with type-II and/or type-III coupling to background field $A$ on spatial manifold $M=S^2\times S^1$, since  $H_{1}(S^2\times S^1,\Z)=\Z$, we may introduce a background field with holonomy ${\bf{b}}\in G$ around this spatial $S^1$. We denote this groundstate as $|GS^q_{S^2\times S^1_{\bf{b}}}\rangle$. Let $\hat{C}_{n,D}(a)$ be an operator implementing a non-local partial rotation on a disc-like region $D\subset S^{2}$ with flux ${\bf{a}} \in G$ inserted. Then we may show that the SPT invariant is
given by the phase of
\begin{align}
\mathcal Z^q[L(n,1)\times S_{\bf{b}}^1,A]=&\; \langle GS^q_{S^2\times S^1_{\bf{b}}}|\hat{C}_{n,D}({\bf{a}})|GS^q_{S^2\times S^1_{\bf{b}}}\rangle \nonumber \\
=&\; \frac{\text{Tr}_{\mathcal H^{q}(D\times S^{1}_{\bf{b}})}\left[\rho_{D\times S^{1}_{\bf{b}}}\hat{C}_{n,D}({\bf{a}})\right] }{\text{Tr}_{\mathcal H^{q}(D\times S^{1}_{\bf{b}})}\left[\rho_{D\times S^{1}_{\bf{b}}}\right]} \nonumber \\
=&\; \frac{\text{Tr}_{\mathcal H^{q}(S^1\times S^{1}_{\bf{b}})}\left[e^{-\xi H_{T^2}}\hat{C}_{n,\partial D}({\bf{a}})\right]}
{\text{Tr}_{\mathcal H^{q}(S^1\times S^{1}_{\bf{b}})}\left[e^{-\xi H_{T^2}}\right]}
\end{align}
where we have traced over disc-like region $\bar{D}$ complement to $D\subset S^2$ and used the fact that the reduced density matrix effectively reduces to the thermal density matrix of the gapless surface on $\partial [D\times S^1]=T^2$ 
\begin{align}
\rho_{D\times S^{1}_{\bf{b}}}\approx \frac{e^{-\xi H_{T^2}}}{\text{Tr}_{\mathcal H^{q}(S^1\times S^1_{\bf{b}})}\left[e^{-\xi H_{T^2}}\right]}
\end{align}
where $\mathcal H^q(S^1\times S^{1}_{\bf{b}})$ is the Hilbert space on the torus with holonomies $0,{\bf{b}}\in G$ along $(\partial D,S^1_{\bf{b}})$ respectively. Let $\tau:=\tau_1+i\tau_2$ denote the modular parameter for the $t-x$ two torus on which the modular matrices $U_2,U^{'}_1$ act as $T,S\in SL(2,\Z)\subset SL(3,\Z)$. Since the $\hat{C}_{n,\partial D}$ acts as a boost along the $x$ direction. The computation is effectively very similar to the $1+1d$ calculation \eqref{lens_sim}, except with holonomy ${\bf{b}}\in G$ inserted along the $S^1$ cycle in the $y$-direction 
\begin{widetext}
\begin{align}
\mathcal Z^q[L(n,1)\times S^1,A]
=&\; \frac{Z^{q}_{({\bf{a}},0,{\bf{b}})}\left(\tau=\frac{i\xi}{L}-\frac{1}{n}\right) }
{Z^q_{(0,0,0)}\left(\tau=\frac{i\xi}{L}\right)} \nonumber \\
=&\; \frac{\sum_{{\bf{c}}}\left(\Gamma \right)_{({\bf{a}},0,{\bf{b}})}^{({\bf{c}}_{\tau},{\bf{c}}_x,{\bf{c}}_y)}Z^q_{({\bf{c}}_{\tau},{\bf{c}}_x,{\bf{c}}_y)} \left(\tau=-\frac{1}{n}+\frac{iL}{\xi n^2}\right)}
{\sum_{{\bf{c}}}\left(U_1^{'}\right)_{(0,0,0)}^{({\bf{c}}_{\tau},{\bf{c}}_x,{\bf{c}}_y)} Z^q_{({\bf{c}}_{\tau},{\bf{c}}_x,{\bf{c}}_y)} \left(\tau= \frac{iL}{\xi}\right)} \nonumber \\
=&\; e^{\sum_{I}\frac{2\pi i}{n} \left(q_{IJ}a^{I}(a^{I}b^{J}-b^{I}a^{J})+q_{IJK}a^{I}(a^{J}b^{K}-b^{J}a^{K})\right)}\frac{ Z^q_{(-{\bf{a}},0,{\bf{b}})}\left(\tau=-\frac{1}{n}+\frac{iL}{\xi n^2}\right)}{ Z^q_{(0,0,0)} \left(\tau=\frac{iL}{\xi}\right)} \nonumber \\
=&\; e^{\sum_{I}\frac{2\pi i}{n} \left(q_{IJ}a^{I}(a^{I}b^{J}-b^{I}a^{J})+q_{IJK}a^{I}(a^{J}b^{K}-b^{J}a^{K})\right)}\left(1+\mathcal O(e^{-L/\xi})+ \cdots \right),
\end{align}
\end{widetext}
where in the 2nd line we have defined the diffeomorphism
$\Gamma=U_1^{'}U_2^nU_1^{'}$ Then by taking the limit $\xi/L\to 0$, we can read
off the SPT invariant.

\section{$d+1$-dimensional topological phases and their $d$-dimensional boundaries}
\label{generalities}
\noindent Several features discussed in the previous sections for $2+1d$ and $3+1d$ bosonic SPTs can be generalized to arbitrary dimensions. Let us consider bosonic SPT phases protected by symmetry $G$ in $d+1$-dimensions where $G$ is a discrete abelian group which for simplicity we shall assume to be $\Z_n^k$. and their $d$-dimensional boundaries. SPT phases with discrete abelian symmetry $G$ are classified by $H_{\text{group}}^{d+1}(G,U(1))$. Then each such SPT phase can be labelled by a group cocycle $[\omega]\in H_{\text{group}}^{d+1}(G,U(1))$. Let us consider a few low dimensional examples of the group cohomology classification
\begin{widetext}
\begin{align}
H^{2}_{\text{group}}[\Z_n^k,U(1)]=&\; (\mathbb Z_{n})^{\left[
\left( \begin{array}{c}
k \\
         2  \end{array} \right)\right]},
  \nonumber \\
H^{3}_{\text{group}}[\mathbb Z_n^k,U(1)]
=&\; (\mathbb Z_{n})^{\left[
\left( \begin{array}{c}
k \\
1  \end{array} \right)
+\left( \begin{array}{c}
k \\
2  \end{array} \right)
+
\left( \begin{array}{c}
k \\
3  \end{array} \right)\right]},
 \nonumber \\
H^{4}_{\text{group}}[\mathbb Z_n^k,U(1)]
=&\; (\mathbb Z_{n})^{\left[
2\left( \begin{array}{c}
k \\
2  \end{array} \right)
+\left( \begin{array}{c}
k \\
3  \end{array} \right)
+
\left( \begin{array}{c}
k \\
         4  \end{array} \right)\right]},
  \nonumber \\
H^{5}_{\text{group}}[\mathbb Z_n^k,U(1)]
=&\; (\mathbb Z_{n})^{\left[
\left( \begin{array}{c}
k \\
1  \end{array} \right)
+
2\left( \begin{array}{c}
k \\
2  \end{array} \right)
4\left( \begin{array}{c}
k \\
3  \end{array} \right)
+
3\left( \begin{array}{c}
k \\
4  \end{array} \right)
+
\left( \begin{array}{c}
k \\
5  \end{array} \right)
\right]}.
\end{align}
\end{widetext}
We can read-off some pattern, notably in odd-dimensions due to the existence of Chern-Simons terms one can build a topological action with a single $\Z_n$ gauge field.  The procedure to build a continuum topological action from a $d+1$-cocycle or vice versa is essentially the same as the lower dimensional analogs. For example in $4+1d$, the Chern-simons like terms $(q_{IJK}/4\pi^2)A^{I}\wedge dA^{J} \wedge dA^{K}$ correspond to the cocycle
\begin{align}
\omega_{q_{IJK}}({\bf{a}},{\bf{b}},{\bf{c}},{\bf{d}},{\bf{e}})=&\;e^{\frac{2\pi q_{IJK}}{n^{3}}a^{I}(b^{J}+c^{J}-[b^{J}+c^{J}])
 (d^{J}+e^{J}-[d^{K}+e^{K}])}
\label{qIJK}
\end{align}
Similarly the topological action of the kind 
\begin{align}
I^{q_{IJKL}}=-\frac{q_{IJKL}n^2}{8\pi^3}\int_{N_{d+1}}A^{I}\wedge A^{J}\wedge A^{K}\wedge dA^{L}
\label{AAAdA}
\end{align}
corresponds to the cocycle
\begin{align}
\omega_{q_{IJKL}}({\bf{a}},{\bf{b}},{\bf{c}},{\bf{d}},{\bf{e}})=e^{\frac{2\pi q_{IJKL}}{n^2}a^{I}b^{J}c^{K}(d^{L}+e^{L}-[d^{L}+e^{L}])}.
\label{qIJKL}
\end{align}
Next, one can design effective actions for SPTs with specific actions of $G$ which imply specific coupling to the flat background $G$ gauge field $A$. Of course these models must have a unique groundstate, no fractional excitations and most importantly furnish the correct topological response theories. By inspection one can realize that all such models can be modeled simply as multicomponent $BF$ theories at `level' 1. In $d+1$-dimensions these take the form 
\begin{align}
\mathcal S=\int_{N_{d+1}}\left[\frac{\delta_{IJ}}{2\pi} b^{I}\wedge da^{J} +\frac{1}{2\pi}A^{I} \wedge (db^{I}+ \dots)\right],
\end{align}
where $b^{I}$ and $a^{I}$ are $d-1$-form and 1-form $U(1)$ connections which satisfy the usual Dirac quantization conditions. $`\dots'$ refers to piece in the coupling to background gauge field that determines the topological response.  
For example for the coupling to background $A$ that gives rise to Chern-Simons like term `$AdAdA$' and `$AAAdA$' type term \eqref{AAAdA} respectively are
\begin{align} 
\mathcal S^q_{cpl}=&\;-\frac{1}{2\pi}\int A^{I} \wedge (db^{I}-\frac{q}{2\pi}da^{J}\wedge da^{K}) \quad  \text{and} \nonumber \\ 
\mathcal S^q_{cpl}=&\;-\frac{1}{2\pi}\int A^{I} \wedge (db^{I}-\frac{q}{4\pi^2}a^{J}\wedge a^{K}\wedge da^{L}).
\end{align}   
The gauging and ungauging procedures too have straightforward generalizations.
The partition function takes the form
\begin{align}
\mathcal Z_{\text{DW}}^{q}[N]
  =\frac{1}{|G|}\sum_{[A]\in H^{1}(N,G)}e^{iI^{q}[N,A]},
\end{align}
where $I^{q}[N,A]$ is the topological response theory corresponding to an SPT
labelled
by cocycle $q\in H^{d+1}_{\text{group}}(G,U(1))$ that is obtained after integrating out $a,b$. This can be ungauged as
\begin{align}
  e^{iI^{q}[N,A]}=\sum_{J_{qp}\in H^{d}(\widehat{G},U(1))}e^{-i\int_{N}J_{qp}\wedge A}
  Z^q_{\text{DW}}[N,J_{qv}]. 
\end{align}
The generalization of boundary physics is more subtle. First we propose a surface theory described by the action
\begin{align}
S =\frac{1}{2\pi}\int_{M_d}d\zeta \wedge d\varphi-H[\zeta,\phi].
\label{boundary_eff_ac}
\end{align}
Such a theory may be derived by enforcing the full $U(1)_{0}\times U(1)_{d-1}$ symmetry of the bulk $BF$ theory\cite{amoretti2012three}, where $U(1)_p$ stands for a $p$-form $U(1)$ symmetry. Let $M_d=X_{d-1}\times S^1$  where $X_{d-1}$ is a compact oriented manifold without boundary. The twisted Hilbert space $\mathcal H_{A}(X_{d-1})$ on $X_{d-1}$ in the presence of background $\Z_n^k$ gauge field $A$ can be derived as before. For example for the 5-cocycles \eqref{qIJK} and \eqref{qIJKL} given above, $\zeta$ is a 2-form $U(1)$ connection and $X$ is a 3-manifold, the twisted Hilbert spaces take the form
\begin{widetext}
\begin{align}
  \mathcal H_{A}^{q_{IJK}}(X_3)&=
                                 \left\{\zeta^{I}(x),\varphi^{I}(x) \Bigg| \oint_{L}\frac{d\varphi^{I}}{2\pi}=\oint_{L}\frac{A^{I}}{2\pi}; \quad \oint_{V}\frac{d\zeta^{I}}{2\pi}=\frac{q_{IJK}}{4\pi^2}\oint_{V}A^{J}\wedge dA^{K}  
                                 \right\},
                                 \nonumber \\
\mathcal H_{A}^{q_{IJKL}}(X_3)&= \left\{\zeta^{I}(x),\varphi^{I}(x) \Bigg| \oint_{L}\frac{d\varphi^{I}}{2\pi}=\oint_{L}\frac{A^{I}}{2\pi}; \quad \oint_{V}\frac{d\zeta^{I}}{2\pi}=\frac{q_{IJKL}}{8\pi^3}\oint_{V}A^{J}\wedge A^{K} \wedge A^{L}  
                                \right\},
\label{twist_hs_general}
\end{align}
\end{widetext}  
where $L\in H_1(X_3,\Z)$ and $V\in H_3(X_3,\Z)$. We note that it is not clear how to implement this procedure for `type-$d+1$' cocycles $\in H^{d+1}(\Z_n^{k},U(1))$. These cocycles take the form 
\begin{align}
  &
    \omega_{q_{I_1I_2I_3 \cdots I_{d+1}}}(a_1^{I_1},a_2^{I_2},\cdots,a_{d+1}^{I_{d+1}})
    \nonumber \\
  &\quad 
    =e^{\frac{2\pi iq_{I_1I_2 \cdots I_{d+1}} }{n}a_1^{I_1}a_2^{I_2}\dots a_{d+1}^{I_{d+1}}}
\end{align} 
and generally give non-abelian topological order upon gauging in the bulk. A quick way to see this is by the fact that these cocycles reduce to non-abelian topological order upon dimensional reduction. Alternately one can check that this kind of cocycle gives rise to an algebra that does not have any non-trivial one-dimensional representations. Since the charges in Dijkgraaf-Witten theories carry a `twisted' representation. This leads to the fact that non-trivial fluxes have quantum dimension $>1$. They cannot be embedded in $U(1)^{k}$, hence we need to go beyond effective field theory of the form \eqref{boundary_eff_ac} to model boundary theories for SPTs protected by such group cocyles.

\bigskip \noindent For all other cocycle types the twisted partition function may be computed on $M_d=X_{d-1}\times S^1$ as
\begin{align}
  Z^{q}[M_{d},A]=
 \text{Tr}_{\mathcal H_A(X_{d-1})}
  \left[\prod_{I}\mathcal G^{I}\left(\oint_{S^1} \frac{A^{I}}{2\pi}\right) e^{2\pi i R_0 H}\right]
\end{align}
where $\mathcal G^{I}$ is the $\Z_n$ symmetry operator corresponding to $I$-th
$\Z_n$ copy and $R_0$ is the radius of $S^1$ along the time direction.
Then we expect $Z^{q}[M_d,A]$ to have a 't-Hooft anomaly that can be cancelled by the response of an SPT on $N_{d+1}|_{\partial N_{d+1}=M_d}$, i.e., together the bulk and boundary partition functions
\begin{align}
Z^{q}[M_d,A]e^{iI^q[N_{d+1},A]}
\end{align}
are gauge invariant and do not suffer from any 't-Hooft anomaly. 

\section{$2+1d$ surface with $U(1)\times \Z_2^{R,T}$ 't-Hooft anomaly}
\label{CP_x_U(1)_SPT}
\noindent
In this section we study a mixed $U(1)\times \Z_{2}^{T,R}$ 't-Hooft anomaly for the following model:
\begin{align}
S=\int_{ M}\left[\frac{1}{2\pi }d\zeta\wedge d\phi-\mathcal H(\zeta,\phi) \right].
\label{CPtheory}
\end{align}
Here, $\mathbb{Z}^{T,R}_2$ represents time-reversal or reflection symmetry,
which can be combined with unitary on-site symmetry.
We show that for different symmetry actions
there may be a $\Z_{2}^{T,R}\times U(1)_0$ or $\Z_{2}^{T,R} \times U(1)_1$
anomaly where $U(1)_p$ refers to $p$-form $U(1)$ global symmetry.
We show that for such a symmetry action, the $\Z_{2}^{T,R}$
projected partition function is not invariant under large $U(1)_{p}$ gauge transformation. 
In the context of fermionic SPT phases, 
similar calculations have been carried out
for the surface theory (gapless $(2+1)d$ Dirac fermion theory) 
of $(3+1)d$ time-reversal or $CR$ 
symmetric topological insulators. 
\cite{hsieh2016global}

\bigskip \noindent 
Details of quantization of \eqref{CPtheory} can be found in
App.\ \ref{2+1_action_appendix}. 
Here we will need the form of the mode expansion 
which decomposes into oscillator and zero-mode parts as
\begin{align}
  \phi(x,y,t)&= \phi^0(x,y,t) + \phi^{osc}(x,y,t),
               \nonumber \\
\zeta_j(x,y,t)&= \zeta_j^0(x,y,t) + \zeta_j^{osc}(x,y,t). 
\end{align}
The zero-mode part takes the form
\begin{align}
  \phi(x,y,t) &= \alpha_0+\frac{\beta_1 x}{R_1}+\frac{\beta_2 y}{R_2}+ \cdots,
                \nonumber \\
\zeta_{j}(x,y,t)&= \frac{\alpha_j}{2\pi R_j}+\frac{\beta_0}{2\pi R_1R_2}x\delta_{j,2}+ \cdots.
\label{zeromode_expansion}
\end{align}
The canonical algebra for this theory implies $\left[\alpha_0,\beta_0\right]=i$ and $\left[\alpha_1,\beta_2\right]=i=-\left[\alpha_2,\beta_1\right]$. We will only be interested in the zero mode part of the mode expansion throughout this section as we seek to diagnose mixed $\Z_{2}^{T,R}\times U(1)_p$ anomaly and $U(1)_p$ only acts on the zeromode part of the mode expansion.
\begin{center}
		\small{\bf $U(1)_0$ and $U(1)_1$ symmetry}\\
	\end{center}
The action \eqref{CPtheory} is invariant under a 0-form and 1-form $U(1)$ symmetry. The 0-form symmetry transformation is 
\begin{align}
\mathcal G^{(0)}(\theta):\phi \mapsto \phi + \theta
\end{align}
explicitly the symmetry operator is $\mathcal G^{(0)}(\theta)=\exp\left\{i \beta_0 \theta \right\}$. To gauge the 0-form $U(1)$ symmetry we introduce a flat 1-form background gauge field $A$, the gauge equivalence 
\begin{align}
  \phi(x)&\mapsto  \phi(x)+\theta(x),
           \nonumber \\
A(x)&\mapsto A(x)- d\theta(x), 
\end{align}
and define the covariant derivative $D_{A}\phi:= d\phi +A$. Then the gauged action is 
\begin{align}
S[\zeta,\phi,A]=\int_{M}\left[\frac{1}{2\pi}d\zeta\wedge D_A\phi -\mathcal H \right]. 
\end{align} 
Notice that $\phi$ satisfies $U(1)_0$ twisted quantization condition
\begin{align}
\oint_{L_i\in H_1(T^2,\Z)}\frac{d\phi}{2\pi}=\oint_{L_i\in H_1(T^2,\Z)}\frac{A}{2\pi}:= \lambda_i.
\end{align} 
Hence we may define the $U(1)_0$ twisted Hilbert space as
\begin{align}
\mathcal H_{\lambda_1,\lambda_2}=\Big\{\phi(x),\zeta(x) \Big|
\oint_{L_i\in H_1(T^2,\Z)}\frac{d\phi}{2\pi}= \lambda_i
\Big\}
\end{align}
Similarly \eqref{CPtheory} is invariant under a global 1-form $U(1)$ symmetry under which acts as
\begin{align}
\mathcal G^{(1)}(\theta \eta):\zeta(x)\mapsto \zeta(x)+\theta \eta(x); \qquad \theta\in \R/2\pi \Z  
\label{1_form_symmetry_operator}
\end{align}
where $\eta$ is a flat bundle. Gauging the 1-form $U(1)$ symmetry implies introducing a flat 2-form background gauge field $B$, and the gauge equivalence 
\begin{align}
  \zeta(x) &\mapsto  \zeta(x)+\theta(x)\eta(x),
             \nonumber \\
B(x) &\mapsto B(x)- d\theta(x) \wedge \eta (x), 
\end{align}
with the covariant derivative $D_{B}\zeta:= d\zeta +B$. The gauged action is 
\begin{align}
S[\zeta,\phi,B]=\int_{M}\left[\frac{1}{2\pi}D_{B}\zeta\wedge d\phi -\mathcal H \right]. 
\end{align} 
The 1-form field $\zeta$ satisfies $U(1)_1$ twisted quantization condition
\begin{align}
\oint_{T^2}\frac{d\zeta}{2\pi}=\oint_{T^2}\frac{B}{2\pi}=: \lambda_0.
\end{align} 
We may define the $U(1)_1$ twisted Hilbert space as
\begin{align}
\mathcal H_{\lambda_0}=\Big\{\phi(x),\zeta(x) \Big|\oint_{T^2}\frac{d\zeta}{2\pi}=\oint_{T^2}\frac{B}{2\pi}= \lambda_0\Big\}
\end{align}

\begin{center}
		\small{\bf $\Z_{2}^{T,R}\times U(1)_0$ anomaly}\\
	\end{center}

\noindent  
Let us consider the following choice of $\Z_2^R$ action implemented by $P_0$ on \eqref{2+1_action},
\begin{align}
  P_0:&\;\phi(t,x,y)\to \phi(t,x,-y),
        \nonumber \\
  :&\;\zeta_1(t,x,y)\to -\zeta_{1}(t,x,-y),
     \nonumber \\
     :&\;\zeta_2(t,x,y)\to \zeta_{2}(t,x,-y)+ \Delta \zeta_2,
     \label{CP2}
\end{align}
where $\Delta\zeta_2=0$ or $\pi$. The zero-mode operators transform under $\Z_2^{R}$ action as
\begin{align}
  P_0:\alpha_0\to&\;  \alpha_0,
                   \nonumber \\
  :\alpha_1\to &\; -\alpha_1,
                 \nonumber \\
  :\alpha_2\to &\; \alpha_2+\Delta \zeta_2 R_2,
                 \nonumber \\
  :\beta_0 \to&\; \beta_0,
                \nonumber \\
  :\beta_1\to &\; \beta_1,
                \nonumber \\
: \beta_2 \to &\;-\beta_2. 
\end{align}
Hence since $\mathcal G^{(0)}(\theta)=e^{i\beta_0\theta}$, we find
$\left[\mathcal G^{(0)}(\theta),P_0\right]=0$. 
We postulate the following $P_0$ action on zeromode vacuum sectors
\begin{align}
P_0|\alpha_0,\alpha_1,\alpha_2\rangle 
=&\; e^{iB_0(\alpha_0,\alpha_1,\alpha_2)} |\alpha_0,-\alpha_1,\alpha_2+\Delta \zeta_2 \rangle,
 \nonumber \\
P_0|\beta_0,\beta_1,\beta_2\rangle =&\; e^{iA_0(\beta_0,\beta_1,\beta_2)} |\beta_0 ,\beta_1,-\beta_2\rangle. 
\label{CPeigen}
\end{align}
The $U(1)$ phase can be read off from the fourier representation of the zero-mode ket
\begin{align}
|\beta_0,\beta_1,\beta_2\rangle=&\; \int \prod_{\mu}d\alpha_{\mu}e^{\left\{i\left(\alpha_0\beta_0+\alpha_1\beta_2-\alpha_2\beta_1\right)\right\}}
|\alpha_0,\alpha_1,\alpha_2\rangle \nonumber \\ 
\end{align}
which implies $A_0(\beta_0,\beta_1,\beta_2)=B_0+\beta_1\Delta\zeta _2$. Writing $\beta_{\mu}=N_{\mu}+\lambda_{\mu}$ where $N_{\mu}\in \Z$ is the untwisted winding mode and $\lambda_{\mu}\in \R/\Z$ is the $U(1)$ twist parameters introduced above. We obtain 
\begin{align}
  P_0|\beta_{0},\beta_1,\beta_2\rangle
  = P[\lambda_1]e^{i N_1\Delta \zeta_2}|\beta_0,\beta_1,-\beta_2\rangle.
\end{align}
If we require that our $\Z_{2}^{R}$ 
action does not depend on $U(1)$ twist\cite{hsieh2014symmetry}, 
we must impose $P[\lambda_1]=1$ 
(i.e., $B=\lambda_1 \Delta \zeta_2$)
\begin{align}
\langle \beta_0,\beta_1,\beta_2|P_0|\beta_0,\beta_1,\beta_2 \rangle =e^{iN_1\Delta\zeta_2 }\delta_{\beta_2,0}.
\end{align}
The $P_0$ twisted partition function in the presence of background $U(1)_0$ flux takes the form
\begin{align}
Z[\mathcal K\times S^1,\lambda_1]=&\;\text{Tr}_{\mathcal H_{\lambda_1}}\left[P_0 e^{-2\pi R_0(H+i\frac{\tau_1}{\tau_2}P_x+(i\frac{\tau_1}{\tau_2}\beta+\gamma)P_y)}\right] \nonumber \\
=&\; Z_{osc}\sum_{N_{0,1}\in\mathbb Z}\exp\Big\{-\frac{\pi \tau_2}{2r^2R_2}N_0^2 \nonumber \\
&\; -2\pi r^2R_2 \tau_2 (N_1+\lambda_1)^2 \nonumber \\ &\;
 +2\pi i\tau_1 N_0(N_1+\lambda_1)+i\Delta \zeta_2 N_1\Big\}.
 \end{align}
Note we cannot insert $\lambda_2$ flux as it is inconsistent with $P_0$
projection. 
For the non-trivial choice of $P_0$ action, i.e., 
$\Delta \zeta_2=\pi$, 
under a large $U(1)_0$ gauge transformation $\lambda_1\to \lambda_1+1$ the parity twisted partition function changes sign 
\begin{align}
  Z[\mathcal K\times S^1,\lambda_1]= - Z[\mathcal K\times S^1, \lambda_1+1].
\end{align}
This is a $\Z_2$ anomaly that signals the existence of a bosonic topological insulator protected by $\mathbb Z_2^{T}\times U(1)_0$ global symmetry.\cite{vishwanath2013physics, kapustin2014bosonic}

\bigskip \noindent
In \cite{kapustin2014bosonic},
it was shown that bosonic SPTs protected by $G=U(1)_0\times \mathbb Z_{2}^{T}$
(or equivalently $U(1)_0 \times \Z_{2}^{R}$)
in $3+1d$ are classified by $\Z_2^4$.
The only mixed term in the response theory takes the form
\begin{align}
I[N,w_1,A]=\int_{N}\frac{n}{2\pi^2}w_1\cup w_1 \cup F
\label{bosonic_paramagnet_resp}
\end{align} 
where $n\in \Z_2$ parametrizes different phases
and $w_1$ is the first Stiefel-Whitney class of the tangent bundle of the
manifold, i.e., $\oint_{L}w_1 =0$ or $\pi $
for any orientation preserving or reversing cycle respectively.
The effective matter theory for such an SPT coupled to background geometry can be modeled as
\begin{align}
\mathcal S=\int_{N}\left[\frac{1}{2\pi} b\cup \delta a + \frac{1}{2\pi} A \cup \delta b +\frac{n}{2\pi^2}w_1\cup w_1 \cup \delta a\right].
\end{align}
Upon integrating out the matter fields $a,b$ using the fact that the cup product is supercommutative upto boundary terms and $\delta$ is a $\Z_2$ graded derivation, we find the correct response \eqref{bosonic_paramagnet_resp}.

\begin{center}
		\small{\bf $\Z_2^{T,R}\times U(1)_1$ anomaly}\\
	\end{center}
  \noindent
  We may consider another distinct $\Z_{2}^{R}$ action given by $P_1$ 
\begin{align}
P_1:&\;\phi(t,x,y)\to \phi(t,x,-y)+\Delta \phi, \nonumber \\
     :&\;\zeta_1(t,x,y)\to -\zeta_{1}(t,x,-y),  \nonumber \\
     :&\;\zeta_2(t,x,y)\to \zeta_{2}(t,x,-y). 
     \label{CP1}
\end{align}
Since $P_1^2=1$, $\Delta \phi=0,\pi$.
We choose non-trivial action, i.e., $\Delta \phi=\pi$. The zero mode operators transform under $P_1$ as
\begin{align}
  P_1:\alpha_0\to&\;  \alpha_0 +\pi,
                   \nonumber \\
  :\alpha_1\to &\; -\alpha_1,
                 \nonumber \\
  :\alpha_2\to &\; \alpha_2,
                 \nonumber \\
  :\beta_0 \to&\; \beta_0,
                \nonumber \\
  :\beta_1\to &\; \beta_1,
                \nonumber \\
: \beta_2 \to &\;-\beta_2. 
\label{CP1}
\end{align}
We postulate the following $P_1$ action on zeromode vacuum sectors
\begin{align}
  P_1|\alpha_0,\alpha_1,\alpha_2\rangle =&\; e^{iB_1(\alpha_0,\alpha_1,\alpha_2)} |\alpha_0 +\Delta \phi,-\alpha_1,\alpha_2\rangle,
                                           \nonumber \\
P_1|\beta_0,\beta_1,\beta_2\rangle =&\; e^{iA_1(\beta_0,\beta_1,\beta_2)} |\beta_0 ,\beta_1,-\beta_2\rangle.
\label{CPeigen}
\end{align}
Similar to the case above for $P_0$, the $U(1)$ phase can be read off from the fourier representation of the zero-mode ket. We find
\begin{align}
P_1 |\beta_0,\beta_1,\beta_2\rangle=&\; \exp\left\{i(B_1-\pi \beta_0)\right\}|\beta_0,\beta_1,-\beta_2\rangle,
 \nonumber \\ 
P_1 |\beta_0,\beta_1,\beta_2\rangle=&\; \exp\left\{-i\pi N_0\right\}|\beta_0,\beta_1,-\beta_2\rangle,
\end{align}
where we have written $\beta_1=N_1+\lambda_1$ and imposed that the $P_1$ eigenvalue does not depend on $U(1)$ twist $\lambda_1$. This implies that $B_1=\pi\lambda_1$. We obtain
\begin{align}
\langle \beta_0,\beta_1,\beta_2|P_1|\beta_0,\beta_1,\beta_2 \rangle =e^{i\pi N_0}\delta_{\beta_2,0}.
\end{align}
The $P_1$ twisted partition function which is the partition function on $\mathcal K\times S^1$ \cite{hsieh2016global, hsieh2014symmetry} takes the form 
\begin{align}
Z[\mathcal K\times S^1,\lambda_0]=&\;\text{Tr}_{\mathcal H_{\lambda_0}}\left[P_1 e^{-2\pi R_0(H+i\frac{\tau_1}{\tau_2}P_x+(i\frac{\tau_1}{\tau_2}\beta+\gamma)P_y)}\right] \nonumber \\
+&\; Z_{osc}\sum_{N_{0,1}\in\mathbb Z}\exp\Big\{-\frac{\pi \tau_2}{2r^2R_2}(N_0+\lambda_0)^2 \nonumber \\
-&\; 2\pi r^2R_2 \tau_2 N_1^2 +2\pi i\tau_1 (N_0+\lambda_0)N_1 +i\pi N_0\Big\}\nonumber \\ 
\end{align}
Under a large gauge transformation $\lambda_0\to \lambda_0+1$ the parity twisted partition function changes sign 
\begin{align}
  Z[\mathcal K\times S^1,\lambda_0]= - Z[\mathcal K\times S^1, \lambda_0+1]
 \end{align}
This is a $\Z_2$ anomaly in the sense that it is cancelled if we take two copies of the theory. This signals the existence of a bosonic topological insulator protected by $\mathbb Z_2^{T}\times U(1)_1$ global symmetry.

\bigskip \noindent We propose the response theory might be 
\begin{align}
I[N,B,w_1]=\int_{N}w_1\cup \delta B
\end{align}
which can be modeled as
\begin{align}
\mathcal S=\int_{N}\left[\frac{1}{2\pi}b \cup \delta a+\frac{1}{2\pi}B \cup \delta a +\frac{1}{2\pi} w_1 \cup \delta B
\right].
\end{align}

\section{Conclusion and outlook}
\noindent In conclusion we have studied a class of invertible topological field theories that admit topologically distinct $G$ actions where $G$ is a discrete abelian group. We study these from complimentary bulk and boundary approaches. In the bulk these model bosonic $G$-SPTs which are labelled by $[\omega]\in H^{d+1}_{\text{group}}(G,U(1))$. Different SPTs furnish distinct responses to background flat gauge field $A$ depending on $\omega $. We explicitly compute these responses on manifolds with field configurations that can distinguish different SPTs. These set of responses supply SPT topological invariants.  Next we describe the gauging procedure and confirm that gauging an SPT gives a topological gauge theory which is none other that Dijkgraaf-Witten theory labelled by $\omega$. We show that Dijkgraaf-Witten theories can be ungauged by gauging a dual symmetry $\widehat{G}$. This is synonymous to condensing the gauge charge. 

\bigskip \noindent In the dual boundary approach, we study bosonic quantum field theories with global $G$ symmetry which suffer from a $G$-'t-Hooft anomaly. For the cases we study, it is shown that these 't-Hooft anomalies can be cancelled by a Dijkgraaf-Witten topological action in one dimension higher signaling that these theories are healthy  on the surface of SPTs. Further we compute SPT invariants directly from the boundary theory and describe a procedure of constructing $G$-characters by orbifolding $G$ on the boundary. These characters can be used to generate modular data for the bulk topological gauge theory. 

\bigskip \noindent Finally we study a quantum field theory in $2+1d$ that suffers from a mixed anomaly between time reversal/reflection and $U(1)$. Depending on how time reversal/reflection acts the $U(1)$ could be a 0-form or 1-form symmetry. We postulate the topological action of the $3+1d$ bulk that cancels such a 't-Hooft anomaly. For 0-form $U(1)\times \Z_2^{T}$ this theory could model the surface of the bosonic SPT phase with this symmetry. For 1-form $U(1)\times \Z_2^{T}$ this signals the existence of an SPT protected by this symmetry. Further we propose an effective field theory and response action for such an SPT.   

\bigskip \noindent We close with a few comments on open issues:
\begin{itemize}
\item In this work we only study gapless surfaces of SPT phases however for bulk spatial dimension $\geq 3$, the boundary can support a gapped QFT with anomalous topological order\cite{vishwanath2013physics, metlitski2015symmetry, wang2013gapped, bonderson2013time}. For onsite symmetry $G$ and in $3+1d$, the SPT invariant can be extracted from the violation of pentagon identity\cite{chen2015anomalous} on the $2+1d$ $G$-equivariant topological order. Moreover the time reversal anomaly can be computed using a recently proposed anomaly indicator by Wang and Levin\cite{wang2016anomaly} however it would be interesting to explore how SPT invariants can be extracted for mixed symmetry groups with both anti-unitary symmetries such as time reversal/ mirror reflection as well as onsite unitary symmetry.  
\item Higher groups have been explored for construction topological gauge theories and higher symmetries have been proposed to protect non-trivial gapped phases of matter\cite{kapustin2017higher}. The surface theories for such gauge theories and phases of matter respectively have been much less explored.  
\item Floquet SPTs\cite{else2016classification, potter2017dynamically, potter2016classification, po2016chiral} or non-trivial dynamical gapped phases of matter as well as several phases of matter protected by certain spacegroup symmetries have not been understood much within the framework of low energy topological field theories. Since TQFT is a robust framework to study phases of matter it is interesting to ask whether such spacetime symmetries can be incorporated within such a framework.    
\item Although we can understand bulk physics for SPTs protected by discrete abelian group $G$ directly by analyzing the boundary. There is a class of cocycles such as Type-III in $2+1d$ and type-IV in $3+1d$ bulk that cannot be captured by our scheme and consequently we cannot study such phases directly from the boundary. This has to do with the fact that upon gauging such SPTs one gets non-abelian topological order that cannot be embedded in $U(1)^k$. In future work we would like to consider a class of models that can admit non-abelian symmetries with the hope that these can model the boundary behavior of type-III or respectively type-IV SPTs. 
\end{itemize}

\acknowledgements{\noindent
  AT would like to thank Michael Stone, Srinidhi Ramamurthy, Michael V. Pak,
  Xueda Wen, Lakshya Bhardwaj and Itziar Ochoa de Alaiza
  for several helpful discussions.
  XC was supported by a postdoctoral fellowship from the Gordon and Betty Moore
  Foundation, under the EPiQS initiative,
  Grant GBMF4304, at the Kavli Institute for Theoretical Physics.
  KS is supported by RIKEN Special Postdoctoral Researcher Program.
  This work is supported in part by the NSF under Grant No.\ DMR-1455296.
}
\appendix

\section{Group cohomology for finite abelian groups}
\label{group_cohomology}
Here we collect some facts about the group cohomology of discrete abelian
groups.
In this paper we use both additive and multiplicative definition of group
action.
When we use additive definition, we will always work with $\R/2\pi \Z$
coefficients,
however, when we work with multiplicative definition, we will work with $U(1)$
coefficients.
Here we define $H_{\text{group}}^{n}(G,\mathbb R/2\pi \mathbb Z)$.
The space of $n$-cochains is defined as the set of homomorphisms 
\begin{align}
C^{n}_{\text{group}}(G,\mathbb R/2\pi \mathbb Z)=\left\{f: G^n\to \mathbb R/2\pi  \mathbb Z\right\}.
\end{align}
$C_{\text{group}}^n$ is an abelian group under pointwise addition: 
\begin{align}
  (f+g)(a_1,a_2,\ldots,a_{n})&=f(a_1,a_2,\ldots,a_{n})
                               \nonumber \\
  &\quad
                               +g(a_1,a_2,\ldots,a_{n}),
\end{align}
where $f,g\in C_{\text{group}}^n$.
Then there exists a coboundary operator $\delta: C_{\text{group}}^{n}\to C_{\text{group}}^{n+1}$ with the action
\begin{align}
  &  (\delta f) (a_1,\ldots,a_{n+1})=
    f(a_2,\ldots,a_{n+1})
    \nonumber \\
  &\qquad 
    +(-1)^{n+1}f(a_1,\ldots, a_{n})
    \nonumber \\
  &\qquad +\sum_{i=1}^{n}(-1)^i f(a_1,\ldots,a_i+a_{i+1},\ldots,a_{n+1})
\end{align} 
$\delta$ satisfies the properties
\begin{align}
  \delta(f+g)& = \delta f +\delta g,
               \nonumber \\
\delta^2 &= 0.
\end{align}
$\delta$ naturally defines two subgroups of $C_{\text{group}}^n$-the group
of $n$-cochains these are $n$-coycles $Z_{\text{group}}^n(G,\mathbb R/2\pi \mathbb Z)$ and $n$-coboundaries $B_{\text{group}}^{n}(G,\mathbb R/2\pi \mathbb Z)$ where $B_{\text{group}}^n\subset Z_{\text{group}}^n \subset C_{\text{group}}^n$
\begin{align}
  Z_{\text{group}}^{n}&=\left\{f \in C_{\text{group}}^n  \ \big| \ \delta f =0 \right\},
                         \nonumber \\
B_{\text{group}}^n&= \left\{f\in C_{\text{group}}^n \ \big| f=\delta h, \ h\in C_{\text{group}}^{n-1}\right\}.
\end{align}
Then the cohomology is defined as usual as
\begin{align}
  H_{\text{group}}^{n}(G,\mathbb R/2\pi \mathbb Z)
  = \frac{Z_{\text{group}}^n(G,\mathbb R/2\pi \mathbb Z)}{ B_{\text{group}}^n(G,\mathbb R/2\pi \mathbb Z)}.
\end{align}
The slant product can be defined,
which lowers the degree by 1
\begin{align}
i_a:C_{\text{group}}^{n}(G,\mathbb R/2\pi \mathbb Z)\to C_{\text{group}}^{n-1}(G,\mathbb R/2\pi \mathbb Z).
\end{align} 
Explicitly, this takes the form
\begin{align}
  &(i_{a}f)(a_1,\ldots,a_{n-1})=
    (-1)^{n-1}f(a,a_1,\ldots,a_{n-1})
    \nonumber \\
  &\quad
    +\sum_{i=1}^{n-1}(-1)^{n-1+i}
    f(a_1,\ldots,a_i,a,a_{i+1},\ldots,a_{n-1}).
\end{align}
Further it can be checked by explicit computation that
$\delta(i_{a}f)=i_a(\delta f)$.
Therefore, if $f\in Z_{\text{group}}^n(G,\R/2\pi \Z)$, then $i_af\in
Z_{\text{group}}^{n-1}(G,\R/2\pi \Z)$,
i.e., $i_a$ establishes a homomorphism
\begin{align}
i_a:H^{n}_{\text{group}}(G,\R/2\pi \Z) \to H^{n-1}_{\text{group}}(G,\R/2\pi \Z).
\end{align}


\section{Orbifolding with discrete torsion and relation to $1+1d$ SPTs}
\label{discrete_torsion}
\noindent 

\begin{figure}[bt]
\centering
\includegraphics[scale=.5]{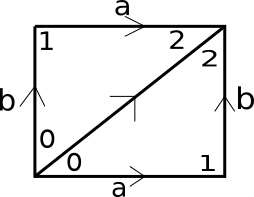}
\\
\caption{
A triangulation of $T^2$ with flux $a,b\in G$ along the two cycles. Dijkgraaf Witten theory labelled by $H^{2}_{\text{group}}(G,U(1))$ associates the $U(1)$ phase $c(a,b)/c(b,a)$ to this assohnment $A$.
}
\label{fig_torus}
\end{figure}

Consider the following partition function on a torus
\begin{align}
  Z_{\text{orb}}(\tau)=\sum_{{\bf{a}},{\bf{b}}\in G}\epsilon({\bf{a}},{\bf{b}})Z_{{\bf{a}},{\bf{b}}}.
\end{align}
Under modular transformations, the twisted sectors transform as
\begin{align}
T:Z_{{\bf{a}},{\bf{b}}}(\tau)\mapsto&\; Z_{{\bf{a}}+{\bf{b}},{\bf{b}}}(\tau),
 \nonumber \\ 
S:Z_{{\bf{a}},{\bf{b}}}(\tau)\mapsto &\; \mathcal Z_{-{\bf{b}},{\bf{a}}}(\tau).
\end{align}
Since the mapping class group of a torus is $SL(2,\mathbb Z)$, a general element may be written as
\begin{align}
U=\left( \begin{array}{cc}
p & q  \\
r  & s \end{array} \right); \quad ps-qr=1
\label{modular_torsion}
\end{align}
Then this implies that 
\begin{align}
\epsilon({\bf{a}}^p{\bf{b}}^q,{\bf{a}}^r{\bf{b}}^s)=\epsilon({\bf{a}},{\bf{b}}).
\end{align}
Further, consider putting the theory on $\Sigma^{2}$, a Riemann surface of genus
2. 
Then $\epsilon: \text{Hom}[\pi_1(\Sigma^2),G]\to U(1)$. 
By modular invariance we demand\cite{vafa1986modular} 
\begin{align}
\epsilon({\bf{a}}_1,{\bf{b}}_1;{\bf{a}}_2,{\bf{b}}_2)=\epsilon({\bf{a}}_1{\bf{b}}_1{\bf{b}}^{-1}_2,{\bf{b}}_1; {\bf{a}}_2{\bf{b}}_2{\bf{b}}^{-1}_{1},{\bf{b}}_2),
\label{condn_1_torsion}
\end{align}
where ${\bf{a}}_{1},{\bf{a}}_{2},{\bf{b}}_{1},{\bf{b}}_{2}$ are the $G$-fluxes
inserted along the non-contractible cycles
$L_{x}^{1},L_{x}^{2},L_{y}^{1},L_{y}^{2}$ respectively. 
Further by the factorization property at genus 2,
\begin{align}
\epsilon({\bf{a}}_1,{\bf{b}}_1;{\bf{a}}_{2},{\bf{b}}_2)=\epsilon({\bf{a}}_1,{\bf{b}}_{1})\epsilon({\bf{a}}_2,{\bf{b}}_2)
\label{condn_2_torsion}
\end{align} 
If we normalize $\epsilon({\bf{1}},{\bf{1}})=1$, then by modular invariance \eqref{modular_torsion}, 
\begin{align}
\epsilon({\bf{g}},{\bf{1}})=\epsilon({\bf{1}},{\bf{g}})=1.
\label{torsion_normalization}
\end{align}
Using these facts and \eqref{condn_1_torsion},\eqref{condn_2_torsion} it can be shown that $\epsilon$ is a 1-dimensional representation of $G$  
\begin{align}
\epsilon({\bf{a}}_1+{\bf{a}}_2,{\bf{b}})
=\epsilon({\bf{a}}_1,{\bf{b}})\epsilon({\bf{a}}_2,{\bf{b}}).
\label{torsion_linear}
\end{align}
It was shown in \cite{vafa1986modular,gaberdiel2000discrete} the set of inequivalent $\epsilon$ that satisfy \eqref{torsion_normalization} and \eqref{torsion_linear} are classified by $[c]\in H_{\text{group}}^{2}(G,U(1))$ and can be written as
\begin{align}
\epsilon({\bf{a}},{\bf{b}})=\frac{c({\bf{a}},{\bf{b}})}{c({\bf{b}},{\bf{a}})}.
\end{align}
Since $[c]\in H_{\text{group}}^{2}(G,U(1))$ it satisfies the cocycle condition 
\begin{align}
c({\bf{a}},{\bf{b}}{\bf{c}})c({\bf{b}},{\bf{c}})=c({\bf{a}}{\bf{b}},{\bf{c}})c({\bf{a}},{\bf{b}}).
\end{align}
Now using this form of $\epsilon$, 
we may verify that the above two properties are satisfied. First
\begin{align}
\frac{\epsilon({\bf{a}}_1{\bf{a}}_2,{\bf{a}}_3)}{\epsilon({\bf{a}}_1,{\bf{a}}_3)\epsilon({\bf{a}}_2,{\bf{a}}_3)}
=&\;\frac{c({\bf{a}}_1{\bf{a}}_2,{\bf{a}}_3)c({\bf{a}}_3,{\bf{a}}_1)c({\bf{a}}_3,{\bf{a}}_2)}{c({\bf{a}}_3,{\bf{a}}_1{\bf{a}}_2)c({\bf{a}}_1,{\bf{a}}_3)c({\bf{a}}_2,{\bf{a}}_3)} \nonumber\\
=&\;\frac{c({\bf{a}}_1{\bf{a}}_2,{\bf{a}}_3)c({\bf{a}}_3,{\bf{a}}_2)}{c({\bf{a}}_1{\bf{a}}_3,{\bf{a}}_2)c({\bf{a}}_1,{\bf{a}}_3)} \nonumber\\
=&\; 1,
 \nonumber \\
\epsilon({\bf{a}}^p{\bf{b}}^q,{\bf{a}}^r{\bf{b}}^s)=&\; \epsilon({\bf{a}}^p,{\bf{a}}^r{\bf{b}}^s)\epsilon({\bf{b}}^q,{\bf{a}}^r{\bf{b}}^s) \nonumber \\
=&\; \epsilon({\bf{a}},{\bf{a}}^r{\bf{b}}^s)^p\epsilon({\bf{b}},{\bf{a}}^r{\bf{b}}^s)^q \nonumber \\
=&\; \epsilon({\bf{a}}^r{\bf{b}}^s,{\bf{a}})^{-p}\epsilon({\bf{a}}^r{\bf{b}}^s,{\bf{b}})^{-q} \nonumber \\
=&\; \epsilon({\bf{b}}^s,{\bf{a}})^{-p}\epsilon({\bf{a}}^r,{\bf{b}})^{-q} \nonumber \\
=&\; \epsilon({\bf{a}},{\bf{b}})^{(ps-qr)}\nonumber \\
=&\; \epsilon({\bf{a}},{\bf{b}})
\end{align}
Furthermore the discrete torsion phase
$\epsilon({\bf{a}},{\bf{b}})=c({\bf{a}},{\bf{b}})/c({\bf{a}},{\bf{b}})$ is
exactly the response of a $1+1d$ SPT protected by $G$ and characterized by
2-cocycle $[c]\in H^{2}_{\text{group}}(G,U(1))$ in the presence of $G$-flux
${\bf{a}},{\bf{b}}$ along the two non-contractible cycles of the torus. 
(See Fig.\ \ref{fig_torus}.) 
To see this recall that given a triangulation $K$ of manifold $M$, Dijkgraaf-Witten theory associates to an assignment $A: H_{1}(K,\Z)\to G$ a $U(1)$ phase i.e the response theory of an SPT classified by $[c]$, explicitly given by
\begin{align}
e^{iI^{c}[K,A]}=\prod_{\sigma\in C_2(K)}\big\langle c(A),\sigma\big\rangle^{o_{\sigma}} 
\end{align}
where $o_{\sigma}=\pm 1$, the orientation of simplex $\sigma$. For a simplex $\sigma[v_0v_1v_2]$ and an assignment $A(v_0v_1)={\bf{a}}$, $A(v_1v_2)={\bf{b}}$, we get $\langle c(A),\sigma\rangle =c({\bf{a}},{\bf{b}})$. Then it is easy to check
\begin{align}
e^{iI^{c}[T^2,A]}=\frac{c({\bf{a}},{\bf{b}})}{c({\bf{b}},{\bf{a}})}=\epsilon({\bf{a}},{\bf{b}})
\end{align}
\section{SPT response theory and group cocycles}
\label{relation}
\begin{figure}[bt]
\centering
\includegraphics[scale=.27]{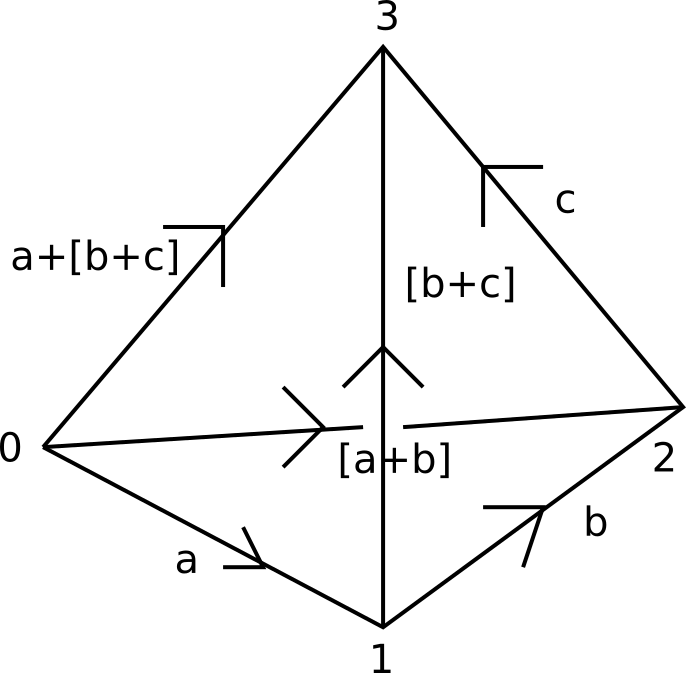}
\\
\caption{
Configuration of a flat $\Z_n$ gauge field on a 3-simplex. $a,b,c\in \Z_n$.
}
\label{tetra}
\end{figure}

\begin{figure}[bt]
\centering
\includegraphics[scale=0.65]{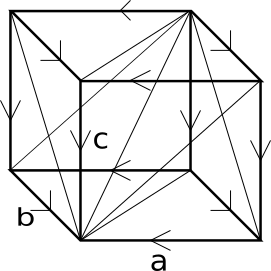}
\\
\caption{
Triangulation of a three-torus containing one 0-simplex, three 1-simplices, three 2-simplices and six 3-simplices.  
}
\label{three_torus}
\end{figure}

\noindent In this appendix we show the relation between the SPT response
theories and the respective group cocycles. We would like to show explicitly
that the SPT response theories written in the main text matches the expression
for the group cocycle. Consider a triangulation of the manifold $N$. 
(See App.\ A in \cite{kapustin2014bosonic} for an introduction to simplicial calculus.) Then a flat $G$ gauge field $[A]\in C^{1}(N,G)$ that satisfies the conditions
\begin{itemize} 
\item $A(\partial f)=0$ for all $f \in C_{2}(N,\Z)$.
\item $A(-e)=A(e)^{-1}$ for all $e\in C_1(N,\Z)$ where $-e$ implies reversing the orientation of edge $e$. .
\end{itemize}
\noindent Let us consider the specific case of $\Z_n$ SPT in $2+1d$. We pick a
triangulation for a 3-manifold $N$. Then a 3-simplex $\sigma^{i}=[v_0v_1v_2v_3]$
comes with an ordering of vertices $0<1<2<3$ that picks an orientation. A choice
of $[A]$ means assigning $A[v_0v_1]=2\pi a/n$. $A[v_1v_2]=2\pi b/n$ and
$A[v_2v_3]=2\pi c/n$ where $a,b,c\in [0,1,..,n-1]$. Then it straightforward to
check that for this choice of flat field $[A]$ 
(see Fig.\ \ref{tetra}).

\begin{align}
\Big\langle\frac{ q}{2\pi}A\cup \delta A,\sigma^{i}\Big\rangle  = \frac{2\pi q}{n^2}a(b+c-[b+c])
\end{align}
The precise meaning of $\delta A$ should be understood as follows. Let $A \in Z^1(M; \frac{2\pi}{n} \Z/\Z)$ be a $\Z_n$ field. 
The coefficient $\frac{2\pi}{n} \Z/\Z$ means $A(01)$ takes values in $\frac{2\pi a}{n} {\rm\ mod\ } 2\pi$ with $a \in \Z$, i.e.\ $A(01) \in \{0,\frac{2\pi}{n}, \dots, \frac{2\pi(n-1)}{n}\}$. 
We shall define the topological action like ``$A \cup \delta A$''. 
To do so, we introduce a lift 
\begin{align}
A \mapsto \wt A \in C^1(M;\frac{2\pi}{n} \Z).
\end{align}
The closed condition of $A$ implies that 
\begin{align}
\delta \wt A \in C^1(M;2\pi \Z), 
\end{align}
i.e.\ $(\delta \wt A)(012)$ takes values in $2\pi \Z$.
A lift $\wt A$ is not unique: an integer valued 1-cochain $a \in C^1(M;2\pi \Z)$ also gives a lift 
\begin{align}
A \mapsto \wt A + a, \qquad 
a \in C^1(M;2\pi \Z).
\end{align}
We define a topological action $S[A]$ of $\Z_n$ fields by 
\begin{align}
I[A] := \frac{q}{2\pi}\int_M \wt A \cup \delta \wt A \in \frac{2\pi \Z}{n}.
\end{align}
This is ill-defined as $\frac{2\pi \Z}{n}$-valued action. 
However, $I[A]$
mod $2\pi \Z$ is well-defined: 
Under a change of lift, the action is changed as 
\begin{widetext}
\begin{align}
\frac{1}{2\pi}\wt A \cup \delta \wt A
&\mapsto 
\frac{1}{2\pi}(\wt A+a) \cup (\delta \wt A + \delta a) \\
&= \frac{1}{2\pi}\left[\wt A \cup \delta \wt A +  a \cup \delta \wt A + \wt A \cup \delta a + a \cup \delta a \right]\\
&= \frac{1}{2\pi}\wt A \cup \delta \wt A + \underbrace{\frac{1}{2\pi}a \cup \delta \wt A}_{2\pi \Z} - \underbrace{\frac{1}{2\pi}\delta (\wt A \cup a)}_{\rm exact} + \underbrace{\frac{1}{2\pi} \delta \wt A \cup a}_{2\pi \Z} + \underbrace{a \cup \delta a}_{2\pi \Z} \\
&= \frac{1}{2\pi}\wt A \cup \delta \wt A - \underbrace{\frac{1}{2\pi}\delta (\wt A \cup a)}_{\rm exact} \qquad ({\rm mod\ 2\pi}\Z).
\end{align}
\end{widetext}
This means $e^{i I[A]}$
serves as a $U(1)$-valued topological action. Similarly for type-II and III cocycle, it is straightforward to check 
\begin{align}
  &\;\Big\langle\frac{ q_{IJ}}{2\pi}A^{I}\cup \delta A^{J},\sigma^{i}\Big\rangle  = \frac{2\pi q_{IJ}}{n^2}a^{I}(b^J+c^J-[b^J+c^J]),
    \nonumber \\
&\;\Big\langle\frac{ q_{IJK}n^2}{4\pi^2}A^{I}\cup A^{J}\cup A_{K},\sigma^{i}\Big\rangle  =\frac{2\pi q_{IJK}}{n}a^{I}b^Jc^K.
\end{align}
Consider a triangulation of a three-torus as shown in Fig.\ \ref{three_torus}. The triangulation has six 3-simplices . Then it is easy to check that the partition function takes the form\cite{propitius1995topological, dijkgraaf1990topological}
\begin{align}
 \mathcal{Z}[T^3,a,b,c]
  &= \frac{1}{|G|}\prod_{\sigma\in Z_3}\langle \omega[A],\sigma \rangle^{o_{\sigma}} \nonumber \\ 
=&\; \frac{\omega(a,b,c)\omega(b,c,a)\omega(c,a,b)}{\omega(a,c,b)\omega(b,a,c),\omega(c,b,a)}
\end{align}
This matches with field theory calculation in \eqref{torus_type_1} and \eqref{torus_type_3}. Furthermore one can compute the SPT or Dijkgraaf Witten theory partition function on lens space $L(n,1)$. This was recently shown in \cite{tantivasadakarn2017dimensional} and we do not repeat the calculation here. The field theory calculation \eqref{lens_space_type_1} matches  the result in \cite{tantivasadakarn2017dimensional}.

\bigskip \noindent 
Similarly for $3+1d$ SPTs for $G=\Z_n^k$, we consider a 4-simplex $\sigma^i=[v_0v_1v_2v_3v_4]$ and a flat $G$ field $[A]$ with the assignment $A^{I}(v_{0}v_{1})=2\pi a^I/n$, $A^{I}(v_{1}v_{2})=2\pi b^I/n$, $A^{I}(v_{2}v_{3})=2\pi c^I/n$ and $A^{I}(v_{3}v_{4})=2\pi d^I/n$. 
\begin{widetext}
\begin{align}
  \Big \langle \frac{2\pi q_{IJ}n}{4\pi^2}A^I\cup A^J \cup \partial A^{J} ,\sigma^{i}\Big \rangle =&\; \frac{2\pi iq_{IJ}}{n^{2}}a^{I}b^{J}\left(c^{J}+d^{J}-[c^J+d^J]\right),
                                                                                                     \nonumber \\
  \Big \langle \frac{2\pi q_{IJK}n}{4\pi^2}A^I\cup A^J \cup \partial A^{K} ,\sigma^{i}\Big \rangle =&\; \frac{2\pi iq_{IJK}}{n^{2}}a^{I}b^{J}\left(c^{K}+d^{K}-[c^K+d^K]\right),
                                                                                                      \nonumber \\
\Big\langle \frac{q_{IJKL}n^3}{8\pi^3}A^{I} \cup A^{J} \cup A^{K} \cup A^{L},\sigma^i \Big \rangle=&\; \frac{2\pi i q_{IJKL}}{n}a^{I}b^{J}c^{K}d^{L}.
\end{align}
\end{widetext}
The computations for partition functions on $T^4$ and $L(n,1)\times S^1$ are more tedious but quite similar to those in 1-dimension lower on $T^3$ and $L(n,1)$ as the latter are dimensionally reduced versions of the former.

\bigskip \noindent Simplicial calculus is naturally analogous to differential
calculus where $p$-cochains map to $p$-forms, cup product maps to wedge product
and the differential $\partial$ maps to the exterior derivative `$d$'. This
matches with the response theories in Eqs.\ \eqref{lens_type_2_resp} and \eqref{torus_type_4_resp} for response theories of SPTs.

\section{Quantization of $2+1d$ surface theory}
\label{2+1_action_appendix}
\noindent The equal time canonical commutation relations for \eqref{2+1_action} are
\begin{align}
\left[\phi({\bf{x}},t),\epsilon^{ij}\partial_{i}\zeta_j({\bf{x}}',t)\right]=2\pi i\delta({\bf{x}}-{\bf{x}}')
\end{align}
The mode expansion decomposes into oscillator and zero mode parts 
\begin{align}
\phi(x,y,t)=&\; \phi^0(x,y,t) + \phi^{osc}(x,y,t) \nonumber \\
\zeta_j(x,y,t)=&\; \zeta_j^0(x,y,t) + \zeta_j^{osc}(x,y,t) 
\end{align}
The zero-mode part takes the form
\begin{align}
\phi(x,y,t)=&\; \alpha_0+\frac{\beta_1 x}{R_1}+\frac{\beta_2 y}{R_2}+... \nonumber \\
\zeta_{j}(x,y,t)=&\; \frac{\alpha_j}{2\pi R_j}+\frac{\beta_0}{2\pi R_1R_2}x\delta_{j,2}+...
\label{zeromode_expansion}
\end{align}
The canonical algebra for this theory implies $\left[\alpha_0,\beta_0\right]=i$ and $ \left[\beta_{1},\alpha_{2}\right]-\left[\beta_{2},\alpha_{1}\right]=i$. One possible choice of commutation relations that satisfy this algebra is 
\begin{align}
\left[\beta_1,\alpha_2\right]=0; \quad \left[\beta_2,\alpha_1\right]=-i
\end{align}
however to quantize $\beta_{\mu}$ we impose 
\begin{align}
\left[\alpha_1,\beta_2\right]=i=-\left[\alpha_2,\beta_1\right]
\end{align}
with this $\beta_{\mu}\in\mathbb Z$. The oscillator part of the mode expansions are
\begin{align}
\phi^{osc}(r)=&\; \frac{1}{\sqrt{R_1R_2}}\sqrt{\frac{1}{2\lambda_1}} \nonumber \\
&\; \times \sum_{k\neq 0}\frac{1}{\omega(k)^{1/2}}\left[\hat{a}(\vec{k})e^{-i\vec{k}.\vec{r}}
+\hat{a}^{\dagger}(\vec{k})e^{i\vec{k}.\vec{r}}
\right] \nonumber \\
\zeta^{osc}_{j}(r)=&\; \frac{1}{\sqrt{R_1R_2}}\sqrt{\frac{\lambda_1}{8\pi^2}} \nonumber \\
&\; \times\sum_{k\neq 0}\frac{-1}{\omega(k)^{3/2}}\epsilon_{jl}k^l \left[\hat{a}(\vec{k})e^{-i\vec{k}.\vec{r}}
+\hat{a}^{\dagger}(\vec{k})e^{i\vec{k}.\vec{r}}
\right]
\end{align}
With the commutator algebra $\left[a(k),a^{\dagger}(k')\right]=\delta_{k,k'}$. The partition function is given by 
\begin{align}
Z=\text{Tr}_{\mathcal H}\left[e^{2\pi iR_0H'}\right]
\end{align}
where $\mathcal H={\otimes_{N_{0,1,2}}}\mathcal H_{N_0,N_1,N_2}$ and $H'=H+i\alpha R_1 P_x/R_0$. The zero-mode part is 
\begin{align}
Z_{0}=&\;\sum_{N_0,N_1,N_2\in \mathbb Z}\exp\left\{-\frac{\pi \tau_2}{2r^2R_2}N_0^2-2r^2\pi R_2\tau_2N_1^2
\right. \nonumber \\
&\; \left.
-\frac{2r^2\pi R_0R_1}{R_2}N_2^2 +2\pi i\tau_1N_0N_1 \right\}
\end{align}
The oscillator part of the partition function is the same as that of free boson.

\bibliography{ref_orb}

\begin{thebibliography}{93}%
\makeatletter
\providecommand \@ifxundefined [1]{%
 \@ifx{#1\undefined}
}%
\providecommand \@ifnum [1]{%
 \ifnum #1\expandafter \@firstoftwo
 \else \expandafter \@secondoftwo
 \fi
}%
\providecommand \@ifx [1]{%
 \ifx #1\expandafter \@firstoftwo
 \else \expandafter \@secondoftwo
 \fi
}%
\providecommand \natexlab [1]{#1}%
\providecommand \enquote  [1]{``#1''}%
\providecommand \bibnamefont  [1]{#1}%
\providecommand \bibfnamefont [1]{#1}%
\providecommand \citenamefont [1]{#1}%
\providecommand \href@noop [0]{\@secondoftwo}%
\providecommand \href [0]{\begingroup \@sanitize@url \@href}%
\providecommand \@href[1]{\@@startlink{#1}\@@href}%
\providecommand \@@href[1]{\endgroup#1\@@endlink}%
\providecommand \@sanitize@url [0]{\catcode `\\12\catcode `\$12\catcode
  `\&12\catcode `\#12\catcode `\^12\catcode `\_12\catcode `\%12\relax}%
\providecommand \@@startlink[1]{}%
\providecommand \@@endlink[0]{}%
\providecommand \url  [0]{\begingroup\@sanitize@url \@url }%
\providecommand \@url [1]{\endgroup\@href {#1}{\urlprefix }}%
\providecommand \urlprefix  [0]{URL }%
\providecommand \Eprint [0]{\href }%
\providecommand \doibase [0]{http://dx.doi.org/}%
\providecommand \selectlanguage [0]{\@gobble}%
\providecommand \bibinfo  [0]{\@secondoftwo}%
\providecommand \bibfield  [0]{\@secondoftwo}%
\providecommand \translation [1]{[#1]}%
\providecommand \BibitemOpen [0]{}%
\providecommand \bibitemStop [0]{}%
\providecommand \bibitemNoStop [0]{.\EOS\space}%
\providecommand \EOS [0]{\spacefactor3000\relax}%
\providecommand \BibitemShut  [1]{\csname bibitem#1\endcsname}%
\let\auto@bib@innerbib\@empty
\bibitem [{\citenamefont {Fradkin}(2013)}]{fradkin2013field}%
  \BibitemOpen
  \bibfield  {author} {\bibinfo {author} {\bibfnamefont {E.}~\bibnamefont
  {Fradkin}},\ }\href@noop {} {\emph {\bibinfo {title} {Field theories of
  condensed matter physics}}}\ (\bibinfo  {publisher} {Cambridge University
  Press},\ \bibinfo {year} {2013})\BibitemShut {NoStop}%
\bibitem [{\citenamefont {Chiu}\ \emph {et~al.}(2016)\citenamefont {Chiu},
  \citenamefont {Teo}, \citenamefont {Schnyder},\ and\ \citenamefont
  {Ryu}}]{chiu2016classification}%
  \BibitemOpen
  \bibfield  {author} {\bibinfo {author} {\bibfnamefont {C.-K.}\ \bibnamefont
  {Chiu}}, \bibinfo {author} {\bibfnamefont {J.~C.}\ \bibnamefont {Teo}},
  \bibinfo {author} {\bibfnamefont {A.~P.}\ \bibnamefont {Schnyder}}, \ and\
  \bibinfo {author} {\bibfnamefont {S.}~\bibnamefont {Ryu}},\ }\href@noop {}
  {\bibfield  {journal} {\bibinfo  {journal} {Reviews of Modern Physics}\
  }\textbf {\bibinfo {volume} {88}},\ \bibinfo {pages} {035005} (\bibinfo
  {year} {2016})}\BibitemShut {NoStop}%
\bibitem [{\citenamefont {Zeng}\ \emph {et~al.}(2015)\citenamefont {Zeng},
  \citenamefont {Chen}, \citenamefont {Zhou},\ and\ \citenamefont
  {Wen}}]{zeng2015quantum}%
  \BibitemOpen
  \bibfield  {author} {\bibinfo {author} {\bibfnamefont {B.}~\bibnamefont
  {Zeng}}, \bibinfo {author} {\bibfnamefont {X.}~\bibnamefont {Chen}}, \bibinfo
  {author} {\bibfnamefont {D.-L.}\ \bibnamefont {Zhou}}, \ and\ \bibinfo
  {author} {\bibfnamefont {X.-G.}\ \bibnamefont {Wen}},\ }\href@noop {}
  {\bibfield  {journal} {\bibinfo  {journal} {arXiv preprint arXiv:1508.02595}\
  } (\bibinfo {year} {2015})}\BibitemShut {NoStop}%
\bibitem [{\citenamefont {Senthil}(2015)}]{senthil2015symmetry}%
  \BibitemOpen
  \bibfield  {author} {\bibinfo {author} {\bibfnamefont {T.}~\bibnamefont
  {Senthil}},\ }\href@noop {} {\bibfield  {journal} {\bibinfo  {journal} {Annu.
  Rev. Condens. Matter Phys.}\ }\textbf {\bibinfo {volume} {6}},\ \bibinfo
  {pages} {299} (\bibinfo {year} {2015})}\BibitemShut {NoStop}%
\bibitem [{\citenamefont {Hasan}\ and\ \citenamefont
  {Kane}(2010)}]{HasanKane10}%
  \BibitemOpen
  \bibfield  {author} {\bibinfo {author} {\bibfnamefont {M.~Z.}\ \bibnamefont
  {Hasan}}\ and\ \bibinfo {author} {\bibfnamefont {C.~L.}\ \bibnamefont
  {Kane}},\ }\href {\doibase 10.1103/RevModPhys.82.3045} {\bibfield  {journal}
  {\bibinfo  {journal} {Rev. Mod. Phys.}\ }\textbf {\bibinfo {volume} {82}},\
  \bibinfo {pages} {3045} (\bibinfo {year} {2010})}\BibitemShut {NoStop}%
\bibitem [{\citenamefont {Qi}\ and\ \citenamefont
  {Zhang}(2011)}]{QiZhangreview11}%
  \BibitemOpen
  \bibfield  {author} {\bibinfo {author} {\bibfnamefont {X.-L.}\ \bibnamefont
  {Qi}}\ and\ \bibinfo {author} {\bibfnamefont {S.-C.}\ \bibnamefont {Zhang}},\
  }\href {\doibase 10.1103/RevModPhys.83.1057} {\bibfield  {journal} {\bibinfo
  {journal} {Rev. Mod. Phys.}\ }\textbf {\bibinfo {volume} {83}},\ \bibinfo
  {pages} {1057} (\bibinfo {year} {2011})}\BibitemShut {NoStop}%
\bibitem [{\citenamefont {Chen}\ \emph {et~al.}(2013)\citenamefont {Chen},
  \citenamefont {Gu}, \citenamefont {Liu},\ and\ \citenamefont
  {Wen}}]{chen2013symmetry}%
  \BibitemOpen
  \bibfield  {author} {\bibinfo {author} {\bibfnamefont {X.}~\bibnamefont
  {Chen}}, \bibinfo {author} {\bibfnamefont {Z.-C.}\ \bibnamefont {Gu}},
  \bibinfo {author} {\bibfnamefont {Z.-X.}\ \bibnamefont {Liu}}, \ and\
  \bibinfo {author} {\bibfnamefont {X.-G.}\ \bibnamefont {Wen}},\ }\href@noop
  {} {\bibfield  {journal} {\bibinfo  {journal} {Physical Review B}\ }\textbf
  {\bibinfo {volume} {87}},\ \bibinfo {pages} {155114} (\bibinfo {year}
  {2013})}\BibitemShut {NoStop}%
\bibitem [{\citenamefont
  {Kapustin}(2014{\natexlab{a}})}]{kapustin2014symmetry}%
  \BibitemOpen
  \bibfield  {author} {\bibinfo {author} {\bibfnamefont {A.}~\bibnamefont
  {Kapustin}},\ }\href@noop {} {\bibfield  {journal} {\bibinfo  {journal}
  {arXiv preprint arXiv:1403.1467}\ } (\bibinfo {year}
  {2014}{\natexlab{a}})}\BibitemShut {NoStop}%
\bibitem [{\citenamefont {Kapustin}(2014{\natexlab{b}})}]{kapustin2014bosonic}%
  \BibitemOpen
  \bibfield  {author} {\bibinfo {author} {\bibfnamefont {A.}~\bibnamefont
  {Kapustin}},\ }\href@noop {} {\bibfield  {journal} {\bibinfo  {journal}
  {arXiv preprint arXiv:1404.6659}\ } (\bibinfo {year}
  {2014}{\natexlab{b}})}\BibitemShut {NoStop}%
\bibitem [{\citenamefont {Schnyder}\ \emph {et~al.}(2008)\citenamefont
  {Schnyder}, \citenamefont {Ryu}, \citenamefont {Furusaki},\ and\
  \citenamefont {Ludwig}}]{schnyder2008classification}%
  \BibitemOpen
  \bibfield  {author} {\bibinfo {author} {\bibfnamefont {A.~P.}\ \bibnamefont
  {Schnyder}}, \bibinfo {author} {\bibfnamefont {S.}~\bibnamefont {Ryu}},
  \bibinfo {author} {\bibfnamefont {A.}~\bibnamefont {Furusaki}}, \ and\
  \bibinfo {author} {\bibfnamefont {A.~W.}\ \bibnamefont {Ludwig}},\
  }\href@noop {} {\bibfield  {journal} {\bibinfo  {journal} {Physical Review
  B}\ }\textbf {\bibinfo {volume} {78}},\ \bibinfo {pages} {195125} (\bibinfo
  {year} {2008})}\BibitemShut {NoStop}%
\bibitem [{\citenamefont {Kitaev}\ \emph {et~al.}(2009)\citenamefont {Kitaev},
  \citenamefont {Lebedev},\ and\ \citenamefont
  {Feigelman}}]{kitaev2009periodic}%
  \BibitemOpen
  \bibfield  {author} {\bibinfo {author} {\bibfnamefont {A.}~\bibnamefont
  {Kitaev}}, \bibinfo {author} {\bibfnamefont {V.}~\bibnamefont {Lebedev}}, \
  and\ \bibinfo {author} {\bibfnamefont {M.}~\bibnamefont {Feigelman}},\ }in\
  \href@noop {} {\emph {\bibinfo {booktitle} {AIP Conference Proceedings}}},\
  Vol.\ \bibinfo {volume} {1134}\ (\bibinfo {organization} {AIP},\ \bibinfo
  {year} {2009})\ pp.\ \bibinfo {pages} {22--30}\BibitemShut {NoStop}%
\bibitem [{\citenamefont {Gu}\ and\ \citenamefont
  {Wen}(2014)}]{gu2014symmetry}%
  \BibitemOpen
  \bibfield  {author} {\bibinfo {author} {\bibfnamefont {Z.-C.}\ \bibnamefont
  {Gu}}\ and\ \bibinfo {author} {\bibfnamefont {X.-G.}\ \bibnamefont {Wen}},\
  }\href@noop {} {\bibfield  {journal} {\bibinfo  {journal} {Physical Review
  B}\ }\textbf {\bibinfo {volume} {90}},\ \bibinfo {pages} {115141} (\bibinfo
  {year} {2014})}\BibitemShut {NoStop}%
\bibitem [{\citenamefont {Kapustin}\ \emph {et~al.}(2015)\citenamefont
  {Kapustin}, \citenamefont {Thorngren}, \citenamefont {Turzillo},\ and\
  \citenamefont {Wang}}]{kapustin2015fermionic}%
  \BibitemOpen
  \bibfield  {author} {\bibinfo {author} {\bibfnamefont {A.}~\bibnamefont
  {Kapustin}}, \bibinfo {author} {\bibfnamefont {R.}~\bibnamefont {Thorngren}},
  \bibinfo {author} {\bibfnamefont {A.}~\bibnamefont {Turzillo}}, \ and\
  \bibinfo {author} {\bibfnamefont {Z.}~\bibnamefont {Wang}},\ }\href@noop {}
  {\bibfield  {journal} {\bibinfo  {journal} {Journal of High Energy Physics}\
  }\textbf {\bibinfo {volume} {2015}},\ \bibinfo {pages} {52} (\bibinfo {year}
  {2015})}\BibitemShut {NoStop}%
\bibitem [{\citenamefont {Freed}\ and\ \citenamefont
  {Hopkins}(2016)}]{freed2016reflection}%
  \BibitemOpen
  \bibfield  {author} {\bibinfo {author} {\bibfnamefont {D.~S.}\ \bibnamefont
  {Freed}}\ and\ \bibinfo {author} {\bibfnamefont {M.~J.}\ \bibnamefont
  {Hopkins}},\ }\href@noop {} {\bibfield  {journal} {\bibinfo  {journal} {arXiv
  preprint arXiv:1604.06527}\ } (\bibinfo {year} {2016})}\BibitemShut {NoStop}%
\bibitem [{\citenamefont {Gaiotto}\ and\ \citenamefont
  {Kapustin}(2016)}]{gaiotto2016spin}%
  \BibitemOpen
  \bibfield  {author} {\bibinfo {author} {\bibfnamefont {D.}~\bibnamefont
  {Gaiotto}}\ and\ \bibinfo {author} {\bibfnamefont {A.}~\bibnamefont
  {Kapustin}},\ }\href@noop {} {\bibfield  {journal} {\bibinfo  {journal}
  {International Journal of Modern Physics A}\ }\textbf {\bibinfo {volume}
  {31}},\ \bibinfo {pages} {1645044} (\bibinfo {year} {2016})}\BibitemShut
  {NoStop}%
\bibitem [{\citenamefont {Wang}\ and\ \citenamefont
  {Gu}(2017)}]{wang2017towards}%
  \BibitemOpen
  \bibfield  {author} {\bibinfo {author} {\bibfnamefont {Q.-R.}\ \bibnamefont
  {Wang}}\ and\ \bibinfo {author} {\bibfnamefont {Z.-C.}\ \bibnamefont {Gu}},\
  }\href@noop {} {\bibfield  {journal} {\bibinfo  {journal} {arXiv preprint
  arXiv:1703.10937}\ } (\bibinfo {year} {2017})}\BibitemShut {NoStop}%
\bibitem [{\citenamefont {Bhardwaj}\ \emph {et~al.}(2017)\citenamefont
  {Bhardwaj}, \citenamefont {Gaiotto},\ and\ \citenamefont
  {Kapustin}}]{bhardwaj2017state}%
  \BibitemOpen
  \bibfield  {author} {\bibinfo {author} {\bibfnamefont {L.}~\bibnamefont
  {Bhardwaj}}, \bibinfo {author} {\bibfnamefont {D.}~\bibnamefont {Gaiotto}}, \
  and\ \bibinfo {author} {\bibfnamefont {A.}~\bibnamefont {Kapustin}},\
  }\href@noop {} {\bibfield  {journal} {\bibinfo  {journal} {Journal of High
  Energy Physics}\ }\textbf {\bibinfo {volume} {2017}},\ \bibinfo {pages} {96}
  (\bibinfo {year} {2017})}\BibitemShut {NoStop}%
\bibitem [{\citenamefont {Cheng}\ \emph {et~al.}(2017)\citenamefont {Cheng},
  \citenamefont {Tantivasadakarn},\ and\ \citenamefont {Wang}}]{cheng2017loop}%
  \BibitemOpen
  \bibfield  {author} {\bibinfo {author} {\bibfnamefont {M.}~\bibnamefont
  {Cheng}}, \bibinfo {author} {\bibfnamefont {N.}~\bibnamefont
  {Tantivasadakarn}}, \ and\ \bibinfo {author} {\bibfnamefont {C.}~\bibnamefont
  {Wang}},\ }\href@noop {} {\bibfield  {journal} {\bibinfo  {journal} {arXiv
  preprint arXiv:1705.08911}\ } (\bibinfo {year} {2017})}\BibitemShut {NoStop}%
\bibitem [{\citenamefont {Wen}(2015)}]{wen2015construction}%
  \BibitemOpen
  \bibfield  {author} {\bibinfo {author} {\bibfnamefont {X.-G.}\ \bibnamefont
  {Wen}},\ }\href@noop {} {\bibfield  {journal} {\bibinfo  {journal} {Physical
  Review B}\ }\textbf {\bibinfo {volume} {91}},\ \bibinfo {pages} {205101}
  (\bibinfo {year} {2015})}\BibitemShut {NoStop}%
\bibitem [{\citenamefont {Shiozaki}\ \emph {et~al.}(2016)\citenamefont
  {Shiozaki}, \citenamefont {Shapourian},\ and\ \citenamefont
  {Ryu}}]{shiozaki2016many}%
  \BibitemOpen
  \bibfield  {author} {\bibinfo {author} {\bibfnamefont {K.}~\bibnamefont
  {Shiozaki}}, \bibinfo {author} {\bibfnamefont {H.}~\bibnamefont
  {Shapourian}}, \ and\ \bibinfo {author} {\bibfnamefont {S.}~\bibnamefont
  {Ryu}},\ }\href@noop {} {\bibfield  {journal} {\bibinfo  {journal} {arXiv
  preprint arXiv:1609.05970}\ } (\bibinfo {year} {2016})}\BibitemShut {NoStop}%
\bibitem [{\citenamefont {Shiozaki}\ and\ \citenamefont
  {Ryu}(2017)}]{shiozaki2017matrix}%
  \BibitemOpen
  \bibfield  {author} {\bibinfo {author} {\bibfnamefont {K.}~\bibnamefont
  {Shiozaki}}\ and\ \bibinfo {author} {\bibfnamefont {S.}~\bibnamefont {Ryu}},\
  }\href@noop {} {\bibfield  {journal} {\bibinfo  {journal} {Journal of High
  Energy Physics}\ }\textbf {\bibinfo {volume} {2017}},\ \bibinfo {pages} {100}
  (\bibinfo {year} {2017})}\BibitemShut {NoStop}%
\bibitem [{\citenamefont {Dijkgraaf}\ \emph {et~al.}(1989)\citenamefont
  {Dijkgraaf}, \citenamefont {Vafa}, \citenamefont {Verlinde},\ and\
  \citenamefont {Verlinde}}]{dijkgraaf1989operator}%
  \BibitemOpen
  \bibfield  {author} {\bibinfo {author} {\bibfnamefont {R.}~\bibnamefont
  {Dijkgraaf}}, \bibinfo {author} {\bibfnamefont {C.}~\bibnamefont {Vafa}},
  \bibinfo {author} {\bibfnamefont {E.}~\bibnamefont {Verlinde}}, \ and\
  \bibinfo {author} {\bibfnamefont {H.}~\bibnamefont {Verlinde}},\ }\href@noop
  {} {\bibfield  {journal} {\bibinfo  {journal} {Communications in Mathematical
  Physics}\ }\textbf {\bibinfo {volume} {123}},\ \bibinfo {pages} {485}
  (\bibinfo {year} {1989})}\BibitemShut {NoStop}%
\bibitem [{\citenamefont {Ginsparg}(1988)}]{ginsparg1988applied}%
  \BibitemOpen
  \bibfield  {author} {\bibinfo {author} {\bibfnamefont {P.}~\bibnamefont
  {Ginsparg}},\ }\href@noop {} {\bibfield  {journal} {\bibinfo  {journal}
  {arXiv preprint hep-th/9108028}\ } (\bibinfo {year} {1988})}\BibitemShut
  {NoStop}%
\bibitem [{\citenamefont {Dijkgraaf}\ and\ \citenamefont
  {Witten}(1990)}]{dijkgraaf1990topological}%
  \BibitemOpen
  \bibfield  {author} {\bibinfo {author} {\bibfnamefont {R.}~\bibnamefont
  {Dijkgraaf}}\ and\ \bibinfo {author} {\bibfnamefont {E.}~\bibnamefont
  {Witten}},\ }\href@noop {} {\bibfield  {journal} {\bibinfo  {journal}
  {Communications in Mathematical Physics}\ }\textbf {\bibinfo {volume}
  {129}},\ \bibinfo {pages} {393} (\bibinfo {year} {1990})}\BibitemShut
  {NoStop}%
\bibitem [{\citenamefont {Levin}\ and\ \citenamefont
  {Gu}(2012)}]{levin2012braiding}%
  \BibitemOpen
  \bibfield  {author} {\bibinfo {author} {\bibfnamefont {M.}~\bibnamefont
  {Levin}}\ and\ \bibinfo {author} {\bibfnamefont {Z.-C.}\ \bibnamefont {Gu}},\
  }\href@noop {} {\bibfield  {journal} {\bibinfo  {journal} {Physical Review
  B}\ }\textbf {\bibinfo {volume} {86}},\ \bibinfo {pages} {115109} (\bibinfo
  {year} {2012})}\BibitemShut {NoStop}%
\bibitem [{\citenamefont {Wang}\ and\ \citenamefont
  {Levin}(2014)}]{wang2014braiding}%
  \BibitemOpen
  \bibfield  {author} {\bibinfo {author} {\bibfnamefont {C.}~\bibnamefont
  {Wang}}\ and\ \bibinfo {author} {\bibfnamefont {M.}~\bibnamefont {Levin}},\
  }\href@noop {} {\bibfield  {journal} {\bibinfo  {journal} {Physical review
  letters}\ }\textbf {\bibinfo {volume} {113}},\ \bibinfo {pages} {080403}
  (\bibinfo {year} {2014})}\BibitemShut {NoStop}%
\bibitem [{\citenamefont {Lin}\ and\ \citenamefont
  {Levin}(2015)}]{lin2015loop}%
  \BibitemOpen
  \bibfield  {author} {\bibinfo {author} {\bibfnamefont {C.-H.}\ \bibnamefont
  {Lin}}\ and\ \bibinfo {author} {\bibfnamefont {M.}~\bibnamefont {Levin}},\
  }\href@noop {} {\bibfield  {journal} {\bibinfo  {journal} {Physical Review
  B}\ }\textbf {\bibinfo {volume} {92}},\ \bibinfo {pages} {035115} (\bibinfo
  {year} {2015})}\BibitemShut {NoStop}%
\bibitem [{\citenamefont {Tiwari}\ \emph {et~al.}(2016)\citenamefont {Tiwari},
  \citenamefont {Chen},\ and\ \citenamefont {Ryu}}]{tiwari2016wilson}%
  \BibitemOpen
  \bibfield  {author} {\bibinfo {author} {\bibfnamefont {A.}~\bibnamefont
  {Tiwari}}, \bibinfo {author} {\bibfnamefont {X.}~\bibnamefont {Chen}}, \ and\
  \bibinfo {author} {\bibfnamefont {S.}~\bibnamefont {Ryu}},\ }\href@noop {}
  {\bibfield  {journal} {\bibinfo  {journal} {arXiv preprint arXiv:1603.08429}\
  } (\bibinfo {year} {2016})}\BibitemShut {NoStop}%
\bibitem [{\citenamefont {Chen}\ \emph {et~al.}(2016)\citenamefont {Chen},
  \citenamefont {Tiwari},\ and\ \citenamefont {Ryu}}]{chen2016bulk}%
  \BibitemOpen
  \bibfield  {author} {\bibinfo {author} {\bibfnamefont {X.}~\bibnamefont
  {Chen}}, \bibinfo {author} {\bibfnamefont {A.}~\bibnamefont {Tiwari}}, \ and\
  \bibinfo {author} {\bibfnamefont {S.}~\bibnamefont {Ryu}},\ }\href@noop {}
  {\bibfield  {journal} {\bibinfo  {journal} {Physical Review B}\ }\textbf
  {\bibinfo {volume} {94}},\ \bibinfo {pages} {045113} (\bibinfo {year}
  {2016})}\BibitemShut {NoStop}%
\bibitem [{\citenamefont {Vafa}(1989)}]{vafa1989quantum}%
  \BibitemOpen
  \bibfield  {author} {\bibinfo {author} {\bibfnamefont {C.}~\bibnamefont
  {Vafa}},\ }\href@noop {} {\bibfield  {journal} {\bibinfo  {journal} {Modern
  Physics Letters A}\ }\textbf {\bibinfo {volume} {4}},\ \bibinfo {pages}
  {1615} (\bibinfo {year} {1989})}\BibitemShut {NoStop}%
\bibitem [{\citenamefont {Bhardwaj}\ and\ \citenamefont
  {Tachikawa}(2017)}]{bhardwaj2017finite}%
  \BibitemOpen
  \bibfield  {author} {\bibinfo {author} {\bibfnamefont {L.}~\bibnamefont
  {Bhardwaj}}\ and\ \bibinfo {author} {\bibfnamefont {Y.}~\bibnamefont
  {Tachikawa}},\ }\href@noop {} {\bibfield  {journal} {\bibinfo  {journal}
  {arXiv preprint arXiv:1704.02330}\ } (\bibinfo {year} {2017})}\BibitemShut
  {NoStop}%
\bibitem [{\citenamefont {Kapustin}\ and\ \citenamefont
  {Thorngren}(2017{\natexlab{a}})}]{kapustin2017fermionic}%
  \BibitemOpen
  \bibfield  {author} {\bibinfo {author} {\bibfnamefont {A.}~\bibnamefont
  {Kapustin}}\ and\ \bibinfo {author} {\bibfnamefont {R.}~\bibnamefont
  {Thorngren}},\ }\href@noop {} {\bibfield  {journal} {\bibinfo  {journal}
  {arXiv preprint arXiv:1701.08264}\ } (\bibinfo {year}
  {2017}{\natexlab{a}})}\BibitemShut {NoStop}%
\bibitem [{\citenamefont {Sule}\ \emph {et~al.}(2013)\citenamefont {Sule},
  \citenamefont {Chen},\ and\ \citenamefont {Ryu}}]{sule2013symmetry}%
  \BibitemOpen
  \bibfield  {author} {\bibinfo {author} {\bibfnamefont {O.~M.}\ \bibnamefont
  {Sule}}, \bibinfo {author} {\bibfnamefont {X.}~\bibnamefont {Chen}}, \ and\
  \bibinfo {author} {\bibfnamefont {S.}~\bibnamefont {Ryu}},\ }\href@noop {}
  {\bibfield  {journal} {\bibinfo  {journal} {Physical Review B}\ }\textbf
  {\bibinfo {volume} {88}},\ \bibinfo {pages} {075125} (\bibinfo {year}
  {2013})}\BibitemShut {NoStop}%
\bibitem [{\citenamefont {Hsieh}\ \emph {et~al.}(2014)\citenamefont {Hsieh},
  \citenamefont {Sule}, \citenamefont {Cho}, \citenamefont {Ryu},\ and\
  \citenamefont {Leigh}}]{hsieh2014symmetry}%
  \BibitemOpen
  \bibfield  {author} {\bibinfo {author} {\bibfnamefont {C.-T.}\ \bibnamefont
  {Hsieh}}, \bibinfo {author} {\bibfnamefont {O.~M.}\ \bibnamefont {Sule}},
  \bibinfo {author} {\bibfnamefont {G.~Y.}\ \bibnamefont {Cho}}, \bibinfo
  {author} {\bibfnamefont {S.}~\bibnamefont {Ryu}}, \ and\ \bibinfo {author}
  {\bibfnamefont {R.~G.}\ \bibnamefont {Leigh}},\ }\href@noop {} {\bibfield
  {journal} {\bibinfo  {journal} {Physical Review B}\ }\textbf {\bibinfo
  {volume} {90}},\ \bibinfo {pages} {165134} (\bibinfo {year}
  {2014})}\BibitemShut {NoStop}%
\bibitem [{\citenamefont {Hsieh}\ \emph {et~al.}(2016)\citenamefont {Hsieh},
  \citenamefont {Cho},\ and\ \citenamefont {Ryu}}]{hsieh2016global}%
  \BibitemOpen
  \bibfield  {author} {\bibinfo {author} {\bibfnamefont {C.-T.}\ \bibnamefont
  {Hsieh}}, \bibinfo {author} {\bibfnamefont {G.~Y.}\ \bibnamefont {Cho}}, \
  and\ \bibinfo {author} {\bibfnamefont {S.}~\bibnamefont {Ryu}},\ }\href@noop
  {} {\bibfield  {journal} {\bibinfo  {journal} {Physical Review B}\ }\textbf
  {\bibinfo {volume} {93}},\ \bibinfo {pages} {075135} (\bibinfo {year}
  {2016})}\BibitemShut {NoStop}%
\bibitem [{\citenamefont {Kapustin}\ and\ \citenamefont
  {Thorngren}(2014{\natexlab{a}})}]{kapustin2014anomalous}%
  \BibitemOpen
  \bibfield  {author} {\bibinfo {author} {\bibfnamefont {A.}~\bibnamefont
  {Kapustin}}\ and\ \bibinfo {author} {\bibfnamefont {R.}~\bibnamefont
  {Thorngren}},\ }\href@noop {} {\bibfield  {journal} {\bibinfo  {journal}
  {Physical review letters}\ }\textbf {\bibinfo {volume} {112}},\ \bibinfo
  {pages} {231602} (\bibinfo {year} {2014}{\natexlab{a}})}\BibitemShut
  {NoStop}%
\bibitem [{\citenamefont {Kapustin}\ and\ \citenamefont
  {Thorngren}(2014{\natexlab{b}})}]{kapustin2014anomalies}%
  \BibitemOpen
  \bibfield  {author} {\bibinfo {author} {\bibfnamefont {A.}~\bibnamefont
  {Kapustin}}\ and\ \bibinfo {author} {\bibfnamefont {R.}~\bibnamefont
  {Thorngren}},\ }\href@noop {} {\bibfield  {journal} {\bibinfo  {journal}
  {arXiv preprint arXiv:1404.3230}\ } (\bibinfo {year}
  {2014}{\natexlab{b}})}\BibitemShut {NoStop}%
\bibitem [{\citenamefont {Han}\ \emph {et~al.}(2017)\citenamefont {Han},
  \citenamefont {Tiwari}, \citenamefont {Hsieh},\ and\ \citenamefont
  {Ryu}}]{han2017boundary}%
  \BibitemOpen
  \bibfield  {author} {\bibinfo {author} {\bibfnamefont {B.}~\bibnamefont
  {Han}}, \bibinfo {author} {\bibfnamefont {A.}~\bibnamefont {Tiwari}},
  \bibinfo {author} {\bibfnamefont {C.-T.}\ \bibnamefont {Hsieh}}, \ and\
  \bibinfo {author} {\bibfnamefont {S.}~\bibnamefont {Ryu}},\ }\href@noop {}
  {\bibfield  {journal} {\bibinfo  {journal} {arXiv preprint arXiv:1704.01193}\
  } (\bibinfo {year} {2017})}\BibitemShut {NoStop}%
\bibitem [{\citenamefont {Hooft}(1980)}]{hooft1980naturalness}%
  \BibitemOpen
  \bibfield  {author} {\bibinfo {author} {\bibfnamefont {G.}~\bibnamefont
  {Hooft}},\ }\href@noop {} {\bibfield  {journal} {\bibinfo  {journal} {Recent
  developments in gauge theories. Proceedings of the NATO Advanced Study
  Institute on recent developments in gauge theories, held in Carg{\`e}se,
  Corsica, August 26-September 8, 1979}\ ,\ \bibinfo {pages} {135}} (\bibinfo
  {year} {1980})}\BibitemShut {NoStop}%
\bibitem [{\citenamefont {Ryu}\ and\ \citenamefont
  {Zhang}(2012)}]{ryu2012interacting}%
  \BibitemOpen
  \bibfield  {author} {\bibinfo {author} {\bibfnamefont {S.}~\bibnamefont
  {Ryu}}\ and\ \bibinfo {author} {\bibfnamefont {S.-C.}\ \bibnamefont
  {Zhang}},\ }\href@noop {} {\bibfield  {journal} {\bibinfo  {journal}
  {Physical Review B}\ }\textbf {\bibinfo {volume} {85}},\ \bibinfo {pages}
  {245132} (\bibinfo {year} {2012})}\BibitemShut {NoStop}%
\bibitem [{\citenamefont {Vishwanath}\ and\ \citenamefont
  {Senthil}(2013)}]{vishwanath2013physics}%
  \BibitemOpen
  \bibfield  {author} {\bibinfo {author} {\bibfnamefont {A.}~\bibnamefont
  {Vishwanath}}\ and\ \bibinfo {author} {\bibfnamefont {T.}~\bibnamefont
  {Senthil}},\ }\href@noop {} {\bibfield  {journal} {\bibinfo  {journal}
  {Physical Review X}\ }\textbf {\bibinfo {volume} {3}},\ \bibinfo {pages}
  {011016} (\bibinfo {year} {2013})}\BibitemShut {NoStop}%
\bibitem [{\citenamefont {Chen}\ \emph {et~al.}(2015)\citenamefont {Chen},
  \citenamefont {Burnell}, \citenamefont {Vishwanath},\ and\ \citenamefont
  {Fidkowski}}]{chen2015anomalous}%
  \BibitemOpen
  \bibfield  {author} {\bibinfo {author} {\bibfnamefont {X.}~\bibnamefont
  {Chen}}, \bibinfo {author} {\bibfnamefont {F.~J.}\ \bibnamefont {Burnell}},
  \bibinfo {author} {\bibfnamefont {A.}~\bibnamefont {Vishwanath}}, \ and\
  \bibinfo {author} {\bibfnamefont {L.}~\bibnamefont {Fidkowski}},\ }\href@noop
  {} {\bibfield  {journal} {\bibinfo  {journal} {Physical Review X}\ }\textbf
  {\bibinfo {volume} {5}},\ \bibinfo {pages} {041013} (\bibinfo {year}
  {2015})}\BibitemShut {NoStop}%
\bibitem [{\citenamefont {Burnell}\ \emph {et~al.}(2014)\citenamefont
  {Burnell}, \citenamefont {Chen}, \citenamefont {Fidkowski},\ and\
  \citenamefont {Vishwanath}}]{burnell2014exactly}%
  \BibitemOpen
  \bibfield  {author} {\bibinfo {author} {\bibfnamefont {F.~J.}\ \bibnamefont
  {Burnell}}, \bibinfo {author} {\bibfnamefont {X.}~\bibnamefont {Chen}},
  \bibinfo {author} {\bibfnamefont {L.}~\bibnamefont {Fidkowski}}, \ and\
  \bibinfo {author} {\bibfnamefont {A.}~\bibnamefont {Vishwanath}},\
  }\href@noop {} {\bibfield  {journal} {\bibinfo  {journal} {Physical Review
  B}\ }\textbf {\bibinfo {volume} {90}},\ \bibinfo {pages} {245122} (\bibinfo
  {year} {2014})}\BibitemShut {NoStop}%
\bibitem [{\citenamefont {Wang}\ and\ \citenamefont
  {Senthil}(2013)}]{wang2013boson}%
  \BibitemOpen
  \bibfield  {author} {\bibinfo {author} {\bibfnamefont {C.}~\bibnamefont
  {Wang}}\ and\ \bibinfo {author} {\bibfnamefont {T.}~\bibnamefont {Senthil}},\
  }\href@noop {} {\bibfield  {journal} {\bibinfo  {journal} {Physical Review
  B}\ }\textbf {\bibinfo {volume} {87}},\ \bibinfo {pages} {235122} (\bibinfo
  {year} {2013})}\BibitemShut {NoStop}%
\bibitem [{\citenamefont {Fidkowski}\ and\ \citenamefont
  {Vishwanath}(2015)}]{fidkowski2015realizing}%
  \BibitemOpen
  \bibfield  {author} {\bibinfo {author} {\bibfnamefont {L.}~\bibnamefont
  {Fidkowski}}\ and\ \bibinfo {author} {\bibfnamefont {A.}~\bibnamefont
  {Vishwanath}},\ }\href@noop {} {\bibfield  {journal} {\bibinfo  {journal}
  {arXiv preprint arXiv:1511.01502}\ } (\bibinfo {year} {2015})}\BibitemShut
  {NoStop}%
\bibitem [{\citenamefont {Witten}(1989)}]{witten1989quantum}%
  \BibitemOpen
  \bibfield  {author} {\bibinfo {author} {\bibfnamefont {E.}~\bibnamefont
  {Witten}},\ }\href@noop {} {\bibfield  {journal} {\bibinfo  {journal}
  {Communications in Mathematical Physics}\ }\textbf {\bibinfo {volume}
  {121}},\ \bibinfo {pages} {351} (\bibinfo {year} {1989})}\BibitemShut
  {NoStop}%
\bibitem [{\citenamefont {Dijkgraaf}\ \emph {et~al.}(1991)\citenamefont
  {Dijkgraaf}, \citenamefont {Pasquier},\ and\ \citenamefont
  {Roche}}]{dijkgraaf1991quasi}%
  \BibitemOpen
  \bibfield  {author} {\bibinfo {author} {\bibfnamefont {R.}~\bibnamefont
  {Dijkgraaf}}, \bibinfo {author} {\bibfnamefont {V.}~\bibnamefont {Pasquier}},
  \ and\ \bibinfo {author} {\bibfnamefont {P.}~\bibnamefont {Roche}},\
  }\href@noop {} {\bibfield  {journal} {\bibinfo  {journal} {Nuclear Physics
  B-Proceedings Supplements}\ }\textbf {\bibinfo {volume} {18}},\ \bibinfo
  {pages} {60} (\bibinfo {year} {1991})}\BibitemShut {NoStop}%
\bibitem [{\citenamefont {Wen}(1992)}]{wen1992theory}%
  \BibitemOpen
  \bibfield  {author} {\bibinfo {author} {\bibfnamefont {X.-G.}\ \bibnamefont
  {Wen}},\ }\href@noop {} {\bibfield  {journal} {\bibinfo  {journal}
  {International journal of modern physics B}\ }\textbf {\bibinfo {volume}
  {6}},\ \bibinfo {pages} {1711} (\bibinfo {year} {1992})}\BibitemShut
  {NoStop}%
\bibitem [{\citenamefont {Hatsugai}(1993)}]{hatsugai1993chern}%
  \BibitemOpen
  \bibfield  {author} {\bibinfo {author} {\bibfnamefont {Y.}~\bibnamefont
  {Hatsugai}},\ }\href@noop {} {\bibfield  {journal} {\bibinfo  {journal}
  {Physical review letters}\ }\textbf {\bibinfo {volume} {71}},\ \bibinfo
  {pages} {3697} (\bibinfo {year} {1993})}\BibitemShut {NoStop}%
\bibitem [{\citenamefont {Cappelli}\ \emph {et~al.}(2002)\citenamefont
  {Cappelli}, \citenamefont {Huerta},\ and\ \citenamefont
  {Zemba}}]{cappelli2002thermal}%
  \BibitemOpen
  \bibfield  {author} {\bibinfo {author} {\bibfnamefont {A.}~\bibnamefont
  {Cappelli}}, \bibinfo {author} {\bibfnamefont {M.}~\bibnamefont {Huerta}}, \
  and\ \bibinfo {author} {\bibfnamefont {G.~R.}\ \bibnamefont {Zemba}},\
  }\href@noop {} {\bibfield  {journal} {\bibinfo  {journal} {Nuclear Physics
  B}\ }\textbf {\bibinfo {volume} {636}},\ \bibinfo {pages} {568} (\bibinfo
  {year} {2002})}\BibitemShut {NoStop}%
\bibitem [{\citenamefont {Cappelli}\ and\ \citenamefont
  {Zemba}(1997)}]{cappelli1997modular}%
  \BibitemOpen
  \bibfield  {author} {\bibinfo {author} {\bibfnamefont {A.}~\bibnamefont
  {Cappelli}}\ and\ \bibinfo {author} {\bibfnamefont {G.~R.}\ \bibnamefont
  {Zemba}},\ }\href@noop {} {\bibfield  {journal} {\bibinfo  {journal} {Nuclear
  Physics B}\ }\textbf {\bibinfo {volume} {490}},\ \bibinfo {pages} {595}
  (\bibinfo {year} {1997})}\BibitemShut {NoStop}%
\bibitem [{\citenamefont {Cappelli}\ \emph {et~al.}(2010)\citenamefont
  {Cappelli}, \citenamefont {Viola},\ and\ \citenamefont
  {Zemba}}]{cappelli2010chiral}%
  \BibitemOpen
  \bibfield  {author} {\bibinfo {author} {\bibfnamefont {A.}~\bibnamefont
  {Cappelli}}, \bibinfo {author} {\bibfnamefont {G.}~\bibnamefont {Viola}}, \
  and\ \bibinfo {author} {\bibfnamefont {G.~R.}\ \bibnamefont {Zemba}},\
  }\href@noop {} {\bibfield  {journal} {\bibinfo  {journal} {Annals of
  Physics}\ }\textbf {\bibinfo {volume} {325}},\ \bibinfo {pages} {465}
  (\bibinfo {year} {2010})}\BibitemShut {NoStop}%
\bibitem [{\citenamefont {Cappelli}\ and\ \citenamefont
  {Viola}(2011)}]{cappelli2011partition}%
  \BibitemOpen
  \bibfield  {author} {\bibinfo {author} {\bibfnamefont {A.}~\bibnamefont
  {Cappelli}}\ and\ \bibinfo {author} {\bibfnamefont {G.}~\bibnamefont
  {Viola}},\ }\href@noop {} {\bibfield  {journal} {\bibinfo  {journal} {Journal
  of Physics A: Mathematical and Theoretical}\ }\textbf {\bibinfo {volume}
  {44}},\ \bibinfo {pages} {075401} (\bibinfo {year} {2011})}\BibitemShut
  {NoStop}%
\bibitem [{\citenamefont {Chen}\ \emph
  {et~al.}(2017{\natexlab{a}})\citenamefont {Chen}, \citenamefont {Tiwari},
  \citenamefont {Nayak},\ and\ \citenamefont {Ryu}}]{chen2017gauging}%
  \BibitemOpen
  \bibfield  {author} {\bibinfo {author} {\bibfnamefont {X.}~\bibnamefont
  {Chen}}, \bibinfo {author} {\bibfnamefont {A.}~\bibnamefont {Tiwari}},
  \bibinfo {author} {\bibfnamefont {C.}~\bibnamefont {Nayak}}, \ and\ \bibinfo
  {author} {\bibfnamefont {S.}~\bibnamefont {Ryu}},\ }\href@noop {} {\bibfield
  {journal} {\bibinfo  {journal} {arXiv preprint arXiv:1706.00560}\ } (\bibinfo
  {year} {2017}{\natexlab{a}})}\BibitemShut {NoStop}%
\bibitem [{\citenamefont {Lu}\ and\ \citenamefont
  {Vishwanath}(2012)}]{lu2012theory}%
  \BibitemOpen
  \bibfield  {author} {\bibinfo {author} {\bibfnamefont {Y.-M.}\ \bibnamefont
  {Lu}}\ and\ \bibinfo {author} {\bibfnamefont {A.}~\bibnamefont
  {Vishwanath}},\ }\href@noop {} {\bibfield  {journal} {\bibinfo  {journal}
  {Physical Review B}\ }\textbf {\bibinfo {volume} {86}},\ \bibinfo {pages}
  {125119} (\bibinfo {year} {2012})}\BibitemShut {NoStop}%
\bibitem [{\citenamefont {Ye}\ and\ \citenamefont
  {Gu}(2016)}]{ye2016topological}%
  \BibitemOpen
  \bibfield  {author} {\bibinfo {author} {\bibfnamefont {P.}~\bibnamefont
  {Ye}}\ and\ \bibinfo {author} {\bibfnamefont {Z.-C.}\ \bibnamefont {Gu}},\
  }\href@noop {} {\bibfield  {journal} {\bibinfo  {journal} {Physical Review
  B}\ }\textbf {\bibinfo {volume} {93}},\ \bibinfo {pages} {205157} (\bibinfo
  {year} {2016})}\BibitemShut {NoStop}%
\bibitem [{\citenamefont {Witten}(2003)}]{witten2003sl}%
  \BibitemOpen
  \bibfield  {author} {\bibinfo {author} {\bibfnamefont {E.}~\bibnamefont
  {Witten}},\ }\href@noop {} {\bibfield  {journal} {\bibinfo  {journal} {arXiv
  preprint hep-th/0307041}\ } (\bibinfo {year} {2003})}\BibitemShut {NoStop}%
\bibitem [{\citenamefont {Propitius}(1995)}]{propitius1995topological}%
  \BibitemOpen
  \bibfield  {author} {\bibinfo {author} {\bibfnamefont {M.~d.~W.}\
  \bibnamefont {Propitius}},\ }\href@noop {} {\bibfield  {journal} {\bibinfo
  {journal} {arXiv preprint hep-th/9511195}\ } (\bibinfo {year}
  {1995})}\BibitemShut {NoStop}%
\bibitem [{\citenamefont {Wen}(2014)}]{wen2014symmetry}%
  \BibitemOpen
  \bibfield  {author} {\bibinfo {author} {\bibfnamefont {X.-G.}\ \bibnamefont
  {Wen}},\ }\href@noop {} {\bibfield  {journal} {\bibinfo  {journal} {Physical
  Review B}\ }\textbf {\bibinfo {volume} {89}},\ \bibinfo {pages} {035147}
  (\bibinfo {year} {2014})}\BibitemShut {NoStop}%
\bibitem [{\citenamefont {Wang}\ \emph {et~al.}(2015)\citenamefont {Wang},
  \citenamefont {Gu},\ and\ \citenamefont {Wen}}]{wang2015field}%
  \BibitemOpen
  \bibfield  {author} {\bibinfo {author} {\bibfnamefont {J.~C.}\ \bibnamefont
  {Wang}}, \bibinfo {author} {\bibfnamefont {Z.-C.}\ \bibnamefont {Gu}}, \ and\
  \bibinfo {author} {\bibfnamefont {X.-G.}\ \bibnamefont {Wen}},\ }\href@noop
  {} {\bibfield  {journal} {\bibinfo  {journal} {Physical review letters}\
  }\textbf {\bibinfo {volume} {114}},\ \bibinfo {pages} {031601} (\bibinfo
  {year} {2015})}\BibitemShut {NoStop}%
\bibitem [{\citenamefont
  {Tantivasadakarn}(2017)}]{tantivasadakarn2017dimensional}%
  \BibitemOpen
  \bibfield  {author} {\bibinfo {author} {\bibfnamefont {N.}~\bibnamefont
  {Tantivasadakarn}},\ }\href@noop {} {\bibfield  {journal} {\bibinfo
  {journal} {arXiv preprint arXiv:1706.09769}\ } (\bibinfo {year}
  {2017})}\BibitemShut {NoStop}%
\bibitem [{\citenamefont {He}\ \emph {et~al.}(2017)\citenamefont {He},
  \citenamefont {Zheng},\ and\ \citenamefont {von Keyserlingk}}]{he2017field}%
  \BibitemOpen
  \bibfield  {author} {\bibinfo {author} {\bibfnamefont {H.}~\bibnamefont
  {He}}, \bibinfo {author} {\bibfnamefont {Y.}~\bibnamefont {Zheng}}, \ and\
  \bibinfo {author} {\bibfnamefont {C.}~\bibnamefont {von Keyserlingk}},\
  }\href@noop {} {\bibfield  {journal} {\bibinfo  {journal} {Physical Review
  B}\ }\textbf {\bibinfo {volume} {95}},\ \bibinfo {pages} {035131} (\bibinfo
  {year} {2017})}\BibitemShut {NoStop}%
\bibitem [{\citenamefont {Bauer}\ \emph {et~al.}(2005)\citenamefont {Bauer},
  \citenamefont {Girardi}, \citenamefont {Stora},\ and\ \citenamefont
  {Thuillier}}]{bauer2005class}%
  \BibitemOpen
  \bibfield  {author} {\bibinfo {author} {\bibfnamefont {M.}~\bibnamefont
  {Bauer}}, \bibinfo {author} {\bibfnamefont {G.}~\bibnamefont {Girardi}},
  \bibinfo {author} {\bibfnamefont {R.}~\bibnamefont {Stora}}, \ and\ \bibinfo
  {author} {\bibfnamefont {F.}~\bibnamefont {Thuillier}},\ }\href@noop {}
  {\bibfield  {journal} {\bibinfo  {journal} {Journal of High Energy Physics}\
  }\textbf {\bibinfo {volume} {2005}},\ \bibinfo {pages} {027} (\bibinfo {year}
  {2005})}\BibitemShut {NoStop}%
\bibitem [{\citenamefont {Bais}\ and\ \citenamefont
  {Slingerland}(2009)}]{bais2009condensate}%
  \BibitemOpen
  \bibfield  {author} {\bibinfo {author} {\bibfnamefont {F.}~\bibnamefont
  {Bais}}\ and\ \bibinfo {author} {\bibfnamefont {J.}~\bibnamefont
  {Slingerland}},\ }\href@noop {} {\bibfield  {journal} {\bibinfo  {journal}
  {Physical Review B}\ }\textbf {\bibinfo {volume} {79}},\ \bibinfo {pages}
  {045316} (\bibinfo {year} {2009})}\BibitemShut {NoStop}%
\bibitem [{\citenamefont {Kong}(2014)}]{kong2014anyon}%
  \BibitemOpen
  \bibfield  {author} {\bibinfo {author} {\bibfnamefont {L.}~\bibnamefont
  {Kong}},\ }\href@noop {} {\bibfield  {journal} {\bibinfo  {journal} {Nuclear
  Physics B}\ }\textbf {\bibinfo {volume} {886}},\ \bibinfo {pages} {436}
  (\bibinfo {year} {2014})}\BibitemShut {NoStop}%
\bibitem [{\citenamefont {Neupert}\ \emph {et~al.}(2016)\citenamefont
  {Neupert}, \citenamefont {He}, \citenamefont {von Keyserlingk}, \citenamefont
  {Sierra},\ and\ \citenamefont {Bernevig}}]{neupert2016boson}%
  \BibitemOpen
  \bibfield  {author} {\bibinfo {author} {\bibfnamefont {T.}~\bibnamefont
  {Neupert}}, \bibinfo {author} {\bibfnamefont {H.}~\bibnamefont {He}},
  \bibinfo {author} {\bibfnamefont {C.}~\bibnamefont {von Keyserlingk}},
  \bibinfo {author} {\bibfnamefont {G.}~\bibnamefont {Sierra}}, \ and\ \bibinfo
  {author} {\bibfnamefont {B.~A.}\ \bibnamefont {Bernevig}},\ }\href@noop {}
  {\bibfield  {journal} {\bibinfo  {journal} {Physical Review B}\ }\textbf
  {\bibinfo {volume} {93}},\ \bibinfo {pages} {115103} (\bibinfo {year}
  {2016})}\BibitemShut {NoStop}%
\bibitem [{\citenamefont {Burnell}(2017)}]{burnell2017anyon}%
  \BibitemOpen
  \bibfield  {author} {\bibinfo {author} {\bibfnamefont {F.}~\bibnamefont
  {Burnell}},\ }\href@noop {} {\bibfield  {journal} {\bibinfo  {journal} {arXiv
  preprint arXiv:1706.04940}\ } (\bibinfo {year} {2017})}\BibitemShut {NoStop}%
\bibitem [{Note1()}]{Note1}%
  \BibitemOpen
  \bibinfo {note} {Large diffeomorphisms are those diffeomorphisms that are not
  path connected to the identity diffeomorphism in the space of
  diffeomorphisms. These form a group known as the mapping class group of $M$
  which we shall abbreviate $MCG(M)$. Then \begin {align} MCG(M)=\pi _0\left
  [\protect \text {Diff}(M)\right ] \end {align}}\BibitemShut {NoStop}%
\bibitem [{\citenamefont {Francesco}\ \emph {et~al.}(2012)\citenamefont
  {Francesco}, \citenamefont {Mathieu},\ and\ \citenamefont
  {S{\'e}n{\'e}chal}}]{francesco2012conformal}%
  \BibitemOpen
  \bibfield  {author} {\bibinfo {author} {\bibfnamefont {P.}~\bibnamefont
  {Francesco}}, \bibinfo {author} {\bibfnamefont {P.}~\bibnamefont {Mathieu}},
  \ and\ \bibinfo {author} {\bibfnamefont {D.}~\bibnamefont
  {S{\'e}n{\'e}chal}},\ }\href@noop {} {\emph {\bibinfo {title} {Conformal
  field theory}}}\ (\bibinfo  {publisher} {Springer Science \& Business
  Media},\ \bibinfo {year} {2012})\BibitemShut {NoStop}%
\bibitem [{\citenamefont {Blumenhagen}\ \emph {et~al.}(2012)\citenamefont
  {Blumenhagen}, \citenamefont {L{\"u}st},\ and\ \citenamefont
  {Theisen}}]{blumenhagen2012introduction}%
  \BibitemOpen
  \bibfield  {author} {\bibinfo {author} {\bibfnamefont {R.}~\bibnamefont
  {Blumenhagen}}, \bibinfo {author} {\bibfnamefont {D.}~\bibnamefont
  {L{\"u}st}}, \ and\ \bibinfo {author} {\bibfnamefont {S.}~\bibnamefont
  {Theisen}},\ }in\ \href@noop {} {\emph {\bibinfo {booktitle} {Basic Concepts
  of String Theory}}}\ (\bibinfo  {publisher} {Springer},\ \bibinfo {year}
  {2012})\ pp.\ \bibinfo {pages} {63--106}\BibitemShut {NoStop}%
\bibitem [{\citenamefont {Vafa}(1986)}]{vafa1986modular}%
  \BibitemOpen
  \bibfield  {author} {\bibinfo {author} {\bibfnamefont {C.}~\bibnamefont
  {Vafa}},\ }\href@noop {} {\bibfield  {journal} {\bibinfo  {journal} {Nuclear
  Physics B}\ }\textbf {\bibinfo {volume} {273}},\ \bibinfo {pages} {592}
  (\bibinfo {year} {1986})}\BibitemShut {NoStop}%
\bibitem [{\citenamefont {Gaberdiel}(2000)}]{gaberdiel2000discrete}%
  \BibitemOpen
  \bibfield  {author} {\bibinfo {author} {\bibfnamefont {M.~R.}\ \bibnamefont
  {Gaberdiel}},\ }\href@noop {} {\bibfield  {journal} {\bibinfo  {journal}
  {Journal of High Energy Physics}\ }\textbf {\bibinfo {volume} {2000}},\
  \bibinfo {pages} {026} (\bibinfo {year} {2000})}\BibitemShut {NoStop}%
\bibitem [{\citenamefont {Chen}\ \emph
  {et~al.}(2017{\natexlab{b}})\citenamefont {Chen}, \citenamefont {Roy},
  \citenamefont {Teo},\ and\ \citenamefont {Ryu}}]{chen2017orbifolding}%
  \BibitemOpen
  \bibfield  {author} {\bibinfo {author} {\bibfnamefont {X.}~\bibnamefont
  {Chen}}, \bibinfo {author} {\bibfnamefont {A.}~\bibnamefont {Roy}}, \bibinfo
  {author} {\bibfnamefont {J.~C.}\ \bibnamefont {Teo}}, \ and\ \bibinfo
  {author} {\bibfnamefont {S.}~\bibnamefont {Ryu}},\ }\href@noop {} {\bibfield
  {journal} {\bibinfo  {journal} {arXiv preprint arXiv:1706.00557}\ } (\bibinfo
  {year} {2017}{\natexlab{b}})}\BibitemShut {NoStop}%
\bibitem [{\citenamefont {Jeffrey}(1992)}]{jeffrey1992chern}%
  \BibitemOpen
  \bibfield  {author} {\bibinfo {author} {\bibfnamefont {L.~C.}\ \bibnamefont
  {Jeffrey}},\ }\href@noop {} {\bibfield  {journal} {\bibinfo  {journal}
  {Communications in mathematical physics}\ }\textbf {\bibinfo {volume}
  {147}},\ \bibinfo {pages} {563} (\bibinfo {year} {1992})}\BibitemShut
  {NoStop}%
\bibitem [{\citenamefont {Qi}\ \emph {et~al.}(2012)\citenamefont {Qi},
  \citenamefont {Katsura},\ and\ \citenamefont {Ludwig}}]{qi2012general}%
  \BibitemOpen
  \bibfield  {author} {\bibinfo {author} {\bibfnamefont {X.-L.}\ \bibnamefont
  {Qi}}, \bibinfo {author} {\bibfnamefont {H.}~\bibnamefont {Katsura}}, \ and\
  \bibinfo {author} {\bibfnamefont {A.~W.}\ \bibnamefont {Ludwig}},\
  }\href@noop {} {\bibfield  {journal} {\bibinfo  {journal} {Physical review
  letters}\ }\textbf {\bibinfo {volume} {108}},\ \bibinfo {pages} {196402}
  (\bibinfo {year} {2012})}\BibitemShut {NoStop}%
\bibitem [{\citenamefont {Wong}(2017)}]{wong2017note}%
  \BibitemOpen
  \bibfield  {author} {\bibinfo {author} {\bibfnamefont {G.}~\bibnamefont
  {Wong}},\ }\href@noop {} {\bibfield  {journal} {\bibinfo  {journal} {arXiv
  preprint arXiv:1706.04666}\ } (\bibinfo {year} {2017})}\BibitemShut {NoStop}%
\bibitem [{\citenamefont {Jiang}\ \emph {et~al.}(2014)\citenamefont {Jiang},
  \citenamefont {Mesaros},\ and\ \citenamefont {Ran}}]{jiang2014generalized}%
  \BibitemOpen
  \bibfield  {author} {\bibinfo {author} {\bibfnamefont {S.}~\bibnamefont
  {Jiang}}, \bibinfo {author} {\bibfnamefont {A.}~\bibnamefont {Mesaros}}, \
  and\ \bibinfo {author} {\bibfnamefont {Y.}~\bibnamefont {Ran}},\ }\href@noop
  {} {\bibfield  {journal} {\bibinfo  {journal} {Physical Review X}\ }\textbf
  {\bibinfo {volume} {4}},\ \bibinfo {pages} {031048} (\bibinfo {year}
  {2014})}\BibitemShut {NoStop}%
\bibitem [{\citenamefont {Wang}\ and\ \citenamefont {Wen}(2015)}]{wang2015non}%
  \BibitemOpen
  \bibfield  {author} {\bibinfo {author} {\bibfnamefont {J.~C.}\ \bibnamefont
  {Wang}}\ and\ \bibinfo {author} {\bibfnamefont {X.-G.}\ \bibnamefont {Wen}},\
  }\href@noop {} {\bibfield  {journal} {\bibinfo  {journal} {Physical Review
  B}\ }\textbf {\bibinfo {volume} {91}},\ \bibinfo {pages} {035134} (\bibinfo
  {year} {2015})}\BibitemShut {NoStop}%
\bibitem [{\citenamefont {Putrov}\ \emph {et~al.}(2017)\citenamefont {Putrov},
  \citenamefont {Wang},\ and\ \citenamefont {Yau}}]{putrov2017braiding}%
  \BibitemOpen
  \bibfield  {author} {\bibinfo {author} {\bibfnamefont {P.}~\bibnamefont
  {Putrov}}, \bibinfo {author} {\bibfnamefont {J.}~\bibnamefont {Wang}}, \ and\
  \bibinfo {author} {\bibfnamefont {S.-T.}\ \bibnamefont {Yau}},\ }\href@noop
  {} {\bibfield  {journal} {\bibinfo  {journal} {Annals of Physics}\ }
  (\bibinfo {year} {2017})}\BibitemShut {NoStop}%
\bibitem [{\citenamefont {Wu}(1991)}]{wu1991topological}%
  \BibitemOpen
  \bibfield  {author} {\bibinfo {author} {\bibfnamefont {S.}~\bibnamefont
  {Wu}},\ }\href@noop {} {\bibfield  {journal} {\bibinfo  {journal}
  {Communications in mathematical physics}\ }\textbf {\bibinfo {volume}
  {136}},\ \bibinfo {pages} {157} (\bibinfo {year} {1991})}\BibitemShut
  {NoStop}%
\bibitem [{\citenamefont {Balachandran}\ and\ \citenamefont
  {Teotonio-Sobrinho}(1993)}]{balachandran1993edge}%
  \BibitemOpen
  \bibfield  {author} {\bibinfo {author} {\bibfnamefont {A.}~\bibnamefont
  {Balachandran}}\ and\ \bibinfo {author} {\bibfnamefont {P.}~\bibnamefont
  {Teotonio-Sobrinho}},\ }\href@noop {} {\bibfield  {journal} {\bibinfo
  {journal} {International Journal of Modern Physics A}\ }\textbf {\bibinfo
  {volume} {8}},\ \bibinfo {pages} {723} (\bibinfo {year} {1993})}\BibitemShut
  {NoStop}%
\bibitem [{\citenamefont {Amoretti}\ \emph {et~al.}(2012)\citenamefont
  {Amoretti}, \citenamefont {Blasi}, \citenamefont {Maggiore},\ and\
  \citenamefont {Magnoli}}]{amoretti2012three}%
  \BibitemOpen
  \bibfield  {author} {\bibinfo {author} {\bibfnamefont {A.}~\bibnamefont
  {Amoretti}}, \bibinfo {author} {\bibfnamefont {A.}~\bibnamefont {Blasi}},
  \bibinfo {author} {\bibfnamefont {N.}~\bibnamefont {Maggiore}}, \ and\
  \bibinfo {author} {\bibfnamefont {N.}~\bibnamefont {Magnoli}},\ }\href@noop
  {} {\bibfield  {journal} {\bibinfo  {journal} {New Journal of Physics}\
  }\textbf {\bibinfo {volume} {14}},\ \bibinfo {pages} {113014} (\bibinfo
  {year} {2012})}\BibitemShut {NoStop}%
\bibitem [{\citenamefont {Moradi}\ and\ \citenamefont
  {Wen}(2015)}]{moradi2015universal}%
  \BibitemOpen
  \bibfield  {author} {\bibinfo {author} {\bibfnamefont {H.}~\bibnamefont
  {Moradi}}\ and\ \bibinfo {author} {\bibfnamefont {X.-G.}\ \bibnamefont
  {Wen}},\ }\href@noop {} {\bibfield  {journal} {\bibinfo  {journal} {Physical
  Review B}\ }\textbf {\bibinfo {volume} {91}},\ \bibinfo {pages} {075114}
  (\bibinfo {year} {2015})}\BibitemShut {NoStop}%
\bibitem [{Note2()}]{Note2}%
  \BibitemOpen
  \bibinfo {note} {We thank Xueda Wen for clarifying this picture}\BibitemShut
  {NoStop}%
\bibitem [{\citenamefont {Metlitski}\ \emph {et~al.}(2015)\citenamefont
  {Metlitski}, \citenamefont {Kane},\ and\ \citenamefont
  {Fisher}}]{metlitski2015symmetry}%
  \BibitemOpen
  \bibfield  {author} {\bibinfo {author} {\bibfnamefont {M.~A.}\ \bibnamefont
  {Metlitski}}, \bibinfo {author} {\bibfnamefont {C.}~\bibnamefont {Kane}}, \
  and\ \bibinfo {author} {\bibfnamefont {M.~P.}\ \bibnamefont {Fisher}},\
  }\href@noop {} {\bibfield  {journal} {\bibinfo  {journal} {Physical Review
  B}\ }\textbf {\bibinfo {volume} {92}},\ \bibinfo {pages} {125111} (\bibinfo
  {year} {2015})}\BibitemShut {NoStop}%
\bibitem [{\citenamefont {Wang}\ \emph {et~al.}(2013)\citenamefont {Wang},
  \citenamefont {Potter},\ and\ \citenamefont {Senthil}}]{wang2013gapped}%
  \BibitemOpen
  \bibfield  {author} {\bibinfo {author} {\bibfnamefont {C.}~\bibnamefont
  {Wang}}, \bibinfo {author} {\bibfnamefont {A.~C.}\ \bibnamefont {Potter}}, \
  and\ \bibinfo {author} {\bibfnamefont {T.}~\bibnamefont {Senthil}},\
  }\href@noop {} {\bibfield  {journal} {\bibinfo  {journal} {Physical Review
  B}\ }\textbf {\bibinfo {volume} {88}},\ \bibinfo {pages} {115137} (\bibinfo
  {year} {2013})}\BibitemShut {NoStop}%
\bibitem [{\citenamefont {Bonderson}\ \emph {et~al.}(2013)\citenamefont
  {Bonderson}, \citenamefont {Nayak},\ and\ \citenamefont
  {Qi}}]{bonderson2013time}%
  \BibitemOpen
  \bibfield  {author} {\bibinfo {author} {\bibfnamefont {P.}~\bibnamefont
  {Bonderson}}, \bibinfo {author} {\bibfnamefont {C.}~\bibnamefont {Nayak}}, \
  and\ \bibinfo {author} {\bibfnamefont {X.-L.}\ \bibnamefont {Qi}},\
  }\href@noop {} {\bibfield  {journal} {\bibinfo  {journal} {Journal of
  Statistical Mechanics: Theory and Experiment}\ }\textbf {\bibinfo {volume}
  {2013}},\ \bibinfo {pages} {P09016} (\bibinfo {year} {2013})}\BibitemShut
  {NoStop}%
\bibitem [{\citenamefont {Wang}\ and\ \citenamefont
  {Levin}(2016)}]{wang2016anomaly}%
  \BibitemOpen
  \bibfield  {author} {\bibinfo {author} {\bibfnamefont {C.}~\bibnamefont
  {Wang}}\ and\ \bibinfo {author} {\bibfnamefont {M.}~\bibnamefont {Levin}},\
  }\href@noop {} {\bibfield  {journal} {\bibinfo  {journal} {arXiv preprint
  arXiv:1610.04624}\ } (\bibinfo {year} {2016})}\BibitemShut {NoStop}%
\bibitem [{\citenamefont {Kapustin}\ and\ \citenamefont
  {Thorngren}(2017{\natexlab{b}})}]{kapustin2017higher}%
  \BibitemOpen
  \bibfield  {author} {\bibinfo {author} {\bibfnamefont {A.}~\bibnamefont
  {Kapustin}}\ and\ \bibinfo {author} {\bibfnamefont {R.}~\bibnamefont
  {Thorngren}},\ }in\ \href@noop {} {\emph {\bibinfo {booktitle} {Algebra,
  Geometry, and Physics in the 21st Century}}}\ (\bibinfo  {publisher}
  {Springer},\ \bibinfo {year} {2017})\ pp.\ \bibinfo {pages}
  {177--202}\BibitemShut {NoStop}%
\bibitem [{\citenamefont {Else}\ and\ \citenamefont
  {Nayak}(2016)}]{else2016classification}%
  \BibitemOpen
  \bibfield  {author} {\bibinfo {author} {\bibfnamefont {D.~V.}\ \bibnamefont
  {Else}}\ and\ \bibinfo {author} {\bibfnamefont {C.}~\bibnamefont {Nayak}},\
  }\href@noop {} {\bibfield  {journal} {\bibinfo  {journal} {Physical Review
  B}\ }\textbf {\bibinfo {volume} {93}},\ \bibinfo {pages} {201103} (\bibinfo
  {year} {2016})}\BibitemShut {NoStop}%
\bibitem [{\citenamefont {Potter}\ and\ \citenamefont
  {Morimoto}(2017)}]{potter2017dynamically}%
  \BibitemOpen
  \bibfield  {author} {\bibinfo {author} {\bibfnamefont {A.~C.}\ \bibnamefont
  {Potter}}\ and\ \bibinfo {author} {\bibfnamefont {T.}~\bibnamefont
  {Morimoto}},\ }\href@noop {} {\bibfield  {journal} {\bibinfo  {journal}
  {Physical Review B}\ }\textbf {\bibinfo {volume} {95}},\ \bibinfo {pages}
  {155126} (\bibinfo {year} {2017})}\BibitemShut {NoStop}%
\bibitem [{\citenamefont {Potter}\ \emph {et~al.}(2016)\citenamefont {Potter},
  \citenamefont {Morimoto},\ and\ \citenamefont
  {Vishwanath}}]{potter2016classification}%
  \BibitemOpen
  \bibfield  {author} {\bibinfo {author} {\bibfnamefont {A.~C.}\ \bibnamefont
  {Potter}}, \bibinfo {author} {\bibfnamefont {T.}~\bibnamefont {Morimoto}}, \
  and\ \bibinfo {author} {\bibfnamefont {A.}~\bibnamefont {Vishwanath}},\
  }\href@noop {} {\bibfield  {journal} {\bibinfo  {journal} {Physical Review
  X}\ }\textbf {\bibinfo {volume} {6}},\ \bibinfo {pages} {041001} (\bibinfo
  {year} {2016})}\BibitemShut {NoStop}%
\bibitem [{\citenamefont {Po}\ \emph {et~al.}(2016)\citenamefont {Po},
  \citenamefont {Fidkowski}, \citenamefont {Morimoto}, \citenamefont {Potter},\
  and\ \citenamefont {Vishwanath}}]{po2016chiral}%
  \BibitemOpen
  \bibfield  {author} {\bibinfo {author} {\bibfnamefont {H.~C.}\ \bibnamefont
  {Po}}, \bibinfo {author} {\bibfnamefont {L.}~\bibnamefont {Fidkowski}},
  \bibinfo {author} {\bibfnamefont {T.}~\bibnamefont {Morimoto}}, \bibinfo
  {author} {\bibfnamefont {A.~C.}\ \bibnamefont {Potter}}, \ and\ \bibinfo
  {author} {\bibfnamefont {A.}~\bibnamefont {Vishwanath}},\ }\href@noop {}
  {\bibfield  {journal} {\bibinfo  {journal} {Physical Review X}\ }\textbf
  {\bibinfo {volume} {6}},\ \bibinfo {pages} {041070} (\bibinfo {year}
  {2016})}\BibitemShut {NoStop}%
\end{thebibliography}%

\end{document}